\RequirePackage{etex}
\documentclass[seceqn,secthm]{elsart}

\usepackage{mathrsfs, epsfig}
\usepackage{dcpic, pictex, pinlabel, rotating} 
\usepackage{array, mathdots, bbm, faktor}

\usepackage{latexsym}
\usepackage{amssymb}
\usepackage{amsmath}
\usepackage{stmaryrd}
\usepackage{graphicx}
\usepackage{pst-all} 
\usepackage{times}

\numberwithin{equation}{section}
\numberwithin{figure}{section}
\numberwithin{table}{section}

\journal{arXiv}

\begin{document}

\begin{frontmatter}

\title{Twisted Interferometry}

\author[StationQ]{Parsa Bonderson},
\author[StationQ]{Lukasz Fidkowski},
\author[StationQ]{Michael Freedman},
\author[StationQ]{Kevin Walker}
\address[StationQ]{Station Q, Microsoft Research, Santa Barbara, California 93106-6105, USA}

\begin{abstract}
We propose and analyze the effect of anyonic interferometers that are designed such that the probe anyons traveling in a given path through the interferometer twist or braid around each other. These ``twisted'' interferometers are found to provide operational utility that may not be available from anyon braiding operations and standard (untwisted) anyonic interferometry measurements. In particular, it enables Ising anyons to generate ``magic states,'' which can be used to implement $\pi/8$-phase gates. We consider the possible implementations of such twisted interferometers in quantum Hall systems, 2D $p_x+ip_y$ superconductors, and 2D Majorana heterostructures, and discuss obstacles and challenges associated with implementation.
\end{abstract}

\date{10 June 2013}

\end{frontmatter}


\section{Introduction}
\label{sec:introduction}

Non-Abelian topological phases of matter support quasiparticle excitations with exotic exchange properties that may be described by non-Abelian braiding statistics~\cite{Leinaas77,Goldin85,Fredenhagen89,Imbo89,Froehlich90,Imbo90,Bais92}. These non-Abelian anyonic quasiparticles possess a degenerate topological state space that is nonlocal. Exchanging quasiparticles acts upon this space via (possibly non-commuting) multi-dimensional representations of the braid group. The nonlocality of the topological state space makes it essentially immune to local perturbations and the exchange transformations acting upon it exact. This gives non-Abelian topological phases great potential for providing an intrinsically fault-tolerant platform for quantum information processing~\cite{Kitaev03,Freedman98,Preskill98,Freedman02a,Freedman02b,Freedman03b,Preskill-lectures,Nayak08}.

Measurement of topological charge is an important primitive operation necessary for topological quantum information processing. This allows one to measure the topological state and hence provide readout of topologically encoded information. It can also be used to generate the anyonic braiding transformations without actually moving the computational quasiparticles~\cite{Bonderson08a,Bonderson08b}. Anyonic interferometry~\cite{Overbosch01,Bonderson07b,Bonderson07c} is a particular form of topological charge measurement that has been proposed for topological systems. Its ability to non-locally and non-demolitionally measure the collective anyonic charge of a group of (non-Abelian) anyons, without decohering their internal state, makes it more powerful than standard local measurements of topological charge. In particular, anyonic interferometry can also be used to generate entangling gates~\cite{Bravyi00,Bravyi06,BondersonWIP} and change between different qubit encodings~\cite{BondersonWIP}. There have been many experimental proposals~\cite{Chamon97,Fradkin98,DasSarma05,Stern06a,Bonderson06a,Bonderson06b,Fidkowski07c,Ardonne08a,Bishara08a,Bishara09,Akhmerov09a,Fu09a,Bonderson11b,Grosfeld11a} to realize and utilize anyonic interferometers, as well as efforts to physically implement them~\cite{Camino05a,Willett09a,Willett09b,McClure12,An11}. Despite the focus anyonic interferometry has received, there is still more to be learned about its potential capabilities.

In this paper, we propose and analyze a novel implementation of anyonic interferometry that we call ``twisted interferometry,'' which can provide capabilities that go beyond that of its standard \emph{untwisted} counterpart. The basic idea of the twisted interferometer is to modify the design of an anyonic interferometer so that the probe anyons traveling in a given path through the interferometer will twist or braid around each other.

The inspiration for twisted interferometry was a series of ideas~\cite{Bravyi00-unpublished,Freedman06a,FNW05b,Bonderson10} going back to the unpublished work of Bravyi and Kitaev for generating topologically protected operations, such as the $\pi / 8$-gate, for a system of Ising non-Abelian anyons using the concept of Dehn surgery on $3$-manifolds.
Indeed, the primary practical motivation for studying twisted interferometry is that it could be used with Ising anyons to generate ``magic states,'' as we will demonstrate. This is significant because, if one only has the ability to perform braiding operations and untwisted anyonic interferometry measurements for Ising anyons, then one can only generate the Clifford group operations, which is not computationally universal and, in fact, can be efficiently simulated on a classical computer~\cite{Gottesman98}. However, if one supplements these operations with magic states, then one can also generate $\pi / 8$-phase gates, which results in a computationally universal gate set~\cite{Boykin99}.

The organization of this paper is as follows. In Section~\ref{sec:review}, we review the tensor category analysis of untwisted anyonic interferometers, following~\cite{Bonderson07b,Bonderson07c}. In Section~\ref{sec:omega_loops}, we introduce $\omega$-loops and their properties which are useful for our analysis. In Section~\ref{sec:twisted_interferometers}, we introduce and analyze twisted interferometers. In Section~\ref{sec:topological_descriptions}, we briefly discuss the topological formulation of (twisted and untwisted) anyonic interferometry in terms of Dehn surgery of $3$-manifolds. This approach is considered in greater detail in a companion paper~\cite{Bonderson13c}. In Section~\ref{sec:Ising}, we explicitly apply our twisted interferometry results to Ising anyons and demonstrate how twisted interferometers can be used to generate magic states. In Section~\ref{sec:fake_twist}, we notice, as an aside inspired by twisted interferometry, that it is possible to use ``partial interferometry,'' i.e. running an (untwisted) interferometer for a fixed number of probes $N$, to generate magic states for Ising anyons, albeit in a topologically unprotected manner. In Section~\ref{sec:engineer_twist}, we consider the possible implementation of twisted interferometers in physical systems, such as quantum Hall states, 2D $p_x+ip_y$ topological superconductors, and 2D Majorana heterostructures, and discuss the significant obstacles and challenges associated with implementation.

\section{Review of Anyonic Interferometers}
\label{sec:review}

Before describing twisted interferometry, it may be useful to review the effects and analysis of ``ordinary'' (untwisted) anyonic interferometry, following and borrowing heavily from~\cite{Bonderson07b,Bonderson07c}. We focus on an idealized Mach-Zehnder type interferometer~\cite{Zehnder1891,Mach1892} for quasiparticles with non-Abelian anyonic braiding statistics. This will serve as a model for realistic interferometry experiments with anyons,
and the methods used in this analysis readily apply to other classes of
interferometers, e.g. (Fabrey-P\'{e}rot) fractional quantum Hall double point-contact interferometers in the weak tunneling limit.

For our analysis, we abstract to an idealized system that supports an arbitrary anyon model, a.k.a. unitary braided tensor category (UBTC), and also allows for a number of desired
manipulations to be effected. Specifically, we posit the experimental abilities to: (1) produce,
isolate, and position desired anyons, (2) provide anyons with some manner of
propulsion to produce a beam of probe anyons, (3) construct lossless
beam-splitters and mirrors, and (4) detect the presence of a probe anyon at the
output legs of the interferometer.

\begin{figure}[t!]
\begin{center}
  \includegraphics[scale=0.6]{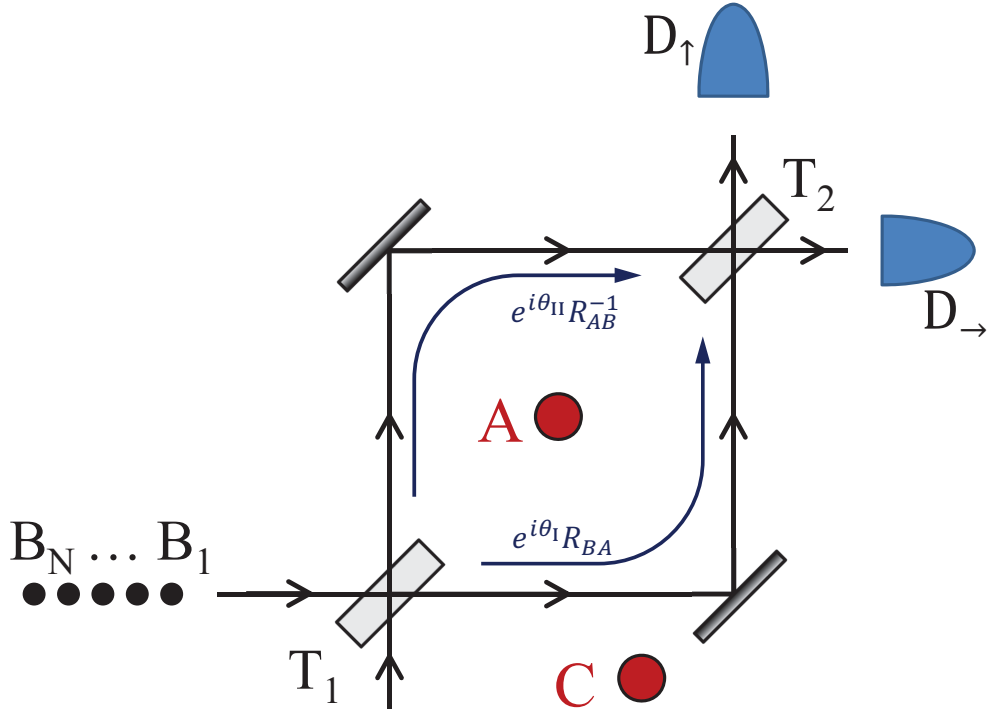}
  \caption{An idealized Mach-Zehnder interferometer for an anyonic system, where $T_{j}$ are beam splitters. The target anyons (collectively denoted $A$) in the central region share entanglement only with the anyon(s) $C$ outside this region. A beam of probe anyons $B_{1},\ldots ,B_{N}$ is sent through the interferometer and detected at one of the two possible outputs by $D_{s}$.}
  \label{fig:interferometer}
\end{center}
\end{figure}

The experimental setup for the anyonic Mach-Zehnder interferometer is shown in Fig.~\ref{fig:interferometer}. The target anyon $A$ is the composite of all the anyons $A_{1},A_{2},\ldots $ located inside the central interferometry region, and so may be in a superposition of states with different total anyonic charges. Since these anyons are treated collectively by the experiment, we ignore their individuality (and possible internal states) and consider them as a single anyon $A$ capable of existing in superposition of different anyonic charges. Anyons outside of the central interferometry region with which the target anyons share entanglement are denoted $C$ and will be similarly treated collectively. (Fig.~\ref{fig:interferometer} only shows such charges below the interferometer, but such anyons may be distributed more generally, in which case it is useful to consider them in groups for each distinct region.)
We will similarly allow the probe anyons, $B_{1},\ldots ,B_{N}$ to be treated as capable of charge superposition (though, for most cases of interest, we can usually restrict our attention to identical probes with a definite value of anyonic charge). The probe anyons are sent as a beam
into the interferometer through two possible input channels. They pass through a
beam splitter $T_{1}$, are reflected by mirrors around the central target
region, pass through a second beam splitter $T_{2}$, and then are detected at
one of the two possible output channels by the detectors $D_{s}$. When a probe
anyon $B$ passes through the bottom path of the interferometer, the state
acquires the phase $e^{i\theta_{\text{I}}}$, which results from background
Aharonov-Bohm interactions~\cite{Aharonov59}, path length differences, phase
shifters, etc., and is also acted upon by the braiding operator $R_{BA}$, which
is strictly due to the braiding statistics between the probe and target anyons.
Similarly, when the probe passes through the top path of the interferometer, the
state acquires the phase $e^{i\theta _{\text{II}}}$ and is acted on by
$R_{AB}^{-1}$.

The effect of running such an interferometry will generally depend on the probe anyons and the variables of the device. However, we can summarize the effect for $N \rightarrow \infty$ identical probes as follows~\cite{Bonderson07b,Bonderson07c}:

\noindent (1) The collective charge of the target anyons $A$ will be projected onto one of the subsets of charge types that the probes are able to distinguish from other subsets via monodromy. (When the probes are able to distinguish between all anyonic charge types, these subsets each contain only one anyonic charge type and projection is onto a state where $A$ has a definite collective charge value.)

\noindent (2) Anyonic entanglement (encoded via connecting anyonic charge lines) across the paths of the probe anyons, i.e. between the target anyons $A$ and other anyons outside the interferometer $C$, will be decohered if the probe anyons are able to detect the charges encoding such entanglement~\cite{Bonderson07a}. (When the probes are able to distinguish between all anyonic charge types, they can detect all charges, except the trivial vacuum charge, so the final state will have no charge lines connecting anyons in the interior of the interferometer with anyons in the exterior.)

\noindent We describe the precise meaning of these statement in more mathematical detail in the following sections.

\subsection{Tensor Category Analysis}
\label{sec:UMTC_analysis}

In this section, we review the Mach-Zehnder interferometer in a general anyonic context. (We refer the reader to~\cite{Bonderson07b,Bonderson07c} for additional details and background on the UBTC formalism used in the analysis.)

\begin{figure}[t!]
\begin{center}
  \includegraphics[scale=0.6]{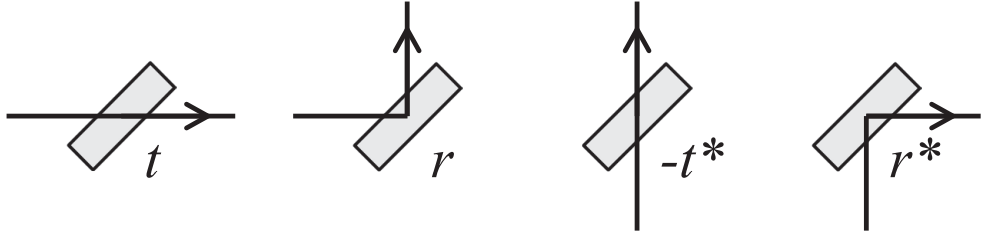}
  \caption{The transmission and reflection coefficients for a beam splitter.}
  \label{fig:splitters}
\end{center}
\end{figure}

Using the two-component vector notation%
\begin{equation}
\left(
\begin{array}{c}
1 \\
0%
\end{array}%
\right) =\left| \shortrightarrow \right\rangle ,\quad \left(
\begin{array}{c}
0 \\
1%
\end{array}%
\right) =\left| \shortuparrow \right\rangle
\end{equation}%
to indicate the direction (horizontal or vertical) a probe anyon is
traveling through the interferometer at any point, the lossless beam
splitters (see Fig.~\ref{fig:splitters}) are represented by%
\begin{equation}
T_{j}=\left[
\begin{array}{cc}
t_{j} & r_{j}^{\ast } \\
r_{j} & -t_{j}^{\ast }%
\end{array}%
\right]
\end{equation}%
(for $j=1,2$), where $\left| t_{j}\right| ^{2}+\left| r_{j}\right| ^{2}=1$.
We note that these matrices could be multiplied by
overall phases without affecting any of the results, since such phases are
not distinguished by the two paths.

The unitary operator representing a probe anyon passing through the
interferometer is given by%
\begin{equation}
U = T_{2}\Sigma T_{1}
\end{equation}
\begin{equation}
\Sigma = \left[
\begin{array}{cc}
0 & e^{i\theta _{\text{II}}}R_{AB}^{-1} \\
e^{i\theta _{\text{I}}}R_{BA} & 0%
\end{array}%
\right] .
\end{equation}%
This can be written diagrammatically as%
\begin{equation}
\label{eq:Udiag}
\pspicture[shift=-0.4](-0.2,0)(1,1)
\rput[tl](0,0){$B_{s'}$}
\rput[tl](0,1.2){$A$}
\rput[tl](0.85,1.2){$B_{s}$}
\rput[tl](0.85,0){$A$}
 \psline[linewidth=0.9pt](0.4,0)(0.4,0.24)
 \psline[linewidth=0.9pt](0.78,0)(0.78,0.24)
 \psline[linewidth=0.9pt](0.4,0.78)(0.4,1)
 \psline[linewidth=0.9pt](0.78,0.78)(0.78,1)
\rput[bl](0.29,0.24){\psframebox{$U$}}
  \endpspicture
=e^{i\theta _{\text{I}}}\left[
\begin{array}{cc}
t_{1}r_{2}^{\ast } & r_{1}^{\ast }r_{2}^{\ast } \\
-t_{1}t_{2}^{\ast } & -r_{1}^{\ast }t_{2}^{\ast }
\end{array}
\right]_{s,s'}
\pspicture[shift=-0.4](-0.2,0)(1.25,1)
  \psset{linewidth=0.9pt,linecolor=black,arrowscale=1.5,arrowinset=0.15}
  \psline(0.92,0.1)(0.2,1)
  \psline{->}(0.92,0.1)(0.28,0.9)
  \psline(0.28,0.1)(1,1)
  \psline[border=2pt]{->}(0.28,0.1)(0.92,0.9)
  \rput[tl]{0}(-0.2,0.2){$B$}
  \rput[tr]{0}(1.25,0.2){$A$}
  \endpspicture
+e^{i\theta _{\text{II}}}\left[
\begin{array}{cc}
r_{1}t_{2} & -t_{1}^{\ast }t_{2} \\
r_{1}r_{2} & -t_{1}^{\ast }r_{2}
\end{array}
\right]_{s,s'}
\pspicture[shift=-0.4](-0.2,0)(1.25,1)
  \psset{linewidth=0.9pt,linecolor=black,arrowscale=1.5,arrowinset=0.15}
  \psline{->}(0.28,0.1)(0.92,0.9)
  \psline(0.28,0.1)(1,1)
  \psline(0.92,0.1)(0.2,1)
  \psline[border=2pt]{->}(0.92,0.1)(0.28,0.9)
  \rput[tl]{0}(-0.2,0.2){$B$}
  \rput[tr]{0}(1.25,0.2){$A$}
  \endpspicture
,
\end{equation}
where we introduce the notation of writing the directional index $s$ of the probe particle as a subscript on its anyonic charge label, e.g. $b_s$.

The position of the anyon $C$ with respect to the other anyons must be
specified, and we will take it to be located below the central
interferometry region and slightly to the right of $A$. (The specification
``slightly to the right'' merely indicates how the diagrams are to be drawn, and
has no physical consequence.) For this choice of positioning, the operator%
\begin{equation}
V=\left[
\begin{array}{cc}
R_{CB}^{-1} & 0 \\
0 & R_{CB}^{-1}%
\end{array}%
\right]
=
\pspicture[shift=-0.4](-0.2,0)(1.25,1)
  \psset{linewidth=0.9pt,linecolor=black,arrowscale=1.5,arrowinset=0.15}
  \psline{->}(0.28,0.1)(0.92,0.9)
  \psline(0.28,0.1)(1,1)
  \psline(0.92,0.1)(0.2,1)
  \psline[border=2pt]{->}(0.92,0.1)(0.28,0.9)
  \rput[tl]{0}(-0.2,0.2){$B$}
  \rput[tr]{0}(1.25,0.2){$C$}
  \endpspicture
\end{equation}%
represents the braiding of $C$ with the probe. We will later discuss the generalizations where the $C$ anyons are located above or both above and below the central interferometry region.

After a probe anyon $B$ passes through the interferometer, it is measured at one of the two detectors and the state undergoes the usual orthogonal measurement collapse with a projection $\Pi _{s}=\left| s\right\rangle \left\langle s\right| $ for the outcome $s = \shortrightarrow$ or $\shortuparrow$.
After the detection of the probe anyon, it no longer interests us, and we remove it from the vicinity of the target anyon system, tracing it out of the post-measurement state. For an initial state $\rho$ of the system (including probe anyons), the state after probe $B$ passes through the interferometry, is measured at detector $D_s$, and is traced out is given by
\begin{equation}
\rho^{\prime} = \frac{1}{\Pr \left( s\right) } \widetilde{\text{Tr}}_{B} \left[ \Pi _{s}  VU  \rho  U^{\dagger }V^{\dagger} \Pi _{s} \right]
,
\label{eq:rhoA_final_measurment_projection}
\end{equation}
where
\begin{equation}
\Pr \left( s\right)  = \widetilde{\text{Tr}}\left[ \Pi _{s}  VU  \rho  U^{\dagger }V^{\dagger} \right]
\end{equation}
is the probability of the measurement having outcome $s$, and the tilde over the traces indicates the of the ``quantum trace'' for anyonic states (defined by diagrammatically by connecting outgoing and incoming lines representing the anyon being traced out).

When considering operations involving non-Abelian anyons, it is important to
keep track of all other anyons with which there is non-trivial entanglement.
Indeed, if these additional particles are not tracked or are
physically inaccessible, one should trace them out of the system, forgoing
the ability to use them to form coherent superpositions of anyonic charge.

We assume that each probe anyon is initially unentangled and sent into the interferometer through the horizontal leg
$s=\shortrightarrow$. In particular, it does not share entanglement with the $A$ or $C$ anyons, nor with the other probe anyons. (This can be be arranged by independently drawing each one from the vacuum together with an antiparticle which is then discarded and traced out.) With this assumption, we can treat the probes as identical quasiparticles, each of which is described by the density matrix
\begin{equation}
\rho ^{B} = \sum\limits_{b} \rho _{b_{\shortrightarrow}}^{B} \frac{1}{d_b} \left| b_{\shortrightarrow} \right\rangle
\left\langle b_{\shortrightarrow} \right|
= \sum\limits_{b} \rho _{b_{\shortrightarrow}}^{B}
\frac{1}{d_{b}}
\pspicture[shift=-1.02](0.15,-1.15)(0.9,1.15)
  \small
  \psset{linewidth=0.9pt,linecolor=black,arrowscale=1.5,arrowinset=0.15}
  \psline(0.35,-0.775)(0.35,0.775)
  \psline{->}(0.35,-0.3)(0.35,0.2)
  \rput[bl]{0}(0.5,-0.1){$b_\shortrightarrow$}
  \endpspicture
,
\end{equation}
where  $\Pr\nolimits_{B} \left( b\right) = \rho _{b_{\shortrightarrow}}^{B}$ is the probability that the probe anyon has charge $b$. We note that the factors of $d_b$, the quantum dimension of anyon charge $b$, appears as normalizing factors in anyonic density matrix.

The target system involves the target anyon $A$ and the anyon $C$ which is
the only one entangled with $A$ that is kept physically accessible. Recall
that these anyons may really represent multiple quasiparticles that are
being treated collectively, but as long as we are not interested in
operations involving the individual quasiparticles, they can be treated as a
single anyon which is allowed to have superpositions of different charge values.
Thus, the density matrix of the target system is%
\begin{eqnarray}
\rho ^{AC} &=&
 \pspicture[shift=-1.4](-1.3,-1.5)(1.1,1.5)
  \small
  \psframe[linewidth=0.9pt,linecolor=black,border=0](-0.75,-0.5)(0.75,0.5)
  \rput[bl]{0}(-0.3,-0.15){$\rho^{AC}$}
  \psset{linewidth=0.9pt,linecolor=black,arrowscale=1.5,arrowinset=0.15}
  \psline(0.5,0.5)(0.5,1.25)
  \psline(-0.5,0.5)(-0.5,1.25)
  \psline(0.5,-0.5)(0.5,-1.25)
  \psline(-0.5,-0.5)(-0.5,-1.25)
  \psline{->}(0.5,0.5)(0.5,1)
  \psline{->}(-0.5,0.5)(-0.5,1)
  \psline{-<}(0.5,-0.5)(0.5,-1)
  \psline{-<}(-0.5,-0.5)(-0.5,-1)
  \rput[bl](0.7,0.95){$C$}
  \rput[bl](-1,0.95){$A$}
  \rput[bl](0.7,-1.2){$C^{\prime}$}
  \rput[bl](-1,-1.2){$A^{\prime}$}
 \endpspicture
=\sum\limits_{a,a^{\prime },c,c^{\prime },f,\mu ,\mu ^{\prime }}
\frac{ \rho_{\left( a,c;f,\mu \right) \left( a^{\prime },c^{\prime
};f,\mu^{\prime}\right) }^{AC} }{ \left( d_{a}d_{a^{\prime
}}d_{c}d_{c^{\prime}}d_{f}^{2} \right) ^{1/4} }
 \pspicture[shift=-0.6](-0.15,-0.45)(1.5,1)
 \small
  \psset{linewidth=0.9pt,linecolor=black,arrowscale=1.5,arrowinset=0.15}
  \psline{->}(0.7,0)(0.7,0.45)
  \psline(0.7,0)(0.7,0.55)
  \psline(0.7,0.55) (0.25,1)
  \psline{->}(0.7,0.55)(0.3,0.95)
  \psline(0.7,0.55) (1.15,1)
  \psline{->}(0.7,0.55)(1.1,0.95)
  \rput[bl]{0}(0.38,0.1){$f$}
  \rput[bl]{0}(1.22,0.8){$c$}
  \rput[bl]{0}(-0.05,0.8){$a$}
  \psline(0.7,0) (0.25,-0.45)
  \psline{-<}(0.7,0)(0.35,-0.35)
  \psline(0.7,0) (1.15,-0.45)
  \psline{-<}(0.7,0)(1.05,-0.35)
  \rput[bl]{0}(1.22,-0.4){$c'$}
  \rput[bl]{0}(-0.05,-0.4){$a'$}
  \scriptsize
  \rput[bl]{0}(0.82,0.38){$\mu$}
  \rput[bl]{0}(0.82,-0.02){$\mu'$}
  \endpspicture
\notag \\
&=& \sum\limits_{a,a^{\prime },c,c^{\prime },f,\mu ,\mu ^{\prime
}}\rho _{\left( a,c;f,\mu \right) \left( a^{\prime },c^{\prime };f,\mu
^{\prime }\right) }^{AC} \frac{1}{d_f} \left| a,c;f,\mu \right\rangle
\left\langle a^{\prime},c^{\prime };f,\mu ^{\prime }\right|
.
\end{eqnarray}

The initial density matrix of the combined target and $N$ unentangled probes system is
\begin{equation}
\rho = \rho^{B_1} \otimes \ldots \otimes \rho^{B_N} \otimes \rho^{AC}
.
\end{equation}
When the probe anyons that are initially unentangled, we can obtain their effect on the target system by considering the effect of each probe individually and iterating the process. Thus, it is straightforward to obtain the many probe results from the single probe analysis.

\subsubsection{Single Probe}
\label{sec:single_probe}

The details of the single probe analysis will help clarify aspects of the twisted interferometry analysis, so we consider it in detail here.
The effect on the target system of a single probe passing through the interferometer and being measured at detector $D_s$ is given by the map
\begin{equation}
\rho^{AC} \mapsto \rho^{AC}\left( s \right) = \frac{1}{\Pr \left( s\right) } \widetilde{\text{Tr}}_{B} \left[ \Pi _{s}  VU  \left( \rho^{B} \otimes \rho^{AC} \right)  U^{\dagger }V^{\dagger} \Pi _{s} \right]
.
\label{eq:rhoA_single_probe}
\end{equation}
To determine the result, we evaluate the corresponding diagram for a single probe with definite anyonic charge $b$ acting upon a specific basis element of the target system, given by
\begin{equation}
\psscalebox{1}{
 \pspicture[shift=-3](0.2,-3.1)(4.0,3.2)
  \small
  \psframe[linewidth=0.9pt,linecolor=black,border=0](0.6,0.9)(1.7,1.4)
  \psframe[linewidth=0.9pt,linecolor=black,border=0](0.6,-0.9)(1.7,-1.4)
  \rput[bl]{0}(1.0,1.01){$U$}
  \rput[tl]{0}(0.95,-0.97){$U^{\dag}$}
  \psset{linewidth=0.9pt,linecolor=black,arrowscale=1.5,arrowinset=0.15}
  \psline(0.8,-1.4)(0.8,-2.8)
  \psline(0.8,1.4)(0.8,2.8)
  \psline(0.8,0.9)(0.8,-0.9)
  \psarc[linewidth=0.9pt,linecolor=black,border=0pt] (1.5,0.9){0.7}{-60}{0}
  \psarc[linewidth=0.9pt,linecolor=black,arrows=->,arrowscale=1.4,
    arrowinset=0.15] (1.5,0.9){0.7}{-60}{-10}
  \psarc[linewidth=0.9pt,linecolor=black,border=0pt]
(2.2,0.9){0.7}{180}{240}
  \psarc[linewidth=0.9pt,linecolor=black,arrows=<-,arrowscale=1.4,
    arrowinset=0.15] (2.2,0.9){0.7}{190}{240}
  \psarc[linewidth=0.9pt,linecolor=black,border=0pt](1.5,-0.9){0.7}{0}{60}
  \psarc[linewidth=0.9pt,linecolor=black,arrows=->,arrowscale=1.4,
    arrowinset=0.15] (1.5,-0.9){0.7}{0}{35}
  \psarc[linewidth=0.9pt,linecolor=black,border=0pt]
(2.2,-0.9){0.7}{120}{180}
  \psarc[linewidth=0.9pt,linecolor=black,arrows=<-,arrowscale=1.4,
    arrowinset=0.15] (2.2,-0.9){0.7}{145}{180}
  \psarc[linewidth=0.9pt,linecolor=black,border=0pt] (1.9,1.3){0.4}{90}{170}
  \psarc[linewidth=0.9pt,linecolor=black,border=0pt] (1.9,-1.3){0.4}{190}{270}
  \psline(1.9,1.7)(2.5,1.7)
  \psline(1.9,-1.7)(2.5,-1.7)
  \psarc[linewidth=0.9pt,linecolor=black,border=0pt](2.5,2.1){0.4}{-90}{0}
  \psarc[linewidth=0.9pt,linecolor=black,border=0pt](2.5,-2.1){0.4}{0}{90}
  \psframe[linewidth=0.9pt,linecolor=black,border=0](2.6,2.1)(3.2,2.6)
  \psframe[linewidth=0.9pt,linecolor=black,border=0](2.6,-2.1)(3.2,-2.6)
  \rput[bl]{0}(2.7,2.18){$\Pi_s$}
  \rput[bl]{0}(2.7,-2.52){$\Pi_s$}
  \psarc[linewidth=0.9pt,linecolor=black,border=0pt](3.3,2.6){0.4}{0}{180}
  \psarc[linewidth=0.9pt,linecolor=black,border=0pt](3.3,-2.6){0.4}{-180}{0}
  \psline(3.7,2.6)(3.7,-2.6)
  \psline(1.85,-0.3)(1.85,0.3)
  \psline[border=2pt](2.2,-0.9)(2.2,-2.8)
  \psline[border=2pt](2.2,0.9)(2.2,2.8)
  \psline{->}(0.8,-2.8)(0.8,-2.1)
  \psline{->}(0.8,-0.9)(0.8,0.12)
  \psline{->}(0.8,1.4)(0.8,2.4)
  \psline{->}(2.2,-2.8)(2.2,-2.1)
  \psline{->}(2.2,1.4)(2.2,2.4)
  \psline{->}(3.7,0.4)(3.7,-0.1)
  \psline{->}(1.9,1.69)(2.02,1.71)
  \psline{<-}(1.8,-1.69)(2.02,-1.7)
  \psline{->}(1.85,-0.3)(1.85,0.12)
  \rput[bl]{0}(0.7,2.9){$a$}
  \rput[tl]{0}(0.7,-2.8){$a'$}
  \rput[bl]{0}(2.1,2.9){$c$}
  \rput[tl]{0}(2.1,-2.8){$c'$}
  \rput[bl]{0}(0.3,-0.1){$b_{\shortrightarrow}$}
  \rput[bl]{0}(3.85,-0.1){$b_{s}$}
  \rput[bl]{0}(1.55,-0.2){$f$}
    \rput[bl]{0}(1.27,0.4){$a$}
  \rput[bl]{0}(1.23,-0.6){$a'$}
  \scriptsize
  \rput[bl]{0}(1.76,0.5){$\mu$}
  \rput[bl]{0}(1.71,-0.7){$\mu'$}
 \endpspicture
}
\end{equation}%

For the outcome $s=\shortrightarrow $, this is%
\begin{eqnarray}
&&
 \pspicture[shift=-1.95](0.3,-2.1)(2.85,1.9)
  \small
  \psframe[linewidth=0.9pt,linecolor=black,border=0](0.8,0.7)(1.7,1.2)
  \psframe[linewidth=0.9pt,linecolor=black,border=0](0.8,-0.7)(1.7,-1.2)
  \rput[bl]{0}(1.1,0.81){$U$}
  \rput[tl]{0}(1.05,-0.77){$U^{\dag}$}
  \psset{linewidth=0.9pt,linecolor=black,arrowscale=1.5,arrowinset=0.15}
  \psline(1,-0.7)(1,0.7)
  \psline(1,-1.2)(1,-1.7)
  \psline(1,1.2)(1,1.7)
  \psarc[linewidth=0.9pt,linecolor=black,border=0pt] (1.8,-0.6){0.3}{0}{180}
  \psarc[linewidth=0.9pt,linecolor=black,border=0pt] (1.8,0.6){0.3}{180}{360}
  \psarc[linewidth=0.9pt,linecolor=black,border=0pt] (1.9,1.1){0.4}{0}{170}
  \psarc[linewidth=0.9pt,linecolor=black,border=0pt] (1.9,-1.1){0.4}{190}{360}
  \psline(1.8,-0.3)(1.8,0.3)
  \psline(1.5,0.6)(1.5,0.7)
  \psline(1.5,-0.6)(1.5,-0.7)
  \psline[border=2pt](2.1,-0.6)(2.1,-1.7)
  \psline[border=2pt](2.1,0.6)(2.1,1.7)
  \psline(2.3,-1.1)(2.3,1.1)
  \psline{->}(1,-0.7)(1,0.12)
  \psline{->}(1,-1.7)(1,-1.3)
  \psline{->}(1,1.2)(1,1.6)
  \psline{->}(2.3,0.4)(2.3,-0.1)
  \psline{->}(2.1,0.6)(2.1,1.0)
  \psline{->}(2.1,-1.1)(2.1,-0.8)
  \psline{->}(1.8,-0.2)(1.8,0.12)
  \psarc[linewidth=0.9pt,linecolor=black,arrows=->,arrowscale=1.4,
    arrowinset=0.15]{<-}(1.8,-0.6){0.3}{120}{180}
  \psarc[linewidth=0.9pt,linecolor=black,arrows=->,arrowscale=1.4,
    arrowinset=0.15]{<-}(1.8,0.6){0.3}{190}{270}
  \rput[bl]{0}(0.9,1.8){$a$}
  \rput[tl]{0}(0.9,-1.7){$a'$}
  \rput[bl]{0}(2.0,1.8){$c$}
  \rput[tl]{0}(2.0,-1.7){$c'$}
  \rput[bl]{0}(0.4,-0.1){$b_{\shortrightarrow}$}
  \rput[bl]{0}(2.45,-0.1){$b_{\shortrightarrow}$}
  \rput[bl]{0}(1.5,-0.2){$f$}
    \rput[bl]{0}(1.25,0.35){$a$}
  \rput[bl]{0}(1.2,-0.55){$a'$}
  \scriptsize
  \rput[bl]{0}(1.75,0.45){$\mu$}
  \rput[bl]{0}(1.71,-0.65){$\mu'$}
 \endpspicture
=\sum\limits_{\left( e,\alpha ,\beta \right) }\left[ \left( F_{a^{\prime
}c^{\prime }}^{ac}\right) ^{-1}\right] _{\left( f,\mu ,\mu ^{\prime
}\right) \left( e,\alpha ,\beta \right) }
\pspicture[shift=-1.95](0.3,-2.1)(3.00,1.9)
 \small
  \psframe[linewidth=0.9pt,linecolor=black,border=0](0.8,0.7)(1.7,1.2)
  \psframe[linewidth=0.9pt,linecolor=black,border=0](0.8,-0.7)(1.7,-1.2)
  \rput[bl]{0}(1.1,0.81){$U$}
  \rput[tl]{0}(1.05,-0.77){$U^{\dag}$}
  \psset{linewidth=0.9pt,linecolor=black,arrowscale=1.5,arrowinset=0.15}
  \psline(1,-0.7)(1,0.7)
  \psline(1,-1.7)(1,-1.2)
  \psline(1,1.2)(1,1.7)
  \psline(1.5,-0.7)(1.5,0.7)
  \psarc[linewidth=0.9pt,linecolor=black,border=0pt] (2.1,1.1){0.4}{0}{90}
  \psline(1.9,1.5)(2.1,1.5)
  \psarc[linewidth=0.9pt,linecolor=black,border=0pt] (1.9,1.1){0.4}{90}{170}
  \psarc[linewidth=0.9pt,linecolor=black,border=0pt] (1.9,-1.1){0.4}{190}{270}
  \psline(1.9,-1.5)(2.1,-1.5)
  \psarc[linewidth=0.9pt,linecolor=black,border=0pt] (2.1,-1.1){0.4}{270}{360}
  \psline(2.1,-1.7)(2.1,1.7)
  \psline[border=2pt](2.1,0.7)(2.1,1.7)
  \psline[border=2pt](2.1,-1.7)(2.1,-0.7)
  \psline(2.5,-1.1)(2.5,1.1)
  \psline(1.5,0.1)(2.1,-0.1)
  \psline{->}(1,-0.7)(1,0.12)
  \psline{->}(1,-1.7)(1,-1.3)
  \psline{->}(1,1.2)(1,1.6)
  \psline{->}(2.5,0.4)(2.5,-0.1)
  \psline{->}(2.1,0.7)(2.1,1.1)
  \psline{->}(2.1,-1.2)(2.1,-0.9)
  \psline{->}(2.1,-0.1)(1.633,0.05)
  \psline{->}(1.5,-0.7)(1.5,-0.2)
  \psline{->}(1.5,0.1)(1.5,0.52)
  \rput[bl]{0}(0.9,1.8){$a$}
  \rput[tl]{0}(0.9,-1.7){$a'$}
  \rput[bl]{0}(2.0,1.8){$c$}
  \rput[tl]{0}(2.0,-1.7){$c'$}
  \rput[bl]{0}(0.4,-0.1){$b_{\shortrightarrow}$}
  \rput[bl]{0}(2.65,-0.1){$b_{\shortrightarrow}$}
  \rput[bl]{0}(1.7,-0.35){$e$}
  \rput[bl]{0}(1.2,0.3){$a$}
  \rput[bl]{0}(1.15,-0.45){$a'$}
  \scriptsize
  \rput[bl]{0}(1.25,0.04){$\alpha$}
  \rput[bl]{0}(2.15,-0.27){$\beta$}
\endpspicture
\nonumber \\
&=&\sum\limits_{\left( e,\alpha ,\beta \right) }\left[ \left( F_{a^{\prime
}c^{\prime }}^{ac}\right) ^{-1}\right] _{\left( f,\mu ,\mu ^{\prime
}\right) \left( e,\alpha ,\beta \right) }
\nonumber \\
&& \times \left\{ \left|
t_{1}\right| ^{2}\left| r_{2}\right| ^{2}
\pspicture[shift=-1.1](0,-1.2)(1.9,1.15)
  \small
  \psellipse[linewidth=0.9pt,linecolor=black,border=0](1.19,0)(0.18,0.4)
  \psset{linewidth=0.9pt,linecolor=black,arrowscale=1.4,arrowinset=0.15}
  \psline{>-}(1.08,-0.30)(1.03,-0.10)
  \psset{linewidth=0.9pt,linecolor=black,arrowscale=1.5,arrowinset=0.15}
  \psline(0.35,-0.775)(0.35,0.775)
  \psline{>-}(0.35,0.3)(0.35,0.6)
  \psline{<-}(0.35,-0.3)(0.35,-0.6)
  \psline(1.55,-0.775)(1.55,0.775)
  \psline{>-}(1.55,0.3)(1.55,0.6)
  \psline{<-}(1.55,-0.3)(1.55,-0.6)
  \psline(1.13,-0.03)(1.55,-0.1)
  \psline[border=2pt](1.13,-0.03)(1.49,-0.09)
   \psline{-<}(0.35,0.1)(0.83,0.02)
   \psline(0.35,0.1)(0.92,0.005)
  \rput[bl]{0}(0.25,0.875){$a$}
  \rput[bl]{0}(1.45,0.875){${c}$}
  \rput[bl]{0}(0.25,-1.075){$a'$}
  \rput[bl]{0}(1.45,-1.075){${c'}$}
  \rput[br]{0}(0.9,0.2){$e$}
  \rput[br]{0}(0.97,-0.52){$b$}
  \scriptsize
  \rput[br]{0}(0.28,0.05){$\alpha$}
  \rput[br]{0}(1.82,-0.25){$\beta$}
  \endpspicture
\right.
+t_{1}r_{1}^{\ast }r_{2}^{\ast }t_{2}^{\ast }e^{i\left( \theta _{\text{I}}-\theta
_{\text{II}}\right) }
 \pspicture[shift=-1.1](-0.25,-1)(1.8,1.35)
  \small
  \psellipse[linewidth=0.9pt,linecolor=black,border=0](0.35,0.6)(0.4,0.18)
  \psset{linewidth=0.9pt,linecolor=black,arrowscale=1.4,arrowinset=0.15}
  \psline{->}(0.1,0.725)(0.3,0.77)
  \psset{linewidth=0.9pt,linecolor=black,arrowscale=1.5,arrowinset=0.15}
  \psline(0.35,-0.6)(0.35,0.69)
  \psline(0.35,0.84)(0.35,0.95)
  \psline[border=3pt]{>-}(0.35,0.3)(0.35,0.6)
  \psline{<-}(0.35,-0.3)(0.35,-0.6)
  \psline(1.35,-0.6)(1.35,0.95)
  \psline{>-}(1.35,0.3)(1.35,0.6)
  \psline{<-}(1.35,-0.3)(1.35,-0.6)
  \psline(0.35,0.1)(1.35,-0.1)
   \psline{->}(1.35,-0.1)(0.7,0.03)
  \rput[bl]{0}(0.25,1.05){$a$}
  \rput[bl]{0}(1.25,1.05){${c}$}
  \rput[bl]{0}(0.25,-0.9){$a'$}
  \rput[bl]{0}(1.25,-0.9){${c'}$}
  \rput[br]{0}(0.95,0.13){$e$}
  \rput[br]{0}(-0.05,0.7){$b$}
  \scriptsize
  \rput[br]{0}(0.28,0.05){$\alpha$}
  \rput[br]{0}(1.62,-0.25){$\beta$}
 \endpspicture
\nonumber \\
&&+t_{1}^{\ast }r_{1}t_{2}r_{2}e^{-i\left( \theta _{\text{I}}-\theta _{\text{II}}\right)
}
 \pspicture[shift=-1.13](-0.25,-1.4)(1.8,1)
  \small
  \psellipse[linewidth=0.9pt,linecolor=black,border=0](0.35,-0.6)(0.4,0.18)
  \psset{linewidth=0.9pt,linecolor=black,arrowscale=1.4,arrowinset=0.15}
  \psline{-<}(0.07,-0.715)(0.27,-0.76)
  \psset{linewidth=0.9pt,linecolor=black,arrowscale=1.5,arrowinset=0.15}
  \psline(0.35,-0.69)(0.35,0.6)
  \psline(0.35,-0.84)(0.35,-0.95)
  \psline{>-}(0.35,0.3)(0.35,0.6)
  \psline[border=3pt]{>-}(0.35,-0.55)(0.35,-0.3)
  \psline(1.35,0.6)(1.35,-0.95)
  \psline{>-}(1.35,0.3)(1.35,0.6)
  \psline{>-}(1.35,-0.55)(1.35,-0.3)
  \psline(0.35,0.1)(1.35,-0.1)
   \psline{->}(1.35,-0.1)(0.7,0.03)
  \rput[bl]{0}(0.25,0.7){$a$}
  \rput[bl]{0}(1.25,0.7){${c}$}
  \rput[bl]{0}(0.25,-1.25){$a'$}
  \rput[bl]{0}(1.25,-1.25){${c'}$}
  \rput[br]{0}(0.95,0.13){$e$}
  \rput[br]{0}(-0.05,-0.9){$b$}
  \scriptsize
  \rput[br]{0}(0.28,0.05){$\alpha$}
  \rput[br]{0}(1.62,-0.25){$\beta$}
 \endpspicture
\left. +\left| r_{1}\right| ^{2}\left| t_{2}\right|
^{2}
\pspicture[shift=-1.1](-0.7,-1.2)(1.75,1.15)
  \small
  \psellipse[linewidth=0.9pt,linecolor=black,border=0](-0.2,0.0)(0.18,0.4)
  \psset{linewidth=0.9pt,linecolor=black,arrowscale=1.4,arrowinset=0.15}
  \psline{->}(-0.358,-0.1)(-0.365,0.1)
  \psset{linewidth=0.9pt,linecolor=black,arrowscale=1.5,arrowinset=0.15}
  \psline(0.35,-0.775)(0.35,0.775)
  \psline{>-}(0.35,0.3)(0.35,0.6)
  \psline{<-}(0.35,-0.3)(0.35,-0.6)
  \psline(1.35,-0.775)(1.35,0.775)
  \psline{>-}(1.35,0.3)(1.35,0.6)
  \psline{<-}(1.35,-0.3)(1.35,-0.6)
  \psline(0.35,0.1)(1.35,-0.1)
   \psline{->}(1.35,-0.1)(0.7,0.03)
  \rput[bl]{0}(0.25,0.875){$a$}
  \rput[bl]{0}(1.25,0.875){${c}$}
  \rput[bl]{0}(0.25,-1.075){$a'$}
  \rput[bl]{0}(1.25,-1.075){${c'}$}
  \rput[br]{0}(0.95,0.13){$e$}
  \rput[br]{0}(-0.5,-0.3){$b$}
  \scriptsize
  \rput[br]{0}(0.28,0.05){$\alpha$}
  \rput[br]{0}(1.62,-0.25){$\beta$}
  \endpspicture
\right\}
\nonumber \\
&=& d_b \sum\limits_{\left( e,\alpha ,\beta \right) }\left[ \left(
F_{a^{\prime
}c^{\prime }}^{ac}\right) ^{-1}\right] _{\left( f,\mu ,\mu ^{\prime
}\right) \left( e,\alpha ,\beta \right) }p_{aa^{\prime }e,b}^{\shortrightarrow
}
\pspicture[shift=-1.1](-0.1,-1.2)(1.8,1.15)
  \small
  \psset{linewidth=0.9pt,linecolor=black,arrowscale=1.5,arrowinset=0.15}
  \psline(0.35,-0.775)(0.35,0.775)
  \psline{>-}(0.35,0.3)(0.35,0.6)
  \psline{<-}(0.35,-0.3)(0.35,-0.6)
  \psline(1.35,-0.775)(1.35,0.775)
  \psline{>-}(1.35,0.3)(1.35,0.6)
  \psline{<-}(1.35,-0.3)(1.35,-0.6)
  \psline(0.35,0.1)(1.35,-0.1)
   \psline{->}(1.35,-0.1)(0.7,0.03)
  \rput[bl]{0}(0.25,0.875){$a$}
  \rput[bl]{0}(1.25,0.875){${c}$}
  \rput[bl]{0}(0.25,-1.075){$a'$}
  \rput[bl]{0}(1.25,-1.075){${c'}$}
  \rput[br]{0}(0.95,0.13){$e$}
  \scriptsize
  \rput[br]{0}(0.28,0.05){$\alpha$}
  \rput[br]{0}(1.62,-0.25){$\beta$}
  \endpspicture
\nonumber \\
&=& d_b \sum\limits_{\substack{ \left( e,\alpha ,\beta \right)  \\ \left(
f^{\prime },\nu ,\nu ^{\prime }\right) }}
\left[ \left( F_{a^{\prime}c^{\prime }}^{ac}\right) ^{-1}\right] _{\left( f,\mu ,\mu ^{\prime}\right) \left( e,\alpha ,\beta \right) }
p_{aa^{\prime }e,b}^{\shortrightarrow}
\left[ F_{a^{\prime }c^{\prime}}^{ac}\right] _{\left( e,\alpha ,\beta \right) \left( f^{\prime },\nu ,\nu^{\prime }\right) }
 \pspicture[shift=-1.1](0,-0.85)(1.3,1.3)
 \small
  \psset{linewidth=0.9pt,linecolor=black,arrowscale=1.5,arrowinset=0.15}
  \psline{->}(0.7,0)(0.7,0.45)
  \psline(0.7,0)(0.7,0.55)
  \psline(0.7,0.55) (0.2,1.05)
  \psline{->}(0.7,0.55)(0.3,0.95)
  \psline(0.7,0.55) (1.2,1.05)
  \psline{->}(0.7,0.55)(1.1,0.95)
  \rput[bl]{0}(0.28,0.1){$f'$}
  \rput[bl]{0}(1.1,1.15){$c$}
  \rput[bl]{0}(0.1,1.15){$a$}
  \psline(0.7,0) (0.2,-0.5)
  \psline{-<}(0.7,0)(0.35,-0.35)
  \psline(0.7,0) (1.2,-0.5)
  \psline{-<}(0.7,0)(1.05,-0.35)
  \rput[bl]{0}(1.1,-0.8){$c'$}
  \rput[bl]{0}(0.1,-0.8){$a'$}
  \scriptsize
  \rput[bl]{0}(0.82,0.45){$\nu$}
  \rput[bl]{0}(0.82,-0.0){$\nu'$}
  \endpspicture
\label{eq:one_probe_analysis}
\end{eqnarray}%
where we have defined%
\begin{eqnarray}
p_{aa^{\prime }e,b}^{\shortrightarrow } &=&\left| t_{1}\right| ^{2}\left|
r_{2}\right| ^{2}M_{eb}+t_{1}r_{1}^{\ast }r_{2}^{\ast }t_{2}^{\ast
}e^{i\left( \theta _{\text{I}}-\theta _{\text{II}}\right) }M_{ab}  \notag \\
&&+t_{1}^{\ast }r_{1}t_{2}r_{2}e^{-i\left( \theta _{\text{I}}-\theta _{\text{II}}\right)
}M_{a^{\prime }b}^{\ast }+\left| r_{1}\right| ^{2}\left| t_{2}\right| ^{2}
.
\end{eqnarray}%
This calculation uses the diagrammatic rule
\begin{equation}
\pspicture[shift=-0.55](-0.25,-0.1)(0.9,1.3)
\small
  \psset{linewidth=0.9pt,linecolor=black,arrowscale=1.5,arrowinset=0.15}
  \psline(0.4,0)(0.4,0.22)
  \psline(0.4,0.45)(0.4,1.2)
  \psellipse[linewidth=0.9pt,linecolor=black,border=0](0.4,0.5)(0.4,0.18)
  \psset{linewidth=0.9pt,linecolor=black,arrowscale=1.4,arrowinset=0.15}
  \psline{->}(0.2,0.37)(0.3,0.34)
\psline[linewidth=0.9pt,linecolor=black,border=2.5pt,arrows=->,arrowscale=1.5,
arrowinset=0.15](0.4,0.5)(0.4,1.1)
  \rput[bl]{0}(-0.15,0.15){$b$}
  \rput[tl]{0}(0.55,1.2){$a$}
  \endpspicture
=\frac{S_{ab}}{S_{0a}}
\pspicture[shift=-0.55](0.05,-0.1)(1,1.3)
\small
  \psset{linewidth=0.9pt,linecolor=black,arrowscale=1.5,arrowinset=0.15}
  \psline(0.4,0)(0.4,1.2)
\psline[linewidth=0.9pt,linecolor=black,arrows=->,arrowscale=1.5,
arrowinset=0.15](0.4,0.5)(0.4,0.9)
  \rput[tl]{0}(0.52,1.0){$a$}
  \endpspicture
\label{eq:loopaway}
\end{equation}
to remove the $b$ loops, and the definitions of the topological $S$-matrix
\begin{equation}
S_{ab}=\mathcal{D}^{-1}\widetilde{\text{Tr}}\left[ R_{ba}R_{ab}\right] =\frac{1}{\mathcal{D}}
\pspicture[shift=-0.4](0.0,0.2)(2.4,1.3)
\small
  \psarc[linewidth=0.9pt,linecolor=black,arrows=<-,arrowscale=1.5,
arrowinset=0.15] (1.6,0.7){0.5}{165}{363}
  \psarc[linewidth=0.9pt,linecolor=black] (0.9,0.7){0.5}{0}{180}
  \psarc[linewidth=0.9pt,linecolor=black,border=3pt,arrows=->,arrowscale=1.5,
arrowinset=0.15] (0.9,0.7){0.5}{180}{375}
  \psarc[linewidth=0.9pt,linecolor=black,border=3pt] (1.6,0.7){0.5}{0}{160}
  \psarc[linewidth=0.9pt,linecolor=black] (1.6,0.7){0.5}{155}{170}
  \rput[bl]{0}(0.15,0.3){$a$}
  \rput[bl]{0}(2.15,0.3){$b$}
  \endpspicture
,
\end{equation}
and the monodromy matrix
\begin{equation}
M_{ab}=\frac{\widetilde{\text{Tr}}\left[
R_{ba}R_{ab}\right]}{\widetilde{\text{Tr}}\mathbb{I}_{ab}}
=\frac{1}{d_{a}d_{b}}
\pspicture[shift=-0.4](0.0,0.2)(2.4,1.3)
\small
  \psarc[linewidth=0.9pt,linecolor=black,arrows=<-,arrowscale=1.5,
arrowinset=0.15] (1.6,0.7){0.5}{165}{363}
  \psarc[linewidth=0.9pt,linecolor=black] (0.9,0.7){0.5}{0}{180}
  \psarc[linewidth=0.9pt,linecolor=black,border=3pt,arrows=->,arrowscale=1.5,
arrowinset=0.15] (0.9,0.7){0.5}{180}{375}
  \psarc[linewidth=0.9pt,linecolor=black,border=3pt] (1.6,0.7){0.5}{0}{160}
  \psarc[linewidth=0.9pt,linecolor=black] (1.6,0.7){0.5}{155}{170}
  \rput[bl]{0}(0.15,0.3){$a$}
  \rput[bl]{0}(2.15,0.3){$b$}
  \endpspicture
=\frac{S_{ab}S_{00}}{S_{0a}S_{0b}}
,
\label{eq:monodromy}
\end{equation}%
which is an important quantity, typically arising in interference terms~\cite{Bonderson06b}.

A similar calculation for
the $s=\shortuparrow $ outcome gives%
\begin{eqnarray}
p_{aa^{\prime }e,b}^{\shortuparrow } &=&\left| t_{1}\right| ^{2}\left|
t_{2}\right| ^{2}M_{eb}-t_{1}r_{1}^{\ast }r_{2}^{\ast }t_{2}^{\ast
}e^{i\left( \theta _{\text{I}}-\theta _{\text{II}}\right) }M_{ab}  \notag \\
&&-t_{1}^{\ast }r_{1}t_{2}r_{2}e^{-i\left( \theta _{\text{I}}-\theta _{\text{II}}\right)
}M_{a^{\prime }b}^{\ast }+\left| r_{1}\right| ^{2}\left| r_{2}\right| ^{2}.
\end{eqnarray}

We obtain the results for general $\rho^{B}$ by simply
replacing $p_{aa^{\prime}e,b}^{s}$ everywhere with%
\begin{equation}
p_{aa^{\prime }e,B}^{s} = \sum\limits_{b}\Pr\nolimits_{B}\left( b\right)
p_{aa^{\prime }e,b}^{s}
\end{equation}%
We will also use the notation $M_{aB}=\sum\nolimits_{b}\Pr\nolimits_{B} \left( b\right) M_{ab}$. When we refer to a probe $B$ being able to distinguish two charges $a$ and $a'$ by monodromy, we mean that $M_{aB} \neq M_{a'B}$, and when we refer to a probe being able to detect a charge $a$ by monodromy, we mean that $M_{aB} \neq 1$.

{}From this, inserting the appropriate coefficients and normalization factors,
we find the reduced density matrix of the target system after a
single probe measurement with outcome $s$ to be
\begin{eqnarray}
\rho ^{AC}\left( s \right) &=&\sum\limits_{\substack{ a,a^{\prime},c,c^{\prime },f,\mu ,\mu ^{\prime } \\ \left( e,\alpha ,\beta \right)
,\left( f^{\prime },\nu ,\nu ^{\prime }\right) }}
\frac{ \rho _{\left(a,c;f,\mu \right) ,\left( a^{\prime },c^{\prime };f,\mu ^{\prime }\right)}^{AC} }{ \left( d_{a}d_{a^{\prime }}d_{c}d_{c^{\prime}}d_{f}^{2} \right)^{1/4}}
\left[ \left( F_{a^{\prime }c^{\prime }}^{ac}\right) ^{-1}\right]_{\left( f,\mu ,\mu ^{\prime }\right) \left( e,\alpha ,\beta \right) }
\nonumber \\
&& \times \frac{p_{aa^{\prime }e,B}^{s}}{\Pr \left( s\right) }
\left[ F_{a^{\prime }c^{\prime }}^{ac}\right] _{\left( e,\alpha ,\beta \right) \left( f^{\prime },\nu ,\nu ^{\prime }\right) }
 \pspicture[shift=-0.75](-0.15,-0.55)(1.5,1.1)
 \small
  \psset{linewidth=0.9pt,linecolor=black,arrowscale=1.5,arrowinset=0.15}
  \psline{->}(0.7,0)(0.7,0.45)
  \psline(0.7,0)(0.7,0.55)
  \psline(0.7,0.55) (0.25,1)
  \psline{->}(0.7,0.55)(0.3,0.95)
  \psline(0.7,0.55) (1.15,1)
  \psline{->}(0.7,0.55)(1.1,0.95)
  \rput[bl]{0}(0.28,0.1){$f'$}
  \rput[bl]{0}(1.22,0.8){$c$}
  \rput[bl]{0}(-0.05,0.8){$a$}
  \psline(0.7,0) (0.25,-0.45)
  \psline{-<}(0.7,0)(0.35,-0.35)
  \psline(0.7,0) (1.15,-0.45)
  \psline{-<}(0.7,0)(1.05,-0.35)
  \rput[bl]{0}(1.22,-0.4){$c'$}
  \rput[bl]{0}(-0.05,-0.4){$a'$}
  \scriptsize
  \rput[bl]{0}(0.82,0.45){$\nu$}
  \rput[bl]{0}(0.82,-0.0){$\nu'$}
  \endpspicture
\nonumber \\
&=&\sum\limits_{\substack{ a,a^{\prime},c,c^{\prime },f,\mu ,\mu ^{\prime } \\ \left( e,\alpha ,\beta \right) ,\left( f^{\prime },\nu ,\nu ^{\prime }\right) }} \frac{\rho _{\left(a,c;f,\mu \right) ,\left( a^{\prime },c^{\prime };f,\mu ^{\prime }\right)}^{AC}}{\left( d_{f} d_{f^{\prime }} \right)^{1/2}}
\left[ \left( F_{a^{\prime },c^{\prime }}^{a,c}\right)^{-1}\right]_{\left( f,\mu ,\mu ^{\prime }\right) \left( e,\alpha ,\beta \right) }
\nonumber \\
& &\times \frac{p_{aa^{\prime }e,B}^{s}}{\Pr \left( s\right) }
\left[F_{a^{\prime },c^{\prime }}^{a,c}\right] _{\left( e,\alpha ,\beta \right)\left( f^{\prime },\nu ,\nu ^{\prime }\right) }
\left| a,c;f^{\prime },\nu \right\rangle \left\langle a^{\prime},c^{\prime };f^{\prime },\nu ^{\prime }\right|
\end{eqnarray}%
where the probability of measurement outcome $s$ is computed by additionally taking
the quantum trace of the target system, which projects onto the $e=0$
components,
giving%
\begin{equation}
\Pr \left( s\right)=\sum\limits_{a,c,f,\mu }\rho _{\left( a,c;f,\mu \right)
,\left( a,c;f,\mu \right) }^{AC} p_{aa0,B}^{s}.
\label{eq:single_probability}
\end{equation}%

We note that%
\begin{eqnarray}
p_{aa0,B}^{\shortrightarrow } &=&\left| t_{1}\right| ^{2}\left| r_{2}\right|
^{2}+\left| r_{1}\right| ^{2}\left| t_{2}\right| ^{2}+2\text{Re}\left\{
t_{1}r_{1}^{\ast }r_{2}^{\ast }t_{2}^{\ast }e^{i\left( \theta _{\text{I}}-\theta
_{\text{II}}\right) }M_{aB}\right\}  \\
p_{aa0,B}^{_{\shortuparrow }} &=&\left| t_{1}\right| ^{2}\left| t_{2}\right|
^{2}+\left| r_{1}\right| ^{2}\left| r_{2}\right| ^{2}-2\text{Re}\left\{
t_{1}r_{1}^{\ast }r_{2}^{\ast }t_{2}^{\ast }e^{i\left( \theta _{\text{I}}-\theta
_{\text{II}}\right) }M_{aB}\right\}
\end{eqnarray}%
give a well-defined probability distribution (i.e. $0\leq p_{aa0,B}^{s}\leq 1
$ and $p_{aa0,B}^{\shortrightarrow }+p_{aa0,B}^{_{\shortuparrow }}=1$).

\subsubsection{Multiple Probes}
\label{sec:multiple_probes}

We can now easily produce the results for multiple probes. If we send $N$ probes through the interferometer, a string of measurement outcomes $\left(s_{1},\ldots ,s_{N}\right)$ occurs with probability%
\begin{equation}
\Pr \left( s_{1},\ldots ,s_{N}\right) =\sum\limits_{a,c,f,\mu}\rho _{\left( a,c;f,\mu \right)
,\left( a,c;f,\mu \right) }^{AC} p_{aa0,B}^{s_{1}}\ldots
p_{aa0,B}^{s_{N}}
\end{equation}%
and results in the measured target anyon reduced density matrix%
\begin{eqnarray}
&& \rho ^{AC}\left( s_{1},\ldots ,s_{N}\right) =
\sum\limits_{\substack{ a,a^{\prime},c,c^{\prime },f,\mu ,\mu ^{\prime } \\ \left( e,\alpha ,\beta \right),\left( f^{\prime },\nu ,\nu ^{\prime }\right) }}
\frac{\rho _{\left(a,c;f,\mu \right) ,\left( a^{\prime },c^{\prime };f,\mu ^{\prime }\right)}^{AC}}{\left( d_{f} d_{f^{\prime }} \right)^{1/2}}
\left[ \left( F_{a^{\prime },c^{\prime }}^{a,c}\right)^{-1}\right]_{\left( f,\mu ,\mu ^{\prime }\right) \left( e,\alpha ,\beta \right) }
\notag \\
&&\times \frac{p_{aa^{\prime }e,B}^{s_{1}}\ldots p_{aa^{\prime }e,B}^{s_{N}}}{\Pr \left( s_{1},\ldots ,s_{N}\right) }
\left[F_{a^{\prime },c^{\prime }}^{a,c}\right] _{\left( e,\alpha ,\beta \right) \left( f^{\prime },\nu ,\nu ^{\prime }\right) }\left| a,c;f^{\prime },\nu \right\rangle \left\langle a^{\prime},c^{\prime };f^{\prime },\nu ^{\prime }\right|
.
\label{eq:rhoA_N}
\end{eqnarray}%
It is clear that the specific order of the measurement outcomes is not
important in the result, but that only the total number of outcomes of each
type matters. Keeping track of only the total numbers leads to a binomial distribution.
We denote the total number of $s_{j}=\rightarrow $ in the string of measurement outcomes as
$n$, and cluster together all results with the same $n$. Defining (for arbitrary
$p$ and $q$)%
\begin{equation}
W_{N}\left( n;p,q\right) =\frac{N!}{n!\left( N-n\right) !}p^{n}q^{N-n}
,
\end{equation}%
the probability of measuring $n$ of the $N$ probes at the horizontal
detector is%
\begin{equation}
\Pr\nolimits_{N}\left( n\right) =\sum\limits_{a,c,f,\mu}\rho _{\left( a,c;f,\mu \right)
,\left( a,c;f,\mu \right) }^{AC}
W_{N}\left(n;p_{aa0,B}^{\rightarrow },p_{aa0,B}^{\uparrow }\right)
\label{eq:prob_n_N}
\end{equation}%
and these measurements produce a resulting target anyon reduced density matrix%
\begin{eqnarray}
&&\rho _{N}^{AC}\left( n\right)  = \sum\limits_{\substack{ a,a^{\prime},c,c^{\prime },f,\mu ,\mu ^{\prime } \\ \left( e,\alpha ,\beta \right),\left( f^{\prime },\nu ,\nu ^{\prime }\right) }}
\frac{\rho _{\left(a,c;f,\mu \right) ,\left( a^{\prime },c^{\prime };f,\mu ^{\prime }\right)}^{AC}}{\left( d_{f} d_{f^{\prime }} \right)^{1/2}}
\left[ \left( F_{a^{\prime },c^{\prime }}^{a,c}\right)^{-1}\right]_{\left( f,\mu ,\mu ^{\prime }\right) \left( e,\alpha ,\beta \right) }
\notag \\
& &\times \frac{W_{N}\left( n;p_{aa^{\prime }e,B}^{\rightarrow },p_{aa^{\prime}e,B}^{\uparrow }\right) }{\Pr\nolimits_{N}\left( n\right) }
\left[F_{a^{\prime },c^{\prime }}^{a,c}\right] _{\left( e,\alpha ,\beta \right) \left( f^{\prime },\nu ,\nu ^{\prime }\right) }\left| a,c;f^{\prime },\nu \right\rangle \left\langle a^{\prime},c^{\prime };f^{\prime },\nu ^{\prime }\right|
.
\label{eq:rho_n}
\end{eqnarray}

The interferometry experiment distinguishes anyonic charges in the target by
their values of $p_{aa1,B}^{s}$, which determine the possible measurement
distributions. Different anyonic charges with the same probability
distributions of probe outcomes are indistinguishable by such probes, and so
should be grouped together into distinguishable subsets. We define $\mathcal{C}_{\kappa }$ to be the maximal disjoint subsets of $\mathcal{C}$, the set of all anyonic charge types, such that $p_{aa0,B}^{\shortrightarrow }=p_{\kappa }$ for all $a\in \mathcal{C}_{\kappa }$,
i.e.
\begin{eqnarray}
\mathcal{C}_{\kappa } &\equiv &\left\{ a\in \mathcal{C}:p_{aa0,B}^{\shortrightarrow }=p_{\kappa }\right\} \\
\mathcal{C}_{\kappa }\cap \mathcal{C}_{\kappa ^{\prime }} &=&\varnothing
\text{ \ \ \ \ for \ }\kappa \neq \kappa ^{\prime } \nonumber \\
\bigcup\limits_{\kappa }\mathcal{C}_{\kappa } &=&\mathcal{C} \nonumber
.
\end{eqnarray}%
Note that $p_{aa0,B}^{\shortrightarrow }=p_{a^{\prime }a^{\prime
}0,B}^{\shortrightarrow }$ (for two different charges $a$ and $a^{\prime }$) iff%
\begin{equation}
\text{Re}\left\{ t_{1}r_{1}^{\ast }r_{2}^{\ast }t_{2}^{\ast }e^{i\left(
\theta_{\text{I}}-\theta_{\text{II}}\right) }M_{aB}\right\} =\text{Re}\left\{
t_{1}r_{1}^{\ast }r_{2}^{\ast }t_{2}^{\ast }e^{i\left( \theta _{\text{I}}-\theta
_{\text{II}}\right) }M_{a^{\prime }B}\right\}
\end{equation}%
which occurs either when:

\noindent
(i) at least one of $t_{1}$, $t_{2}$, $r_{1}$, or $r_{2}$ is zero, or

\noindent
(ii) $\left| M_{aB}\right| \cos \left( \theta +\varphi _{a}\right) =\left|
M_{a^{\prime }B}\right| \cos \left( \theta +\varphi _{a^{\prime }}\right) $,
where $\theta =\arg \left( t_{1}r_{1}^{\ast }r_{2}^{\ast }t_{2}^{\ast
}e^{i\left( \theta _{\text{I}}-\theta _{\text{II}}\right) }\right) $ and $\varphi
_{a}=\arg \left( M_{aB}\right) $.

If condition (i) is satisfied, then there is no interference and $\mathcal{C}%
_{0}=\mathcal{C}$ (all target anyonic charges give the same probe
measurement distribution). Condition (ii) is generically\footnote{%
The term ``generic'' is used in this paper only in reference to the
collection of interferometer parameters $t_{j}$, $r_{j}$, $\theta _{\text{I}}$, and
$\theta _{\text{II}}$.} only satisfied when $M_{aB}=M_{a^{\prime }B}$, but can also be satisfied
with the fine-tuned condition $\theta =-\arg \left\{
M_{aB}-M_{a^{\prime }B}\right\} \pm \frac{\pi}{2} $.

With this notation, we may rewrite the probabilities in the convenient form%
\begin{eqnarray}
\Pr\nolimits_{N}\left( n\right) &=&\sum\limits_{\kappa
}\Pr\nolimits_{A}\left( \kappa \right) W_{N}\left( n;p_{\kappa },1-p_{\kappa
}\right) \\
\Pr\nolimits_{A}\left( \kappa \right) &=& \widetilde{\text{Tr}} \left[ \rho^{AC} \Pi^{A}_{\mathcal{C}_{\kappa }} \right] = \sum\limits_{a\in \mathcal{C}%
_{\kappa },c,f,\mu }\rho _{\left( a,c;f,\mu \right) ,\left( a,c;f,\mu
\right) }^{AC}
,
\end{eqnarray}%
where
\begin{equation}
\Pi^{A}_{\mathcal{C}_{\kappa }} = \sum\limits_{a \in \mathcal{C}_{\kappa }} \Pi^{A}_{a}
\end{equation}
for $\Pi^{A}_{a}$ the projector of anyon(s) $A$ onto (collective) anyonic charge $a$.

The projector onto collective topological charge $a$ of $n$ anyons (collectively denoted as $A$) of definite charges $a_1,\ldots,a_n$ is given by
\begin{equation}
\Pi_{a}^{A} = \sum_{\substack{ c_{2},\ldots,c_{n-1} \\ \mu_{2},\ldots,\mu_{n}}} \sqrt{\frac{d_{a}}{d_{a_{1}} \ldots d_{a_{n}} }}
 \pspicture[shift=-2](-0.35,-2)(2.5,2.4)
  \small
  \psset{linewidth=0.9pt,linecolor=black,arrowscale=1.5,arrowinset=0.15}
  \psline(0.0,1.75)(1,0.5)
  \psline(2.0,1.75)(1,0.5)
  \psline(0.4,1.25)(0.8,1.75)
   \psline{->}(0.4,1.25)(0.1,1.625)
   \psline{->}(0.4,1.25)(0.7,1.625)
   \psline{->}(1,0.5)(1.9,1.625)
   \psline{->}(1,0.5)(0.5,1.125)
   \rput[bl]{0}(-0.15,1.85){$a_1$}
   \rput[bl]{0}(0.75,1.85){$a_2$}
   \rput[bl]{0}(1.95,1.85){$a_n$}
\rput[bl](1.25,1.85){$\cdots$}
\rput{-45}(0.9,1.05){$\cdots$}
   \rput[bl]{0}(0.25,0.65){$c_2$}
  \psset{linewidth=0.9pt,linecolor=black,arrowscale=1.5,arrowinset=0.15}
  \psline(0.0,-1.45)(1,-0.2)
  \psline(2.0,-1.45)(1,-0.2)
  \psline(0.4,-0.95)(0.8,-1.45)
   \psline{-<}(0.4,-0.95)(0.1,-1.325)
   \psline{-<}(0.4,-0.95)(0.7,-1.325)
   \psline{-<}(1,-0.2)(1.9,-1.325)
   \psline{-<}(1,-0.2)(0.5,-0.825)
   \rput[bl]{0}(-0.15,-1.85){$a_1$}
   \rput[bl]{0}(0.75,-1.85){$a_2$}
   \rput[bl]{0}(1.95,-1.85){$a_n$}
   \rput[bl]{0}(0.25,-0.6){$c_2$}
   \rput[bl](1.25,-1.85){$\cdots$}
   \rput{45}(0.9,-0.75){$\cdots$}
  \psline(1,-0.2)(1,0.5)
  \psline{->}(1,-0.2)(1,0.3)
   \rput[bl]{0}(1.15,0.0){$a$}
\scriptsize
   \rput[bl]{0}(0,1.0){$\mu_2$}
   \rput[bl]{0}(0.55,0.35){$\mu_n$}
   \rput[bl]{0}(0,-0.95){$\mu_2$}
   \rput[bl]{0}(0.55,-0.3){$\mu_n$}
\endpspicture
.
\label{eq:PIan}
\end{equation}

We can now take the limit as $N\rightarrow \infty $, to determine the asymptotic behavior of interferometry when many probe anyons are sent through the interferometer. The fraction $r=n/N$ of probes measured in the $s=\shortrightarrow$ detector will be found to go to $r=p_{\kappa}$ with probability $\Pr\nolimits_{A}\left( \kappa \right) $, and the target
anyon density matrix will generically collapse onto the corresponding ``fixed states'' given by%
\begin{eqnarray}
&& \rho _{\kappa }^{AC} =\sum\limits_{\substack{ a,a^{\prime },c,c^{\prime},f,\mu ,\mu ^{\prime } \\ \left( e,\alpha ,\beta \right) ,\left( f^{\prime},\nu ,\nu ^{\prime }\right) }}
\frac{\rho _{\left(a,c;f,\mu \right) ,\left( a^{\prime },c^{\prime };f,\mu ^{\prime }\right)}^{AC}}{\left( d_{f}d_{f^{\prime }} \right) ^{1/2} }
\left[ \left( F_{a^{\prime }c^{\prime }}^{ac}\right) ^{-1}\right]_{\left( f,\mu ,\mu ^{\prime }\right) \left( e,\alpha ,\beta \right) }
\notag \\
&&\times
\Delta _{aa^{\prime}e,B}\left( p_{\kappa }\right)
\left[ F_{a^{\prime }c^{\prime }}^{ac}\right] _{\left( e,\alpha ,\beta \right)\left( f^{\prime },\nu ,\nu ^{\prime }\right) }
\left| a,c;f^{\prime },\nu\right\rangle \left\langle a^{\prime },c^{\prime };f^{\prime },\nu ^{\prime}\right|
\end{eqnarray}%
where%
\begin{equation}
\Delta _{aa^{\prime }e,B}\left( p_{\kappa }\right) =\left\{
\begin{array}{cc}
\frac{1}{\Pr\nolimits_{A}\left( \kappa \right) } & \text{if }p_{aa^{\prime
}e,B}^{\shortrightarrow }=1-p_{aa^{\prime }e,B}^{\shortuparrow }=p_{\kappa }\text{ and
}a,a^{\prime }\in \mathcal{C}_{\kappa } \\
0 & \text{otherwise}%
\end{array}%
\right. .
\end{equation}%
(Fixed state density matrices are left unchanged by probe measurements.)
We emphasize that the condition: $p_{aa^{\prime }e,B}^{\shortrightarrow
}=1-p_{aa^{\prime }e,B}^{\shortuparrow }=p_{\kappa }$ and $a,a^{\prime }\in
\mathcal{C}_{\kappa }$ is equivalent to $M_{eB}=1$ (which also implies
$M_{aB}=M_{a^{\prime}B}$).

We note that if the probes can distinguish between all charge types, then each $\mathcal{C}_{\kappa }$ contains a single element and $M_{eB}=1$ iff $e=0$. The fixed states in this case are given
by
\begin{equation}
\rho _{\kappa_{a} }^{AC} = \sum\limits_{c} \frac{\Pr\nolimits _{A} \left(c | a \right)}{ d_{a}d_{c} } \,\, \mathbb{I}_{ac}
= \sum\limits_{c,f^{\prime},\nu} \frac{\Pr\nolimits _{A} \left(c | a \right)}{ d_{a}d_{c} } \,\, \left| a,c;f^{\prime },\nu \right\rangle \left\langle a,c;f^{\prime },\nu \right|
\end{equation}
where
\begin{equation}
\Pr\nolimits _{A} \left(c | a \right) = \frac{ \sum\limits_{f,\mu}\rho _{\left(
a,c;f,\mu \right) \left( a,c;f,\mu \right)}^{A} }{\sum\limits_{c,f,\mu}\rho _{\left(
a,c;f,\mu \right) \left( a,c;f,\mu \right)}^{A}}
,
\end{equation}%
for which the target anyon $A$ has definite charge and no entanglement with $C$.

This calculation shows that asymptotic operation of a generically tuned anyonic interferometer selects a charge sector $\kappa$ with probability $\Pr_{A}(\kappa)$  and then: (1) projects the anyonic state onto the subspace where the $A$ anyons have collective anyonic charge in $\mathcal{C}_\kappa$, and (2) decoheres all anyonic entanglement between subsystem $A$ and $C$ that the probes can detect. The sector $\kappa$ may be a single charge or a collection of charges with identical monodromy elements with the probes, i.e. $M_{a,B} = M_{a^{\prime},B}$ for $a,a^{\prime} \in \mathcal{C}_{\kappa}$. The anyonic entanglement between $A$ and $C$ is described in the form of anyonic charge lines connecting these subsystems, i.e. the charge lines label by charge $e$ in the preceding analysis, where the contribution of a diagram to the density matrix will be removed if it has $e \notin \mathcal{C}_0$ (i.e. $M_{eB}\neq1$).

\subsubsection{Generalized Target System}
\label{sec:generalized_target}

\begin{figure}[t!]
\begin{center}
  \includegraphics[scale=0.6]{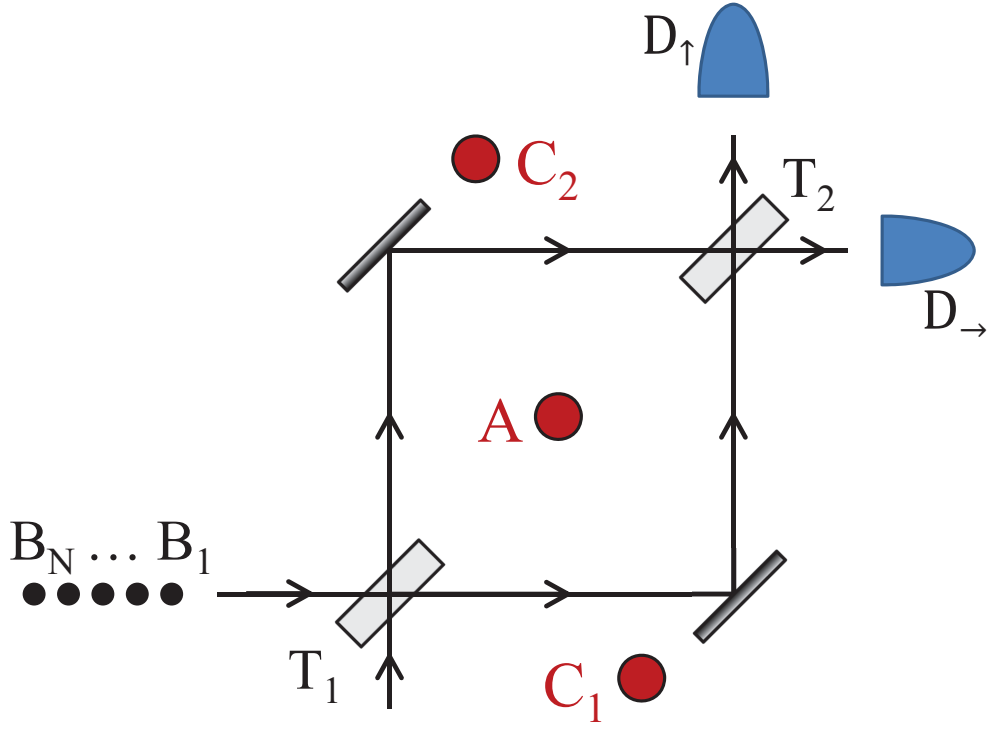}
  \caption{An idealized Mach-Zehnder interferometer for which the anyons $C$ that are entangled with the target anyon(s) $A$ are split into two groups of anyons $C_1$ and $C_2$ located below and above the interferometer, respectively.}
  \label{fig:Untwisted_int2}
\end{center}
\end{figure}

This is a convenient place to consider in more detail a modest generalization of this analysis that was mentioned briefly in~\cite{Bonderson07c}, where the complementary anyons $C$ (outside of the interferometry loop) are divided into two groups $C_1$ and $C_2$ located on the two different sides of the interferometer (below and above), as shown in Fig.~\ref{fig:Untwisted_int2}. In this circumstance, the decoherence effect (2) applies not just to anyonic entanglement lines connecting $A$ with $C_1$ and $C_2$, but also to anyonic entanglement lines connecting $C_1$ with $C_2$, since these groups of anyons are separated by the paths of the probe anyons.

More specifically, we start with a density matrix of the form
\begin{eqnarray}
\rho^{AC} &=&
 \pspicture[shift=-1.9](-2.3,-2)(1.1,1.5)
  \small
  \psframe[linewidth=0.9pt,linecolor=black,border=0](-1.75,-0.5)(0.75,0.5)
  \rput[bl]{0}(-0.8,-0.15){$\rho^{AC}$}
  \psset{linewidth=0.9pt,linecolor=black,arrowscale=1.5,arrowinset=0.15}
  \psline(0.5,0.5)(0.5,1.25)
  \psline(-0.5,0.5)(-0.5,1.25)
  \psline(0.5,-0.5)(0.5,-1.25)
  \psline(-0.5,-0.5)(-0.5,-1.25)
  \psline(-1.5,0.5)(-1.5,1.25)
  \psline(-1.5,-0.5)(-1.5,-1.25)
  \psline{->}(0.5,0.5)(0.5,1)
  \psline{->}(-0.5,0.5)(-0.5,1)
  \psline{-<}(0.5,-0.5)(0.5,-1)
  \psline{-<}(-0.5,-0.5)(-0.5,-1)
  \psline{->}(-1.5,0.5)(-1.5,1)
  \psline{-<}(-1.5,-0.5)(-1.5,-1)
  \rput[bl](0.4,1.3){$C_1$}
  \rput[bl](-0.6,1.35){$A$}
  \rput[bl](0.4,-1.7){$C_1^{\prime}$}
  \rput[bl](-0.6,-1.6){$A^{\prime}$}
  \rput[bl](-1.7,1.3){$C_2$}
  \rput[bl](-1.7,-1.7){$C_2^{\prime}$}
 \endpspicture
\notag \\
&=& \sum\limits_{ \substack{ a, c_1, c_2, g, f \\ a^{\prime}, c_1^{\prime}, c_2^{\prime}, g^{\prime}} }
\frac{\rho^{AC}_{\left(c_2, a, g, c_1, f \right) \left(c_2^{\prime},a^{\prime},g^{\prime},c_1^{\prime},f \right) }}{\left( d_{a} d_{a^{\prime}} d_{c_{1}} d_{c_{1}^{\prime}} d_{c_{2}} d_{c_{2}^{\prime}} d_{f}^2 \right)^{1/4}}
 \pspicture[shift=-2](-1.2,-1.8)(1.2,1.8)
  \small
  \psset{linewidth=0.9pt,linecolor=black,arrowscale=1.5,arrowinset=0.15}
  \psline(0.0,0.5)(0.8,1.5)
  \psline(0.0,0.5)(-0.8,1.5)
  \psline(-0.4,1)(0.0,1.5)
    \psline{->}(-0.4,1)(-0.1,1.375)
    \psline{->}(0.4,1.0)(0.7,1.375)
    \psline{->}(0,0.5)(-0.3,0.875)
    \psline{->}(0,0.5)(-0.7,1.375)
  \psset{linewidth=0.9pt,linecolor=black,arrowscale=1.5,arrowinset=0.15}
  \psline(0.0,0.0)(0.8,-1)
  \psline(0.0,0.0)(-0.8,-1)
  \psline(-0.4,-0.5)(0.0,-1)
    \psline{-<}(-0.4,-0.5)(-0.1,-0.875)
    \psline{-<}(0,0.0)(0.7,-0.875)
    \psline{-<}(0,0.0)(-0.3,-0.375)
    \psline{-<}(0,0.0)(-0.7,-0.875)
  \psline(0,0.0)(0,0.5)
  \psline{->}(0,0.0)(0,0.4)
  \rput[bl]{0}(0.15,0.05){$f$}
  \rput[bl]{0}(-0.5,0.5){$g$}
  \rput[bl]{0}(-0.95,1.6){$c_{2}$}
  \rput[bl]{0}(-0.1,1.6){$a_{\phantom{1}}$}
  \rput[bl]{0}(0.7,1.6){$c_{1}$}
  \rput[bl]{0}(-0.55,-0.3){$g^{\prime}$}
  \rput[bl]{0}(-0.95,-1.4){$c_{2}^{\prime}$}
  \rput[bl]{0}(-0.1,-1.4){$a_{\phantom{1}}^{\prime}$}
  \rput[bl]{0}(0.7,-1.4){$c_{1}^{\prime}$}
 \endpspicture
,
\label{eq:rho3}
\end{eqnarray}
where the second line is written in the standard basis, and the Greek indices labeling the internal states of the fusion/splitting spaces are left implicit to reduce clutter.

Applying a similar single probe analysis as before, the results involve sums of the following four diagrams
\begin{eqnarray}
&&
\pspicture[shift=-1.9](-2.3,-2)(1.1,1.5)
  \small
  \psline(-1.75,-0.5)(-1.75,0.5)
  \psline(-1.75,-0.5)(0.75,-0.5)
  \psline(-1.75,0.5)(0.75,0.5)
  \psline(0.75,-0.5)(0.75,0.5)
  \psset{linewidth=0.9pt,linecolor=black,arrowscale=1.5,arrowinset=0.15}
  \psline(0.5,0.5)(0.5,1.25)
  \psline(-0.5,0.5)(-0.5,1.25)
  \psline(0.5,-0.5)(0.5,-1.25)
  \psline(-0.5,-0.5)(-0.5,-1.25)
  \psline(-1.5,0.5)(-1.5,1.25)
  \psline(-1.5,-0.5)(-1.5,-1.25)
  \psline{->}(0.5,0.5)(0.5,1)
  \psline{->}(-0.5,0.5)(-0.5,1)
  \psline{-<}(0.5,-0.5)(0.5,-1)
  \psline{-<}(-0.5,-0.5)(-0.5,-1)
  \psline{->}(-1.5,0.5)(-1.5,1)
  \psline{-<}(-1.5,-0.5)(-1.5,-1)
  \rput[bl](0.4,1.3){$C_1$}
  \rput[bl](-0.6,1.35){$A$}
  \rput[bl](0.4,-1.7){$C_1^{\prime}$}
  \rput[bl](-0.6,-1.6){$A^{\prime}$}
  \rput[bl](-1.7,1.3){$C_2$}
  \rput[bl](-1.7,-1.7){$C_2^{\prime}$}
  \psellipse[linewidth=0.9pt,linecolor=black,border=0.05](0,0)(0.3,1.0)
  \psline{->}(0.279,-0.13)(0.28,-0.1)
  \psframe[linewidth=0.9pt,linecolor=white,border=0.1,fillcolor=white,fillstyle=solid](-0.3,-0.5)(-0.2,0.5)
  \psline(0,0.5)(-0.5,0.5)
  \psline(0,-0.5)(-0.5,-0.5)
  \rput[bl]{0}(-0.8,-0.15){$\rho^{AC}$}
  \scriptsize
  \rput[bl](0.35,-0.1){$B$}
 \endpspicture
, \qquad
 \pspicture[shift=-1.9](-2.3,-2)(1.1,1.5)
  \small
  \psline(-1.75,-0.5)(-1.75,0.5)
  \psline(-1.75,-0.5)(0.75,-0.5)
  \psline(-1.75,0.5)(0.75,0.5)
  \psline(0.75,-0.5)(0.75,0.5)
  \psset{linewidth=0.9pt,linecolor=black,arrowscale=1.5,arrowinset=0.15}
  \psline(0.5,0.5)(0.5,1.25)
  \psline(-0.5,0.5)(-0.5,1.25)
  \psline(0.5,-0.5)(0.5,-1.25)
  \psline(-0.5,-0.5)(-0.5,-1.25)
  \psline(-1.5,0.5)(-1.5,1.25)
  \psline(-1.5,-0.5)(-1.5,-1.25)
  \psline{->}(0.5,0.5)(0.5,1)
  \psline{->}(-0.5,0.5)(-0.5,1)
  \psline{-<}(0.5,-0.5)(0.5,-1)
  \psline{-<}(-0.5,-0.5)(-0.5,-1)
  \psline{->}(-1.5,0.5)(-1.5,1)
  \psline{-<}(-1.5,-0.5)(-1.5,-1)
  \rput[bl](0.4,1.3){$C_1$}
  \rput[bl](-0.6,1.35){$A$}
  \rput[bl](0.4,-1.7){$C_1^{\prime}$}
  \rput[bl](-0.6,-1.6){$A^{\prime}$}
  \rput[bl](-1.7,1.3){$C_2$}
  \rput[bl](-1.7,-1.7){$C_2^{\prime}$}
  \psellipse[linewidth=0.9pt,linecolor=black,border=0.05](-0.5,0)(0.7,1)
  \psline{->}(-1.175,-0.13)(-1.18,-0.1)
  \psframe[linewidth=0.9pt,linecolor=white,border=0.1,fillcolor=white,fillstyle=solid](0.3,-0.5)(0,0.5)
  \psline(-0.5,0.5)(0.5,0.5)
  \psline(-0.5,-0.5)(0.5,-0.5)
  \psline[border=0.1](-0.5,-0.6)(-0.5,-1.25)
  \psline{-<}(-0.5,-0.5)(-0.5,-1)
  \rput[bl]{0}(-0.8,-0.15){$\rho^{AC}$}
  \scriptsize
  \rput[bl](-1.55,-0.1){$B$}
 \endpspicture
, \notag \\
&&
\pspicture[shift=-1.9](-2.3,-2)(1.1,1.5)
  \small
  \psline(-1.75,-0.5)(-1.75,0.5)
  \psline(-1.75,-0.5)(0.75,-0.5)
  \psline(-1.75,0.5)(0.75,0.5)
  \psline(0.75,-0.5)(0.75,0.5)
  \psset{linewidth=0.9pt,linecolor=black,arrowscale=1.5,arrowinset=0.15}
  \psline(0.5,0.5)(0.5,1.25)
  \psline(-0.5,0.5)(-0.5,1.25)
  \psline(0.5,-0.5)(0.5,-1.25)
  \psline(-0.5,-0.5)(-0.5,-1.25)
  \psline(-1.5,0.5)(-1.5,1.25)
  \psline(-1.5,-0.5)(-1.5,-1.25)
  \psline{->}(0.5,0.5)(0.5,1)
  \psline{->}(-0.5,0.5)(-0.5,1)
  \psline{-<}(0.5,-0.5)(0.5,-1)
  \psline{-<}(-0.5,-0.5)(-0.5,-1)
  \psline{->}(-1.5,0.5)(-1.5,1)
  \psline{-<}(-1.5,-0.5)(-1.5,-1)
  \rput[bl](0.4,1.3){$C_1$}
  \rput[bl](-0.6,1.35){$A$}
  \rput[bl](0.4,-1.7){$C_1^{\prime}$}
  \rput[bl](-0.6,-1.6){$A^{\prime}$}
  \rput[bl](-1.7,1.3){$C_2$}
  \rput[bl](-1.7,-1.7){$C_2^{\prime}$}
  \psellipse[linewidth=0.9pt,linecolor=black,border=0.05](-0.5,0)(0.7,1)
  \psline{->}(-1.175,-0.13)(-1.18,-0.1)
  \psframe[linewidth=0.9pt,linecolor=white,border=0.1,fillcolor=white,fillstyle=solid](0.3,-0.5)(0,0.5)
  \psline(-0.5,0.5)(0.5,0.5)
  \psline(-0.5,-0.5)(0.5,-0.5)
  \psline[border=0.1](-0.5,0.6)(-0.5,1.25)
  \psline{->}(-0.5,0.5)(-0.5,1)
  \rput[bl]{0}(-0.8,-0.15){$\rho^{AC}$}
  \scriptsize
  \rput[bl](-1.55,-0.1){$B$}
 \endpspicture
, \qquad
\pspicture[shift=-1.9](-2.3,-2)(1.1,1.5)
  \small
  \psline(-1.75,-0.5)(-1.75,0.5)
  \psline(-1.75,-0.5)(0.75,-0.5)
  \psline(-1.75,0.5)(0.75,0.5)
  \psline(0.75,-0.5)(0.75,0.5)
  \psset{linewidth=0.9pt,linecolor=black,arrowscale=1.5,arrowinset=0.15}
  \psline(0.5,0.5)(0.5,1.25)
  \psline(-0.5,0.5)(-0.5,1.25)
  \psline(0.5,-0.5)(0.5,-1.25)
  \psline(-0.5,-0.5)(-0.5,-1.25)
  \psline(-1.5,0.5)(-1.5,1.25)
  \psline(-1.5,-0.5)(-1.5,-1.25)
  \psline{->}(0.5,0.5)(0.5,1)
  \psline{->}(-0.5,0.5)(-0.5,1)
  \psline{-<}(0.5,-0.5)(0.5,-1)
  \psline{-<}(-0.5,-0.5)(-0.5,-1)
  \psline{->}(-1.5,0.5)(-1.5,1)
  \psline{-<}(-1.5,-0.5)(-1.5,-1)
  \rput[bl](0.4,1.3){$C_1$}
  \rput[bl](-0.6,1.35){$A$}
  \rput[bl](0.4,-1.7){$C_1^{\prime}$}
  \rput[bl](-0.6,-1.6){$A^{\prime}$}
  \rput[bl](-1.7,1.3){$C_2$}
  \rput[bl](-1.7,-1.7){$C_2^{\prime}$}
  \psellipse[linewidth=0.9pt,linecolor=black,border=0.05](-1,0)(0.3,1.0)
  \psline{->}(-1.279,-0.13)(-1.28,-0.1)
  \psframe[linewidth=0.9pt,linecolor=white,border=0.1,fillcolor=white,fillstyle=solid](-0.8,-0.5)(-0.7,0.5)
  \psline(-1,0.5)(-0.5,0.5)
  \psline(-1,-0.5)(-0.5,-0.5)
  \rput[bl]{0}(-0.8,-0.15){$\rho^{AC}$}
  \scriptsize
  \rput[bl](-1.65,-0.1){$B$}
 \endpspicture
\label{eq:target_gen_probed}
\end{eqnarray}
weighted by the amplitudes for each configuration of the probe loop, corresponding to how the probe passes through the interferometer.

In order to evaluate the terms corresponding to these probe loop configurations, we must apply a more complicated sequence of $F$-moves to the target density matrix. We will not explicitly write out this sequence of $F$-moves, because it is cumbersome, but the steps should be clear from analogy with the previous analysis. We display here the most relevant intermediate stages of the diagrams in this sequence of $F$-moves:
\begin{equation}
\pspicture[shift=-2](-1.2,-1.6)(1.2,1.8)
  \small
  \psset{linewidth=0.9pt,linecolor=black,arrowscale=1.5,arrowinset=0.15}
  \psline(0.8,-1.0)(0.8,1.5)
  \psline(-0.8,-1.0)(-0.8,1.5)
  \psline(0.0,-1.0)(0.0,1.5)
    \psline{->}(-0.8,1.0)(-0.8,1.375)
    \psline{-<}(-0.8,-0.5)(-0.8,-0.875)
    \psline{->}(0.8,1.0)(0.8,1.375)
    \psline{-<}(0.8,-0.5)(0.8,-0.875)
    \psline{->}(0.0,1.0)(0.0,1.375)
    \psline{-<}(0.0,-0.5)(0.0,-0.875)
    \psline{->}(0.0,0.0)(0.0,0.375)
  \psset{linewidth=0.9pt,linecolor=black,arrowscale=1.5,arrowinset=0.15}
  \psline(0.0,0.0)(0.8,-0.5)
  \psline(0.0,0.5)(-0.8,1.0)
    \psline{->}(0,0.5)(-0.6,0.875)
    \psline{->}(0.8,-0.5)(0.2,-0.125)
  \rput[bl]{0}(0.15,0.1){$h_1$}
  \rput[bl]{0}(-0.6,0.3){$e_2$}
  \rput[bl]{0}(-0.95,1.6){$c_{2}$}
  \rput[bl]{0}(-0.1,1.6){$a_{\phantom{1}}$}
  \rput[bl]{0}(0.7,1.6){$c_{1}$}
  \rput[bl]{0}(0.2,-0.7){$e_1$}
  \rput[bl]{0}(-0.95,-1.4){$c_{2}^{\prime}$}
  \rput[bl]{0}(-0.1,-1.4){$a_{\phantom{1}}^{\prime}$}
  \rput[bl]{0}(0.7,-1.4){$c_{1}^{\prime}$}
 \endpspicture
\qquad \qquad \text{and} \qquad \qquad
\pspicture[shift=-2](-1.2,-1.6)(1.2,1.8)
  \small
  \psset{linewidth=0.9pt,linecolor=black,arrowscale=1.5,arrowinset=0.15}
  \psline(0.8,-1.0)(0.8,1.5)
  \psline(-0.8,-1.0)(-0.8,1.5)
  \psline(0.0,-1.0)(0.0,1.5)
    \psline{->}(-0.8,1.0)(-0.8,1.375)
    \psline{-<}(-0.8,-0.5)(-0.8,-0.875)
    \psline{->}(0.8,1.0)(0.8,1.375)
    \psline{-<}(0.8,-0.5)(0.8,-0.875)
    \psline{->}(0.0,1.0)(0.0,1.375)
    \psline{-<}(0.0,-0.5)(0.0,-0.875)
    \psline{->}(0.0,0.0)(0.0,0.375)
  \psset{linewidth=0.9pt,linecolor=black,arrowscale=1.5,arrowinset=0.15}
  \psline(0.0,1.0)(0.8,0.5)
  \psline(0.0,-0.5)(-0.8,0.0)
    \psline{->}(0.0,-0.5)(-0.6,-0.125)
    \psline{->}(0.8,0.5)(0.2,0.875)
  \rput[bl]{0}(-0.5,0.1){$h_2$}
  \rput[bl]{0}(-0.6,-0.7){$e_2$}
  \rput[bl]{0}(-0.95,1.6){$c_{2}$}
  \rput[bl]{0}(-0.1,1.6){$a_{\phantom{1}}$}
  \rput[bl]{0}(0.7,1.6){$c_{1}$}
  \rput[bl]{0}(0.2,0.3){$e_1$}
  \rput[bl]{0}(-0.95,-1.4){$c_{2}^{\prime}$}
  \rput[bl]{0}(-0.1,-1.4){$a_{\phantom{1}}^{\prime}$}
  \rput[bl]{0}(0.7,-1.4){$c_{1}^{\prime}$}
 \endpspicture
.
\label{eq:rho3_F_moved}
\end{equation}

The resulting factors multiplying the corresponding components of the density matrix are
\begin{eqnarray}
p_{h_1 h_2 e_1 e_2 ,b}^{\shortrightarrow } &=&\left| t_{1}\right| ^{2}\left|
r_{2}\right| ^{2}M_{e_1 b}+t_{1}r_{1}^{\ast }r_{2}^{\ast }t_{2}^{\ast
}e^{i\left( \theta _{\text{I}}-\theta _{\text{II}}\right) }M_{h_1 b }  \notag \\
&&+t_{1}^{\ast }r_{1}t_{2}r_{2}e^{-i\left( \theta _{\text{I}}-\theta _{\text{II}}\right)
}M_{h_2 b}^{\ast }+\left| r_{1}\right| ^{2}\left| t_{2}\right| ^{2} M_{e_2 b}
\end{eqnarray}%
\begin{eqnarray}
p_{h_1 h_2 e_1 e_2 ,b}^{\shortuparrow } &=&\left| t_{1}\right| ^{2}\left|
t_{2}\right| ^{2}M_{e_1 b}-t_{1}r_{1}^{\ast }r_{2}^{\ast }t_{2}^{\ast
}e^{i\left( \theta _{\text{I}}-\theta _{\text{II}}\right) }M_{h_1 b}  \notag \\
&&-t_{1}^{\ast }r_{1}t_{2}r_{2}e^{-i\left( \theta _{\text{I}}-\theta _{\text{II}}\right)
}M_{h_2 b}^{\ast }+\left| r_{1}\right| ^{2}\left| r_{2}\right| ^{2} M_{e_2 b}
,
\end{eqnarray}
where the anyonic charges $h_1$, $h_2$, $e_1$, and $e_2$ label the fusion channels indicated in the diagrams of Eq.~(\ref{eq:rho3_F_moved}). The diagrams in Eq.~(\ref{eq:rho3_F_moved}) represent the steps (within the sequence of $F$-moves) at which one can apply Eq.~(\ref{eq:loopaway}) to remove the four configurations of the probe loop shown in Eq.~(\ref{eq:target_gen_probed}). These four configurations, where the probe loop is linked on the $e_1$, $h_1$, $h_2$, and $e_2$ lines, respectively, give rise to the corresponding four terms in the expressions for $p_{h_1 h_2 e_1 e_2 ,b}^{s}$. When the probe anyons are allowed to carry different charge values, we can again simply replace these factors with their expectation values, which we denote as
\begin{equation}
p_{h_1 h_2 e_1 e_2 ,B}^{s} = \sum_{b} \Pr\nolimits_B(b) p_{h_1 h_2 e_1 e_2 ,b}^{s}
.
\end{equation}

A similar multi-probe analysis can be used to obtain the state resulting from sending $N$ probes through the interferometer. The asymptotic effect ($N \rightarrow \infty$) of running the interferometer is given by the anyonic charge sets
\begin{equation}
\mathcal{C}_{\kappa } \equiv \left\{ a\in \mathcal{C}:p_{aa00,B}^{\shortrightarrow }=p_{\kappa }\right\}
,
\end{equation}
the probability
\begin{equation}
\Pr\nolimits_{A}\left( \kappa \right) = \widetilde{\text{Tr}} \left[ \rho^{AC} \Pi^{A}_{\mathcal{C}_{\kappa }} \right]
\end{equation}
that the interferometry measurement will correspond to outcome $\kappa$ (i.e., that the collective charge of anyon(s) $A$ is in $\mathcal{C}_{\kappa }$), and the quantity
\begin{equation}
\Delta _{h_1 h_2 e_1 e_2,B}\left( p_{\kappa }\right) =\left\{
\begin{array}{cc}
\frac{1}{\Pr\nolimits_{A}\left( \kappa \right) } & \text{if }p_{h_1 h_2 e_1 e_2,B}^{\shortrightarrow }=1-p_{h_1 h_2 e_1 e_2 ,B}^{\shortuparrow }=p_{\kappa }\text{ and
}h_1,h_2 \in \mathcal{C}_{\kappa } \\
0 & \text{otherwise}%
\end{array}%
\right.
,
\end{equation}%
which determines the components of the target anyons' density matrix that survive after the interferometry measurement.
We emphasize that $h_1$ and $h_2$ are generally not the same as $a$ and $a^{\prime}$. However, the condition that $p_{h_1 h_2 e_1 e_2,B}^{\shortrightarrow }=1-p_{h_1 h_2 e_1 e_2 ,B}^{\shortuparrow }=p_{\kappa }$ and $h_1,h_2 \in \mathcal{C}_{\kappa }$ is equivalent to the condition that $M_{e_1 B}=M_{e_2 B}=1$, which also implies that $M_{h_1 B} = M_{h_2 B} = M_{a B} = M_{a^{\prime} B}$ and $a,a^{\prime} \in \mathcal{C}_{\kappa }$. When the probes can distinguish between all charge types, then each $\mathcal{C}_{\kappa }$ contains a single element and $M_{e_1 B}=M_{e_2 B}=1$ iff $e_1=e_2=0$. Thus, the (generically tuned) anyonic interferometer in the asymptotic limit selects a charge sector $\kappa$ with probability $\Pr_{A}(\kappa)$ and then: (1) projects the anyonic state onto the subspace where the $A$ anyons have collective anyonic charge in $\mathcal{C}_\kappa$, and (2) decoheres all anyonic entanglement pairwise between subsystems $A$, $C_1$, and $C_2$ that the probes can detect.

\section{$\omega$-Loops}
\label{sec:omega_loops}

We now consider $\omega$-loops and their specific properties which will be useful for the analysis in this paper. For this and the rest of the paper, we restrict modular tensor categories (MTCs), which are anyon models/UBTCs whose $S$-matrix is unitary and which correspond to a TQFT. We begin by recalling the definition of an $\omega_a$-loop in a MTC
\begin{equation}
\pspicture[shift=-0.55](-0.25,-0.1)(0.9,1.3)
\small
  \psset{linewidth=0.9pt,linecolor=black,arrowscale=1.5,arrowinset=0.15}
  \psellipse[linewidth=0.9pt,linecolor=black,border=0](0.4,0.5)(0.4,0.18)
  \psset{linewidth=0.9pt,linecolor=black,arrowscale=1.4,arrowinset=0.15}
  \psline{->}(0.2,0.37)(0.3,0.34)
  \rput[bl]{0}(0.0,0.0){$\omega_a$}
  \endpspicture
=
\pspicture[shift=-0.55](-0.25,-0.1)(0.9,1.3)
\small
  \psset{linewidth=0.9pt,linecolor=black,arrowscale=1.5,arrowinset=0.15}
  \psellipse[linewidth=0.9pt,linecolor=black,border=0](0.4,0.5)(0.4,0.18)
  \psset{linewidth=0.9pt,linecolor=black,arrowscale=1.4,arrowinset=0.15}
  \psline{-<}(0.2,0.37)(0.3,0.34)
  \rput[bl]{0}(0.0,0.0){$\omega_{\bar{a}}$}
  \endpspicture
= \sum_{x} S_{0a} S^{\ast}_{ax}
\pspicture[shift=-0.55](-0.25,-0.1)(0.9,1.3)
\small
  \psset{linewidth=0.9pt,linecolor=black,arrowscale=1.5,arrowinset=0.15}
  \psellipse[linewidth=0.9pt,linecolor=black,border=0](0.4,0.5)(0.4,0.18)
  \psset{linewidth=0.9pt,linecolor=black,arrowscale=1.4,arrowinset=0.15}
  \psline{->}(0.2,0.37)(0.3,0.34)
  \rput[bl]{0}(0.0,0.05){$x$}
  \endpspicture
\label{eq:omega_a}
\end{equation}
which (given the unitarity of the $S$-matrix in a MTC) acts as a projector on the total collective charge of anyonic charge lines passing through the loop, i.e.
\begin{equation}
\pspicture[shift=-0.55](-0.25,-0.1)(0.9,1.3)
\small
  \psset{linewidth=0.9pt,linecolor=black,arrowscale=1.5,arrowinset=0.15}
  \psline(0.4,0)(0.4,0.22)
  \psline(0.4,0.45)(0.4,1.2)
  \psellipse[linewidth=0.9pt,linecolor=black,border=0](0.4,0.5)(0.4,0.18)
  \psset{linewidth=0.9pt,linecolor=black,arrowscale=1.4,arrowinset=0.15}
  \psline{->}(0.2,0.37)(0.3,0.34)
\psline[linewidth=0.9pt,linecolor=black,border=2.5pt,arrows=->,arrowscale=1.5,
arrowinset=0.15](0.4,0.5)(0.4,1.1)
  \rput[bl]{0}(-0.2,0.0){$\omega_a$}
  \rput[tl]{0}(0.55,1.2){$b$}
  \endpspicture
= \delta_{ab}
\pspicture[shift=-0.55](0.05,-0.1)(1,1.3)
\small
  \psset{linewidth=0.9pt,linecolor=black,arrowscale=1.5,arrowinset=0.15}
  \psline(0.4,0)(0.4,1.2)
\psline[linewidth=0.9pt,linecolor=black,arrows=->,arrowscale=1.5,
arrowinset=0.15](0.4,0.5)(0.4,0.9)
  \rput[tl]{0}(0.52,1.0){$b$}
  \endpspicture
\label{eq:omega_a_projection}
.
\end{equation}
In other words, we can write the projector of $n$ anyons of (possibly indefinite) charges $A_1,\ldots,A_n$ onto definite collective topological charge $a$ by enclosing the charge lines of these anyons with an $\omega_a$ loop
\begin{equation}
\Pi_{a}^{(1 \ldots n)} =
 \pspicture[shift=-2.5](-0.6,-2)(2.8,2.4)
  \small
  \psset{linewidth=0.9pt,linecolor=black,arrowscale=1.5,arrowinset=0.15}
  \psline(0.0,1.75)(0.0,-1.0)
  \psline(2.0,1.75)(2.0,-1.0)
  \psline(0.8,1.75)(0.8,-1.0)
   \psline{->}(0.0,1.25)(0.0,1.625)
   \psline{->}(0.8,1.25)(0.8,1.625)
   \psline{->}(2.0,1.25)(2.0,1.625)
   \rput[bl]{0}(-0.2,1.85){$A_1$}
   \rput[bl]{0}(0.6,1.85){$A_2$}
   \rput[bl]{0}(1.25,1.85){$\ldots$}
   \rput[bl]{0}(1.8,1.85){$A_n$}
  \psellipse[linewidth=0.9pt,linecolor=black,border=0.1](1.0,0.5)(1.5,0.3)
  \psline[linewidth=0.9pt,linecolor=black,border=0.1](0.0,0.5)(0.0,1.0)
  \psline[linewidth=0.9pt,linecolor=black,border=0.1](0.8,0.5)(0.8,1.0)
  \psline[linewidth=0.9pt,linecolor=black,border=0.1](2.0,0.5)(2.0,1.0)
  \psset{linewidth=0.9pt,linecolor=black,arrowscale=1.4,arrowinset=0.15}
  \psline{->}(-0.2,0.33)(-0.1,0.3)
  \rput[bl]{0}(-0.6,0.0){$\omega_a$}
\endpspicture
.
\label{eq:omega_a_projector}
\end{equation}

We define a $\omega_{\mathcal{B}}$-loop, which projects onto the subset $\mathcal{B}$ of charges, in the obvious way by summing over $\omega$-loops
\begin{equation}
\pspicture[shift=-0.55](-0.25,-0.1)(0.9,1.3)
\small
  \psset{linewidth=0.9pt,linecolor=black,arrowscale=1.5,arrowinset=0.15}
  \psellipse[linewidth=0.9pt,linecolor=black,border=0](0.4,0.5)(0.4,0.18)
  \psset{linewidth=0.9pt,linecolor=black,arrowscale=1.4,arrowinset=0.15}
  \psline{->}(0.2,0.37)(0.3,0.34)
  \rput[bl]{0}(-0.1,-0.05){$\omega_{\mathcal{B}}$}
  \endpspicture
= \sum_{a \in \mathcal{B}}
\pspicture[shift=-0.55](-0.25,-0.1)(0.9,1.3)
\small
  \psset{linewidth=0.9pt,linecolor=black,arrowscale=1.5,arrowinset=0.15}
  \psellipse[linewidth=0.9pt,linecolor=black,border=0](0.4,0.5)(0.4,0.18)
  \psset{linewidth=0.9pt,linecolor=black,arrowscale=1.4,arrowinset=0.15}
  \psline{->}(0.2,0.37)(0.3,0.34)
  \rput[bl]{0}(0.0,0.0){$\omega_{a}$}
  \endpspicture
\label{eq:omega_B}
\end{equation}

\subsection{Effect of the Anyonic Interferometer}

Using $\omega$-loops, the effect of running the anyonic interferometer on the target system can be expressed in a compact form, which provides a clarifying visual understanding of the effect. In the limit where the number of probes sent through the interferometer $N \rightarrow \infty$, the fraction $r=n/N$ of probes measured in the $s=\shortrightarrow$ detector will be found to go to $r=p_{\kappa}$ with probability $\text{Pr}_{AC}(\kappa)$, and the target system will have resulting density matrix
\begin{equation}
{\rho}_{\kappa}^{AC} = \frac{1}{{\Pr}_{AC}(\kappa)}
\pspicture[shift=-2.9](-2.8,-3)(2.8,2.5)
  \small
  \psline(-2.5,-0.5)(-2.5,0.5)
  \psline(-2.5,-0.5)(2.5,-0.5)
  \psline(-2.5,0.5)(2.5,0.5)
  \psline(2.5,-0.5)(2.5,0.5)
  \psset{linewidth=0.9pt,linecolor=black,arrowscale=1.5,arrowinset=0.15}
  \psline(2.0,0.5)(2.0,2)
  \psline(0.0,0.5)(0.0,2)
  \psline(2.0,-0.5)(2.0,-2)
  \psline(0.0,-0.5)(0.0,-2)
  \psline(-2.0,0.5)(-2.0,2)
  \psline(-2.0,-0.5)(-2.0,-2)
  \psline{->}(2.0,1.5)(2.0,1.75)
  \psline{->}(0.0,1.5)(0.0,1.75)
  \psline{-<}(2.0,-1.5)(2.0,-1.75)
  \psline{-<}(0.0,-1.5)(0.0,-1.75)
  \psline{->}(-2.0,1.5)(-2.0,1.75)
  \psline{-<}(-2.0,-1.5)(-2.0,-1.75)
  \rput[bl](1.9,2.05){$C_1$}
  \rput[bl](-0.1,2.1){$A$}
  \rput[bl](1.9,-2.45){$C_1^{\prime}$}
  \rput[bl](-0.1,-2.35){$A^{\prime}$}
  \rput[bl](-2.3,2.05){$C_2$}
  \rput[bl](-2.3,-2.45){$C_2^{\prime}$}
  \psbezier[linewidth=0.9pt,linecolor=black,border=0.05](-0.5,0.0)(-1.25,1.5)(-1.0,2.0)(-0.25,0.5)
  \psbezier[linewidth=0.9pt,linecolor=black,border=0.05](-0.5,0.0)(0.25,-1.5)(1.25,-2.5)(0.25,-0.5)
   \psline{<-}(-0.5,0.0)(-0.55,0.1)
  \psline[linewidth=0.9pt,linecolor=black,border=0.1](-0.4,0.5)(-0.2,0.5)
  \psline[linewidth=0.9pt,linecolor=black,border=0.1](0.4,-0.5)(0.2,-0.5)
  \psbezier[linewidth=0.9pt,linecolor=black,border=0.05](0.5,0.0)(1.5,2.0)(1.0,2.0)(0.1,0.5)
  \psbezier[linewidth=0.9pt,linecolor=black,border=0.05](0.5,0.0)(-0.5,-2.0)(-1.3,-2.0)(-0.4,-0.5)
  \psline[linewidth=0.9pt,linecolor=black,border=0.1](0.4,0.5)(0.05,0.5)
  \psline[linewidth=0.9pt,linecolor=black,border=0.1](-0.5,-0.5)(-0.35,-0.5)
   \psline{->}(0.5,0.0)(0.55,0.1)
  \psellipse[linewidth=0.9pt,linecolor=black,border=0.05](1.5,0.0)(0.3,1.5)
  \psline{->}(1.215,-0.13)(1.215,-0.1)
  \psframe[linewidth=0.9pt,linecolor=white,border=0.1,fillcolor=white,fillstyle=solid](1.7,-0.5)(1.8,0.5)
  \psline(2.0,0.5)(1.5,0.5)
  \psline(2.0,-0.5)(1.5,-0.5)
  \psellipse[linewidth=0.9pt,linecolor=black,border=0.05](-1.5,0.0)(-0.3,1.5)
  \psline{->}(-1.19,-0.13)(-1.19,-0.1)
  \psframe[linewidth=0.9pt,linecolor=white,border=0.1,fillcolor=white,fillstyle=solid](-1.7,-0.5)(-1.8,0.5)
  \psline(-2.0,0.5)(-1.5,0.5)
  \psline(-2.0,-0.5)(-1.5,-0.5)
  \rput[bl]{0}(-0.3,-0.15){$\rho^{AC}$}
  \scriptsize
  \rput[bl](1.35,-0.4){$\omega_{\mathcal{B}_0}$}
  \rput[bl](0.6,-0.1){$\omega_{\mathcal{C}_\kappa}$}
  \rput[bl](-1.8,-0.4){$\omega_{\mathcal{B}_0}$}
  \rput[bl](-1.0,-0.1){$\omega_{\mathcal{C}_\kappa}$}
 \endpspicture
\label{eq:target_projected_omega}
\end{equation}
where $\mathcal{B}_0 = \left\{ a \in \mathcal{C} : M_{aB} = 1 \right\}$.
Note that the four $\omega$-loops in this expression (none of which are linked with each other) are in exactly the same four probe loop configurations from Eq.~(\ref{eq:target_gen_probed}).

\subsection{Braiding}

We can also use $\omega$-loops to reexpress certain braiding operations. For example, the pure braid of two anyons is given by
\begin{eqnarray}
P_{ab} &=& R_{ba} R_{ab} = \sum_{c,\mu} \frac{\theta_{c}}{\theta_a \theta_b} \left| a,b;c,\mu \right\rangle \left\langle a,b;c,\mu \right| = \sum_{c} \frac{\theta_{c}}{\theta_a \theta_b} \Pi_{c}^{(ab)}
\notag \\
&=&
\pspicture[shift=-1.3](-0.2,-1.2)(1.2,1.6)
  \psset{linewidth=0.9pt,linecolor=black,arrowscale=1.5,arrowinset=0.15}
  \psline{->}(1,0)(0.1,0.9)
  \psline(0,0)(1,1)
  \psline(1,0)(0,1)
  \psline(0,-1)(1,0)
  \psline(1,-1)(0,0)
  \psline[border=2pt](0,-1)(.7,-0.3)
  \psline[border=2pt]{->}(0.3,0.3)(0.9,0.9)
  \rput[tl]{0}(-0.1,1.3){$a$}
  \rput[tr]{0}(1.1,1.4){$b$}
  \endpspicture
= \sum_{c,\mu} \frac{\theta_{c}}{\theta_a \theta_b} \sqrt{ \frac{d_c}{d_a d_b} }
 \pspicture[shift=-1](0,-0.9)(1.3,1.3)
 \small
  \psset{linewidth=0.9pt,linecolor=black,arrowscale=1.5,arrowinset=0.15}
  \psline{->}(0.7,0)(0.7,0.45)
  \psline(0.7,0)(0.7,0.55)
  \psline(0.7,0.55) (0.25,1)
  \psline{->}(0.7,0.55)(0.3,0.95)
  \psline(0.7,0.55) (1.15,1)
  \psline{->}(0.7,0.55)(1.1,0.95)
  \rput[bl]{0}(0.38,0.2){$c$}
  \rput[bl]{0}(1.05,1.1){$b$}
  \rput[bl]{0}(0.15,1.1){$a$}
  \psline(0.7,0) (0.25,-0.45)
  \psline{-<}(0.7,0)(0.35,-0.35)
  \psline(0.7,0) (1.15,-0.45)
  \psline{-<}(0.7,0)(1.05,-0.35)
  \rput[bl]{0}(1.05,-0.8){$b$}
  \rput[bl]{0}(0.15,-0.8){$a$}
  \scriptsize
  \rput[bl]{0}(0.82,0.38){$\mu$}
  \rput[bl]{0}(0.82,-0.02){$\mu$}
  \endpspicture
= \sum_{c} \frac{\theta_{c}}{\theta_a \theta_b}
\pspicture[shift=-1](-0.5,-0.9)(1.7,1.3)
 \small
  \psset{linewidth=0.9pt,linecolor=black,arrowscale=1.5,arrowinset=0.15}
  \psline(1.15,-0.45)(1.15,1)
  \psline(0.25,-0.45)(0.25,1)
  \psline{->}(0.25,0.7)(0.25,0.9)
  \rput[bl]{0}(1.05,1.1){$b$}
  \rput[bl]{0}(0.15,1.1){$a$}
  \psellipse[linewidth=0.9pt,linecolor=black,border=0.1](0.7,0.3)(.8,0.2)
  \psline[linewidth=0.9pt,linecolor=black,arrows=->,arrowscale=1.5,arrowinset=0.15,border=0.1](1.15,0.3)(1.15,0.9)
  \psline[linewidth=0.9pt,linecolor=black,arrows=->,arrowscale=1.5,arrowinset=0.15,border=0.1](0.25,0.3)(0.25,0.9)
  \psline{->}(0.2,0.15)(0.3,0.13)
  \rput[bl]{0}(-0.3,-0.1){$\omega_{c}$}
  \endpspicture
\label{eq:purebraid}
,
\end{eqnarray}
which expresses the ``ribbon property,'' relating the pure braid to the topological spins (a.k.a. twist factors)
\begin{equation}
\theta _{a}=\theta _{\bar{a}}=d_{a}^{-1}\widetilde{\text{Tr}} \left[ R_{aa} \right]
=\sum\limits_{c,\mu } \frac{d_{c}}{d_{a}}\left[ R_{c}^{aa}\right] _{\mu \mu }
= \frac{1}{d_{a}}
\pspicture[shift=-0.5](-1.3,-0.6)(1.5,0.6)
\small
  \psset{linewidth=0.9pt,linecolor=black,arrowscale=1.5,arrowinset=0.15}
  \psarc[linewidth=0.9pt,linecolor=black] (0.7071,0.0){0.5}{-135}{135}
  \psarc[linewidth=0.9pt,linecolor=black] (-0.7071,0.0){0.5}{45}{315}
  \psline(-0.3536,0.3536)(0.3536,-0.3536)
  \psline[border=2.3pt](-0.3536,-0.3536)(0.3536,0.3536)
  \psline[border=2.3pt]{->}(-0.3536,-0.3536)(0.0,0.0)
  \rput[bl]{0}(-0.2,-0.5){$a$}
  \endpspicture
,
\end{equation}
which are equal to roots of unity.

This expression can be generalized to the operation that braids $n$ anyons completely around each other (and back to their original positions)
\begin{eqnarray}
P_{a_1 \ldots a_n} &=& \sum_{a} \frac{\theta_{a}}{\theta_{a_1} \ldots \theta_{a_n}} \Pi_{a}^{(a_1 \ldots a_n)}
\notag \\
&=&
\pspicture[shift=-2.5](-0.2,-2.4)(2.3,2.7)
  \small
  \psset{linewidth=0.9pt,linecolor=black,arrowscale=1.5,arrowinset=0.15}
  \psline(0.0,-2.0)(2.0,0.0)
  \psline(0.5,-2.0)(2.0,-0.5)
  \psline(1.5,-2.0)(2.0,-1.5)
  \psline(2.0,-2.0)(0.0,0.0)
  \psline(2.0,0.0)(0.0,2.0)
  \psline(2.0,-0.5)(0.0,1.5)
  \psline(2.0,-1.5)(0.0,0.5)
  \psline(0.0,0.0)(2.0,2.0)
  \psline(0.0,1.5)(0.5,2.0)
  \psline(0.0,0.5)(1.5,2.0)
   \psline{->}(1.5,1.5)(1.8,1.8)
   \psline{->}(1.0,1.5)(1.3,1.8)
   \psline{->}(0.5,-2.0)(0.9,-1.6)
   \psline{->}(0.0,-2.0)(0.4,-1.6)
 \psline[linewidth=0.9pt,linecolor=black,border=0.1](0.8,-1.2)(1.8,-0.2)
 \psline[linewidth=0.9pt,linecolor=black,border=0.1](1.0,-1.5)(1.8,-0.7)
 \psline[linewidth=0.9pt,linecolor=black,border=0.1](1.7,-1.8)(1.8,-1.7)
 \psline[linewidth=0.9pt,linecolor=black,border=0.1](0.2,0.2)(1.5,1.5)
 \psline[linewidth=0.9pt,linecolor=black,border=0.1](0.2,0.7)(1.0,1.5)
 \psline[linewidth=0.9pt,linecolor=black,border=0.1](0.2,1.7)(0.3,1.8)
   \rput[bl]{0}(-0.2,2.25){$a_1$}
   \rput[bl]{0}(0.3,2.25){$a_2$}
   \rput[bl]{0}(1.1,2.25){$\ldots$}
   \rput[bl]{0}(0.9,-2.0){$\ldots$}
   \rput[bl]{0}(1.8,2.25){$a_n$}
\endpspicture
= \sum_{a} \frac{\theta_{a}}{\theta_{a_1} \ldots \theta_{a_n}}
\pspicture[shift=-2.5](-0.9,-2)(2.8,2.4)
  \small
  \psset{linewidth=0.9pt,linecolor=black,arrowscale=1.5,arrowinset=0.15}
  \psline(0.0,1.75)(0.0,-1.0)
  \psline(2.0,1.75)(2.0,-1.0)
  \psline(0.5,1.75)(0.5,-1.0)
  \psline(1.5,1.75)(1.5,-1.0)
   \psline{->}(0.0,1.25)(0.0,1.625)
   \psline{->}(0.5,1.25)(0.5,1.625)
   \psline{->}(2.0,1.25)(2.0,1.625)
   \psline{->}(1.5,1.25)(1.5,1.625)
   \rput[bl]{0}(-0.2,1.85){$a_1$}
   \rput[bl]{0}(0.3,1.85){$a_2$}
   \rput[bl]{0}(1.1,1.85){$\ldots$}
   \rput[bl]{0}(0.8,-1.0){$\ldots$}
   \rput[bl]{0}(1.8,1.85){$a_n$}
  \psellipse[linewidth=0.9pt,linecolor=black,border=0.1](1.0,0.5)(1.5,0.3)
  \psline[linewidth=0.9pt,linecolor=black,border=0.1](0.0,0.5)(0.0,1.0)
  \psline[linewidth=0.9pt,linecolor=black,border=0.1](0.5,0.5)(0.5,1.0)
  \psline[linewidth=0.9pt,linecolor=black,border=0.1](1.5,0.5)(1.5,1.0)
  \psline[linewidth=0.9pt,linecolor=black,border=0.1](2.0,0.5)(2.0,1.0)
  \psset{linewidth=0.9pt,linecolor=black,arrowscale=1.4,arrowinset=0.15}
  \psline{->}(-0.2,0.33)(-0.1,0.3)
  \rput[bl]{0}(-0.7,0.1){$\omega_a$}
\endpspicture
\label{eq:n_strand_purebraid}
.
\end{eqnarray}
This $n$ strand ribbon property can be obtained iteratively from the $2$ strand ribbon property. We note that the direction of the arrow of the $\omega$-loop is arbitrary in this expression, since $\theta_a = \theta_{\bar{a}}$. The inverse pure braid is obtained by conjugating the topological spins, i.e.
\begin{equation}
P^{-1}_{a_1 \ldots a_n} = P^{\dagger}_{a_1 \ldots a_n} = \sum_{a} \theta_{a}^{-1} \theta_{a_1} \ldots \theta_{a_n} \Pi_{a}^{(a_1 \ldots a_n)}
.
\end{equation}

We can also write the ``twist'' operator $\Theta$ for $n$ anyons in terms of $\omega$-loops as
\begin{eqnarray}
\Theta_{a_1 \ldots a_n} &=& \sum_{a} \theta_{a} \Pi_{a}^{(a_1 \ldots a_n)}
\notag \\
&=&
\pspicture[shift=-2.2](-0.5,-2.2)(3.8,2.6)
  \small
  \psset{linewidth=0.9pt,linecolor=black,arrowscale=1.5,arrowinset=0.15}
  \psline(0.0,2.0)(0.0,1.5)
  \psline(0.5,2.0)(0.5,1.5)
  \psline(1.5,2.0)(1.5,1.5)
   \psline{->}(0.0,1.5)(0.0,1.8)
   \psline{->}(0.5,1.5)(0.5,1.8)
   \psline{->}(1.5,1.5)(1.5,1.8)
  \psline(0.0,-2.0)(0.0,-1.5)
  \psline(0.5,-2.0)(0.5,-1.5)
  \psline(1.5,-2.0)(1.5,-1.5)
   \rput[bl]{0}(-0.2,2.1){$a_1$}
   \rput[bl]{0}(0.3,2.1){$a_2$}
   \rput[bl]{0}(0.8,2.1){$\ldots$}
   \rput[bl]{0}(0.8,-2.0){$\ldots$}
   \rput[bl]{0}(1.3,2.1){$a_n$}
  \psbezier[linewidth=0.9pt,linecolor=black,border=0.1](1.5,1.5)(1.5,-0.5)(2.5,-0.5)(2.5,0.0)
  \psbezier[linewidth=0.9pt,linecolor=black,border=0.1](0.5,1.5)(0.5,-1.5)(3.0,-1.5)(3.0,0.0)
  \psbezier[linewidth=0.9pt,linecolor=black,border=0.1](0.0,1.5)(0.0,-2.0)(3.5,-2.0)(3.5,0.0)
  \psbezier[linewidth=0.9pt,linecolor=black,border=0.1](1.5,-1.5)(1.5,0.5)(2.5,0.5)(2.5,0.0)
  \psbezier[linewidth=0.9pt,linecolor=black,border=0.1](0.5,-1.5)(0.5,1.5)(3.0,1.5)(3.0,0.0)
  \psbezier[linewidth=0.9pt,linecolor=black,border=0.1](0.0,-1.5)(0.0,2.0)(3.5,2.0)(3.5,0.0)
\endpspicture
= \sum_{a} \theta_{a}
\pspicture[shift=-2.5](-0.9,-2)(2.8,2.4)
  \small
  \psset{linewidth=0.9pt,linecolor=black,arrowscale=1.5,arrowinset=0.15}
  \psline(0.0,1.75)(0.0,-1.0)
  \psline(2.0,1.75)(2.0,-1.0)
  \psline(0.8,1.75)(0.8,-1.0)
   \psline{->}(0.0,1.25)(0.0,1.625)
   \psline{->}(0.8,1.25)(0.8,1.625)
   \psline{->}(2.0,1.25)(2.0,1.625)
   \rput[bl]{0}(-0.2,1.85){$a_1$}
   \rput[bl]{0}(0.6,1.85){$a_2$}
   \rput[bl]{0}(1.25,1.85){$\ldots$}
   \rput[bl]{0}(1.25,-1.0){$\ldots$}
   \rput[bl]{0}(1.8,1.85){$a_n$}
  \psellipse[linewidth=0.9pt,linecolor=black,border=0.1](1.0,0.5)(1.5,0.3)
  \psline[linewidth=0.9pt,linecolor=black,border=0.1](0.0,0.5)(0.0,1.0)
  \psline[linewidth=0.9pt,linecolor=black,border=0.1](0.8,0.5)(0.8,1.0)
  \psline[linewidth=0.9pt,linecolor=black,border=0.1](2.0,0.5)(2.0,1.0)
  \psset{linewidth=0.9pt,linecolor=black,arrowscale=1.4,arrowinset=0.15}
  \psline{->}(-0.2,0.33)(-0.1,0.3)
  \rput[bl]{0}(-0.7,0.1){$\omega_a$}
\endpspicture
\label{eq:n_strand_twist}
.
\end{eqnarray}
The direction of the arrow of the $\omega$-loop is also arbitrary in this expression, and the inverse twist is obtained by conjugating the topological spins, i.e.
\begin{equation}
\Theta^{-1}_{a_1 \ldots a_n} =\Theta^{\dagger}_{a_1 \ldots a_n} = \sum_{a} \theta_{a}^{-1}  \Pi_{a}^{(a_1 \ldots a_n)}
.
\end{equation}
The $n$ strand ribbon property can also be obtained from this relation, as one can see that
\begin{equation}
P_{a_1 \ldots a_n} = \Theta_{a_1}^{-1} \ldots \Theta_{a_n}^{-1} \Theta_{a_1 \ldots a_n},
\end{equation}
where the twist operator acting on a single anyon is equivalent to multiplying by the topological spin, i.e. $\Theta_{a_j} = \theta_{a_j}$. The twist operator here is clearly related to the topological $T$-matrix, which has matrix elements $T_{ab} = \theta_{a} \delta_{ab}$. For a modular theory, the topological $S$ and $T$ matrices of the MTC are the corresponding TQFT's (projective) representations of the $S$ and $T$ generators of modular transformations.

We can write the twist operators more compactly by defining $\tau^{m}$-loops to be loops that have the effect of $m$ applications of the twist operator for all anyons whose charge lines pass through the loop, which can be written in terms of $\omega$-loops as
\begin{equation}
\pspicture[shift=-0.55](-0.25,-0.1)(0.9,1.3)
\small
  \psset{linewidth=0.9pt,linecolor=black,arrowscale=1.5,arrowinset=0.15}
  \psellipse[linewidth=0.9pt,linecolor=black,border=0](0.4,0.5)(0.4,0.18)
  \psset{linewidth=0.9pt,linecolor=black,arrowscale=1.4,arrowinset=0.15}
  \psline{->}(0.2,0.37)(0.3,0.34)
  \rput[bl]{0}(-0.4,0.0){$\tau^{m}$}
\endpspicture
= \sum_{a} \theta_a^{m}
\pspicture[shift=-0.55](-0.25,-0.1)(0.9,1.3)
\small
  \psset{linewidth=0.9pt,linecolor=black,arrowscale=1.5,arrowinset=0.15}
  \psellipse[linewidth=0.9pt,linecolor=black,border=0](0.4,0.5)(0.4,0.18)
  \psset{linewidth=0.9pt,linecolor=black,arrowscale=1.4,arrowinset=0.15}
  \psline{->}(0.2,0.37)(0.3,0.34)
  \rput[bl]{0}(0.0,0.0){$\omega_a$}
\endpspicture
= \sum_{x} \left[ \tau^{m} \right]_{x}
\pspicture[shift=-0.55](-0.25,-0.1)(0.9,1.3)
\small
  \psset{linewidth=0.9pt,linecolor=black,arrowscale=1.5,arrowinset=0.15}
  \psellipse[linewidth=0.9pt,linecolor=black,border=0](0.4,0.5)(0.4,0.18)
  \psset{linewidth=0.9pt,linecolor=black,arrowscale=1.4,arrowinset=0.15}
  \psline{->}(0.2,0.37)(0.3,0.34)
  \rput[bl]{0}(0.0,0.05){$x$}
  \endpspicture
\label{eq:T_loop}
,
\end{equation}
where
\begin{equation}
\left[ \tau^{m} \right]_{x} = \sum_{a} \theta_a^{m} S_{0a} S^{\ast}_{ax} = \left[ S T^{m} S^{\dagger} \right]_{0x}
,
\end{equation}
and the direction of the arrow of a $\tau$-loop is also arbitrary.
With this definition, we can write the $m$-twist operator as
\begin{equation}
\Theta^{m}_{a_1 \ldots a_n} =
\pspicture[shift=-2.5](-0.9,-2)(2.8,2.4)
  \small
  \psset{linewidth=0.9pt,linecolor=black,arrowscale=1.5,arrowinset=0.15}
  \psline(0.0,1.75)(0.0,-1.0)
  \psline(2.0,1.75)(2.0,-1.0)
  \psline(0.8,1.75)(0.8,-1.0)
   \psline{->}(0.0,1.25)(0.0,1.625)
   \psline{->}(0.8,1.25)(0.8,1.625)
   \psline{->}(2.0,1.25)(2.0,1.625)
   \rput[bl]{0}(-0.2,1.85){$a_1$}
   \rput[bl]{0}(0.6,1.85){$a_2$}
   \rput[bl]{0}(1.25,1.85){$\ldots$}
   \rput[bl]{0}(1.25,-1.0){$\ldots$}
   \rput[bl]{0}(1.8,1.85){$a_n$}
  \psellipse[linewidth=0.9pt,linecolor=black,border=0.1](1.0,0.5)(1.5,0.3)
  \psline[linewidth=0.9pt,linecolor=black,border=0.1](0.0,0.5)(0.0,1.0)
  \psline[linewidth=0.9pt,linecolor=black,border=0.1](0.8,0.5)(0.8,1.0)
  \psline[linewidth=0.9pt,linecolor=black,border=0.1](2.0,0.5)(2.0,1.0)
  \psset{linewidth=0.9pt,linecolor=black,arrowscale=1.4,arrowinset=0.15}
  \psline{->}(-0.2,0.33)(-0.1,0.3)
  \rput[bl]{0}(-0.8,0.0){$\tau^{m}$}
\endpspicture
\label{eq:n_strand_twist_tau}
\end{equation}
and the $m$-pure braid operator as
\begin{equation}
P^{m}_{a_1 \ldots a_n} = \theta_{a_1}^{-m} \ldots \theta_{a_n}^{-m}
\pspicture[shift=-2.5](-0.9,-2)(2.8,2.4)
  \small
  \psset{linewidth=0.9pt,linecolor=black,arrowscale=1.5,arrowinset=0.15}
  \psline(0.0,1.75)(0.0,-1.0)
  \psline(2.0,1.75)(2.0,-1.0)
  \psline(0.8,1.75)(0.8,-1.0)
   \psline{->}(0.0,1.25)(0.0,1.625)
   \psline{->}(0.8,1.25)(0.8,1.625)
   \psline{->}(2.0,1.25)(2.0,1.625)
   \rput[bl]{0}(-0.2,1.85){$a_1$}
   \rput[bl]{0}(0.6,1.85){$a_2$}
   \rput[bl]{0}(1.25,1.85){$\ldots$}
   \rput[bl]{0}(1.25,-1.0){$\ldots$}
   \rput[bl]{0}(1.8,1.85){$a_n$}
  \psellipse[linewidth=0.9pt,linecolor=black,border=0.1](1.0,0.5)(1.5,0.3)
  \psline[linewidth=0.9pt,linecolor=black,border=0.1](0.0,0.5)(0.0,1.0)
  \psline[linewidth=0.9pt,linecolor=black,border=0.1](0.8,0.5)(0.8,1.0)
  \psline[linewidth=0.9pt,linecolor=black,border=0.1](2.0,0.5)(2.0,1.0)
  \psset{linewidth=0.9pt,linecolor=black,arrowscale=1.4,arrowinset=0.15}
  \psline{->}(-0.2,0.33)(-0.1,0.3)
  \rput[bl]{0}(-0.8,0.0){$\tau^{m}$}
\endpspicture
\label{eq:n_strand_purebraid_tau}
\end{equation}

\section{Twisted Interferometers}
\label{sec:twisted_interferometers}

\begin{figure}[t!]
\begin{center}
  \includegraphics[scale=0.5]{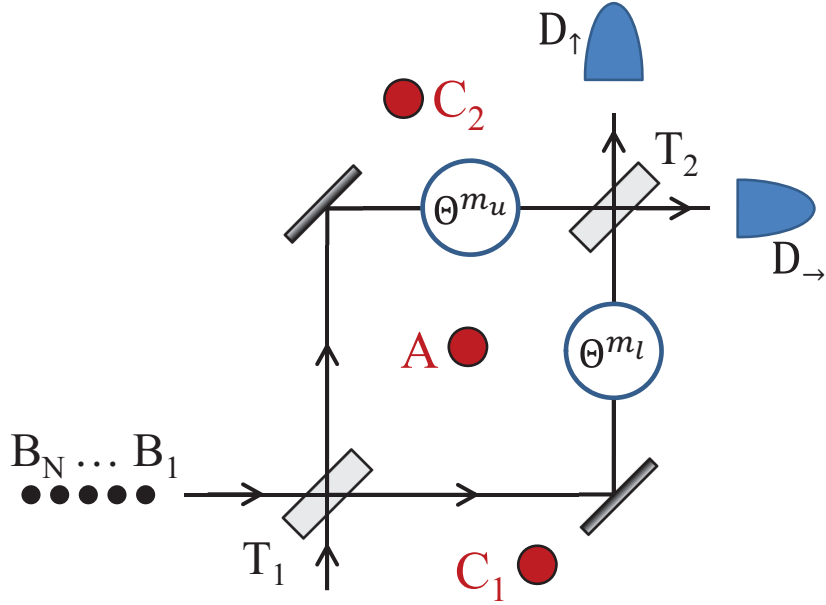}
  \caption{An idealized Mach-Zehnder twisted interferometer for an anyonic system. The target anyons (collectively denoted $A$) in the central region share entanglement only with the anyon(s) $C_1$ in the region below the interferometer and $C_2$ in the region above the interferometer. The probe anyons traveling the lower path through the interferometer twist with each other $m_l$ times and those traveling through the upper path twist with each other $m_u$ times, as represented by the multiple twist operators $\Theta^{m_l}$ and $\Theta^{m_u}$, respectively. The probe anyons $B_{1},\ldots ,B_{N}$ sent through the interferometer are detected at one of the two possible outputs by $D_{s}$.}
  \label{fig:twisted_int1}
\end{center}
\end{figure}

\begin{figure}[t!]
\begin{center}
  \includegraphics[scale=0.6]{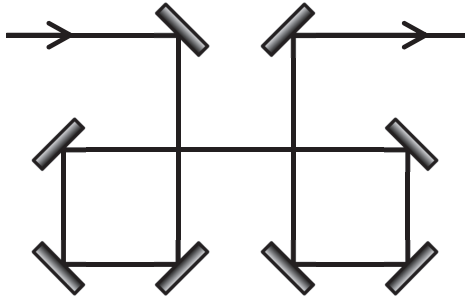}
  \caption{A doubly twisted path that may be used to implement a double twist $\Theta^2$ or pure-braid $P^2$ operation for probe anyons in an idealized anyonic system.}
  \label{fig:Double_Twist}
\end{center}
\end{figure}

We now consider a generalization of anyonic interferometers where the probe anyons that travel through the lower path in the interferometer execute an integer $m_l$ number of (counter-clockwise) twists around each other, and the probe anyons that travel through the upper path execute $m_r$ twists around each other (and there is no mutual twisting between probe anyons that travel through different paths). We call this interferometric operation involving probe anyons twisting around each other as part of the interference process: ``twisted interferometry.'' We will show that such twisting crucially modifies the effect of running an interferometer. We represent this twisted interferometer schematically in Fig.~\ref{fig:twisted_int1} by introducing elements into the lower and upper paths of probes through the interferometer which generate the multiple twist operators, $\Theta^{m_l}$ and $\Theta^{m_u}$, respectively.
We emphasize that each probe anyon will have some amplitude for passing through each path, so these twisting operations of probe anyons are not performed deterministically. To accomplish this, the interferometer must include some sort of twists in the two paths of the probe anyons through the interferometer, after the first beam splitter. Such twist operations can be implemented by appropriately modifying the paths. An example of a doubly twisted path is shown in Fig.~\ref{fig:Double_Twist}. Moreover, these twisted paths must allow \emph{all} of the probes traveling though each path to wind around each other, and all of the probes must be sent through the interferometer in rapid enough succession so as to ensure that this indeed occurs. In this section, we will focus on evaluating the effect of this interferometer on the target system, and defer further discussion of physical implementation issues to Section~\ref{sec:engineer_twist}.

The analysis can be initiated in the same way as for the ``untwisted'' interferometers. However, in twisted interferometers, the effect of sending $N$ probe anyons through the interferometer cannot similarly be obtained by simply using the product $\prod_{j=1}^{N} p^{s_j}_{h_1 h_2 e_1 e_2 ,b_j}$ of factors corresponding to each probe anyon, as in Eq.~(\ref{eq:rhoA_N}). This is because the twisting of probe anyons with each other creates correlations which prevent the factorization of the probe anyons' effect, i.e. the resulting probe anyon loops are linked with each other and so each one's effect cannot be individually evaluated. At first sight, the resulting configurations of probe loops may seem hopelessly complicated for evaluation, since there is a sum over $4^N$ possible probe loop configurations (each probe can take one of two paths through the interferometer, and this is doubled in the diagrams, as the two possibilities may occur for each of the bra and the ket of each probe anyon in the diagrammatic evaluation), and each configuration involves different probe anyons being twisted with each other.

However, by utilizing the expression in Eq.~(\ref{eq:n_strand_twist_tau}), we can replace the probe anyons' pure twist operations in each of these configurations with their corresponding sums over $\tau$-loops encircling \emph{untwisted} probe anyons' charge lines. Written in this way, it is clear that the effect of a twisted interferometer will be the same as an untwisted interferometer acting on a twisted basis of anyonic states for the target system. In particular, by thinking of the $\tau$-loops as part of the target system, the probe anyons' charge lines are no longer twisted with each other and, hence, can be evaluated in the same way as in the untwisted case. The $\tau$-loops modify the effect of the untwisted interferometry upon the target system, essentially ``twisting'' the basis of the target system in which the interferometer acts.

\begin{figure}[t!]
\begin{center}
  \includegraphics[scale=0.5]{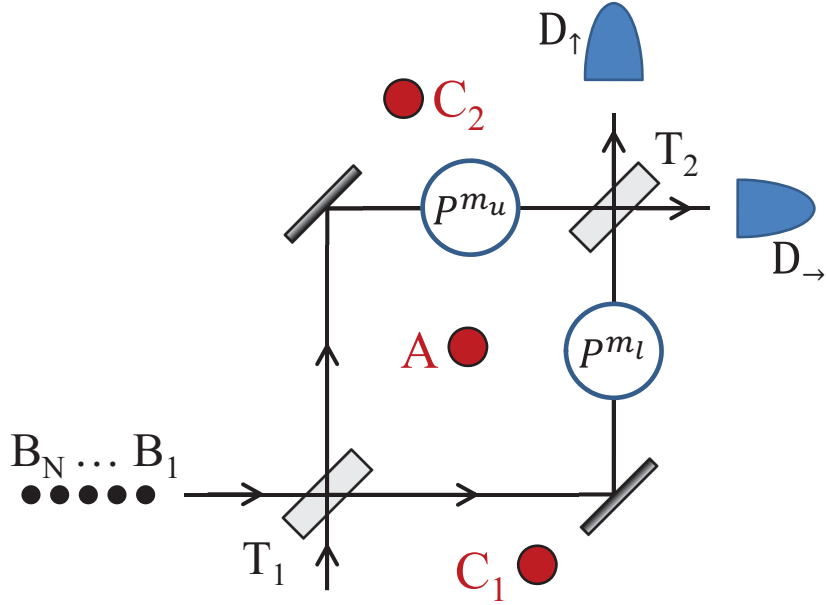}
  \caption{An idealized Mach-Zehnder twisted interferometer for an anyonic system where the probe anyons execute multiple pure braid operators $P^{m_l}$ and $P^{m_u}$ with each other (rather than twists) as they pass through the lower and upper paths, respectively.}
  \label{fig:twisted_int2}
\end{center}
\end{figure}

One might question whether it is physically possible to implement the twisting operations we wish to use. Comparing twist operators with pure braid operators, we see that the twist operator on a collection of anyons is equivalent to a pure braid operator on those anyons together with twists acting individually on each of the anyons. This means that implementation of the twist operators requires some way of twisting individual anyons. One might envision this as naturally occurring if the anyon has some oblong shape and is forced to travel through a narrow twisting track. However, there are situations where such a twisting does not seem possible, such as when the anyons are point-like or edge excitations. In this case, we instead implement the twisted interferometer using pure braid operators, as shown in Fig.~\ref{fig:twisted_int2}. We will still call this a ``twisted interferometer,'' since the twist operator (through its relation to the pure braid operator) still plays a crucial role and the effect of running the interferometer is similar, but with modified $p^{s_j}_{h_1 h_2 e_1 e_2 ,b_j}$ factors. In particular, the topological spin factors of the individual probe anyons factorize and so are absorbed in the $p^{s_j}_{h_1 h_2 e_1 e_2 ,b_j}$ factors (given explicitly later). For the $j$th probe anyon, there will be a factor of $\theta_{b_j}^{-m_l + m_u}$ for the term corresponding to this anyon taking the lower path through the interferometer in the ket and the upper path in the bra. Similarly, there will be a factor of $\theta_{b_j}^{m_l - m_u}$ for the term where this anyon takes the upper path in the ket and the lower path in the bra, and no factor for two terms where the probe anyon takes the same path in the ket and bra (since the topological spin factors cancel in these cases). The asymptotic ($N\rightarrow \infty$) effect of running the twisted interferometer on the target system is actually identical in the two cases. The only difference is the probabilities of the measurement outcomes.

We now consider the analysis of these twisted interferometers in explicit detail.

\subsection{Tensor Category Analysis}

We begin by considering the case of a twisted interferometer that only has probe anyon twisting in the lower path ($m_l=m$ and $m_u=0$) and when the anyons entangled with the target anyons are all below the interferometer ($C_1 = C$ and $C_2 =0$). In order to determine the effect of the twisted interferometer on the target system's density matrix, we evaluate its action on a specific basis element $\left| a,c;f,\mu \right\rangle \left\langle a^{\prime},c^{\prime};f,\mu^{\prime} \right|$. This is done by projecting the post-interferometry basis element onto the basis element $\left| a,c;f^{\prime},\nu \right\rangle \left\langle a^{\prime},c^{\prime};f^{\prime},\nu^{\prime} \right|$ (similar to the evaluation of the untwisted interferometer), which we represent diagrammatically by
\begin{equation}
\pspicture[shift=-2.0](-2.2,-4.0)(1.2,4.0)
  \small
  \psset{linewidth=0.9pt,linecolor=black,arrowscale=1.5,arrowinset=0.15}
  \psline(0.0,-0.5)(0.0,0.5)
  \psline(0.0,0.5)(1.0,1.5)
  \psline(0.0,0.5)(-1.0,1.5)
  \psline(1.0,1.5)(0.0,2.5)
  \psline(-1.0,1.5)(0.0,2.5)
  \psline(0.0,2.5)(0.0,3.0)
  \psline(0.0,3.0)(0.5,3.5)
  \psline(0.0,3.0)(-0.5,3.5)
  \psarc[linewidth=0.9pt,linecolor=black]{<-}(0.0,1.5){0.6}{-110}{270}
  \psline(0.0,-0.5)(1.0,-1.5)
  \psline(0.0,-0.5)(-1.0,-1.5)
  \psline(1.0,-1.5)(0.0,-2.5)
  \psline(-1.0,-1.5)(0.0,-2.5)
  \psline(0.0,-2.5)(0.0,-3.0)
  \psline(0.0,-3.0)(0.5,-3.5)
  \psline(0.0,-3.0)(-0.5,-3.5)
  \psarc[linewidth=0.9pt,linecolor=black]{<-}(0.0,-1.5){0.6}{-110}{270}
  \psline{->}(0.0,-0.5)(0.0,0.2)
  \psline{->}(0.0,0.5)(0.8,1.3)
  \psline{->}(0.0,0.5)(-0.8,1.3)
  \psline{-<}(0.0,-0.5)(0.8,-1.3)
  \psline{-<}(0.0,-0.5)(-0.8,-1.3)
  \psline{->}(0.0,2.5)(0.0,2.9)
  \psline{->}(0.0,-3.0)(0.0,-2.6)
  \psline{->}(0.0,3.0)(0.4,3.4)
  \psline{->}(0.0,3.0)(-0.4,3.4)
  \psline{-<}(0.0,-3.0)(0.4,-3.4)
  \psline{-<}(0.0,-3.0)(-0.4,-3.4)
  \rput[bl]{0}(-0.6,3.6){$a$}
  \rput[bl]{0}(0.4,3.6){$c$}
  \rput[bl]{0}(-0.6,-3.8){$a^{\prime}$}
  \rput[bl]{0}(0.4,-3.8){$c^{\prime}$}
  \rput[bl]{0}(-1.0,0.9){$a$}
  \rput[bl]{0}(0.8,0.9){$c$}
  \rput[bl]{0}(-1.0,-1.1){$a^{\prime}$}
  \rput[bl]{0}(0.8,-1.1){$c^{\prime}$}
 \rput[bl]{0}(-0.4,-0.1){$f$}
  \rput[bl]{0}(-0.5,-2.9){$f^{\prime}$}
  \rput[bl]{0}(-0.5,2.6){$f^{\prime}$}
  \rput[bl]{0}(-0.15,1.4){$\otimes$}
  \rput[bl]{0}(-0.15,-1.6){$\otimes$}
  \rput[bl]{0}(-2.15,1.4){$\otimes$}
  \rput[bl]{0}(-2.15,-1.6){$\otimes$}
\scriptsize
  \rput[bl]{0}(0.1,0.3){$\mu$}
  \rput[bl]{0}(0.1,-0.5){$\mu^{\prime}$}
  \rput[bl]{0}(0.1,2.5){$\nu$}
  \rput[bl]{0}(0.1,2.9){$\nu$}
  \rput[bl]{0}(0.1,-2.6){$\nu^{\prime}$}
  \rput[bl]{0}(0.1,-3.0){$\nu^{\prime}$}
  \rput[bl]{0}(-0.3,1.1){$\tau^{m}$}
  \rput[bl]{0}(-0.3,-1.9){$\tau^{-m}$}
\endpspicture
\label{eq:twisted_int_start}
\end{equation}
where, in order to reduce the clutter that would occur by including all the probe lines in the diagram, we use the $\otimes$ symbols as placeholders to indicate where the probe anyon lines pass through [recall Eqs.~(\ref{eq:one_probe_analysis}) and (\ref{eq:target_gen_probed})]. As such, we must be careful to remember that we cannot pass lines across these $\otimes$ symbols when applying diagrammatic manipulations, as it would involve passing lines through probe anyon lines. The $\tau^{m}$-loop encircling the upper-right $\otimes$ symbol in this diagram represents an application of the $m$-twist operator to all the anyon lines corresponding to probe anyons passing through the lower path of the interferometer. Similarly, the $\tau^{-m}$-loop encircling the lower-right $\otimes$ symbol represents the same operator acting in the conjugate portion of the diagram (i.e. the conjugate $m$-twist operator $\Theta^{-m}$ is applied to the bra, when the $m$-twist operator $\Theta^{m}$ is applied to the ket). The method of evaluation we use is to absorb the $\tau$-loops into the target system's diagrams, at which point the probe anyons' lines are untwisted, but acting upon a modified target system. We break this evaluation into several steps.

First, through a series of $F$-moves, we have the relation
\begin{equation}
\psscalebox{.75}{
\pspicture[shift=-1.6](-1.2,-0.2)(1.2,3.2)
  \small
  \psset{linewidth=0.9pt,linecolor=black,arrowscale=1.5,arrowinset=0.15}
  \psline(0.0,0.0)(0.0,0.5)
  \psline(0.0,0.5)(1.0,1.5)
  \psline(0.0,0.5)(-1.0,1.5)
  \psline(1.0,1.5)(0.0,2.5)
  \psline(-1.0,1.5)(0.0,2.5)
  \psline(0.0,2.5)(0.0,3.0)
  \psarc[linewidth=0.9pt,linecolor=black]{<-}(0.0,1.5){0.6}{-110}{270}
  \psline{->}(0.0,0.0)(0.0,0.4)
  \psline{->}(0.0,0.5)(0.8,1.3)
  \psline{->}(0.0,0.5)(-0.8,1.3)
  \psline{->}(0.0,2.5)(0.0,2.9)
  \rput[bl]{0}(-1.0,0.9){$a$}
  \rput[bl]{0}(0.8,0.9){$c$}
 \rput[bl]{0}(-0.4,0.0){$f$}
  \rput[bl]{0}(-0.5,2.6){$f^{\prime}$}
  \rput[bl]{0}(-0.15,1.4){$\otimes$}
\scriptsize
  \rput[bl]{0}(0.1,0.3){$\mu$}
  \rput[bl]{0}(0.1,2.5){$\nu$}
  \rput[bl]{0}(-0.1,1.1){$x$}
\endpspicture
}
= \sum_{\substack{\hat{a},\hat{c} \\ \hat{\mu},\hat{\nu} \\ \lambda,\gamma} } \left( \frac{d_a d_{\hat{c}} }{d_{\hat{a}} d_c } \right)^{\frac{1}{2}} \left[ F^{a x \hat{c} }_{f} \right]^{\ast}_{(\hat{a},\gamma, \hat{\mu}) (c, \lambda, \mu) } \left[ F^{a x \hat{c} }_{f^{\prime}} \right]_{(\hat{a},\gamma, \hat{\nu}) (c, \lambda, \nu)}
\psscalebox{.75}{
\pspicture[shift=-1.6](-1.2,-0.2)(1.2,3.2)
  \small
  \psset{linewidth=0.9pt,linecolor=black,arrowscale=1.5,arrowinset=0.15}
  \psline(0.0,0.0)(0.0,0.5)
  \psline(0.0,0.5)(1.0,1.5)
  \psline(0.0,0.5)(-1.0,1.5)
  \psline(1.0,1.5)(0.0,2.5)
  \psline(-1.0,1.5)(0.0,2.5)
  \psline(0.0,2.5)(0.0,3.0)
  \psline{->}(0.0,0.0)(0.0,0.4)
  \psline{->}(0.0,0.5)(0.8,1.3)
  \psline{->}(0.0,0.5)(-0.8,1.3)
  \psline{->}(0.0,2.5)(0.0,2.9)
  \rput[bl]{0}(-1.0,0.9){$\hat{a}$}
  \rput[bl]{0}(0.8,0.9){$\hat{c}$}
 \rput[bl]{0}(-0.4,0.0){$f$}
  \rput[bl]{0}(-0.5,2.6){$f^{\prime}$}
  \rput[bl]{0}(-0.15,1.4){$\otimes$}
\scriptsize
  \rput[bl]{0}(0.1,0.3){$\hat{\mu}$}
  \rput[bl]{0}(0.1,2.5){$\hat{\nu}$}
\endpspicture
}
\label{eq:twisted_int_loop_remove}
\end{equation}
We use the circumflexes to indicate charges and state indices that have been ``twisted.''

Applying this and a similar relation for the lower portion of Eq.~(\ref{eq:twisted_int_start}), the $\tau$-loops are incorporated in the target system's diagram, giving a form for which the probe loops are no longer twisted, but now acting on the symbols with circumflexes over them. We can now apply the analysis of the untwisted interferometer to the basis elements which have absorbed the $\tau$-loops, since the probe lines are no longer twisted, giving
\begin{eqnarray}
\psscalebox{.75}{
\pspicture[shift=-3](-2.2,-3.2)(1.4,3.2)
  \small
  \psset{linewidth=0.9pt,linecolor=black,arrowscale=1.5,arrowinset=0.15}
  \psline(0.0,-0.5)(0.0,0.5)
  \psline(0.0,0.5)(1.0,1.5)
  \psline(0.0,0.5)(-1.0,1.5)
  \psline(1.0,1.5)(0.0,2.5)
  \psline(-1.0,1.5)(0.0,2.5)
  \psline(0.0,2.5)(0.0,3.0)
  \psline(0.0,-0.5)(1.0,-1.5)
  \psline(0.0,-0.5)(-1.0,-1.5)
  \psline(1.0,-1.5)(0.0,-2.5)
  \psline(-1.0,-1.5)(0.0,-2.5)
  \psline(0.0,-2.5)(0.0,-3.0)
  \psline{->}(0.0,-0.5)(0.0,0.2)
  \psline{->}(0.0,0.5)(0.8,1.3)
  \psline{->}(0.0,0.5)(-0.8,1.3)
  \psline{-<}(0.0,-0.5)(0.8,-1.3)
  \psline{-<}(0.0,-0.5)(-0.8,-1.3)
  \psline{->}(0.0,2.5)(0.0,2.9)
  \psline{->}(0.0,-3.0)(0.0,-2.6)
  \rput[bl]{0}(-1.0,0.9){$\hat{a}$}
  \rput[bl]{0}(0.8,0.9){$\hat{c}$}
  \rput[bl]{0}(-1.0,-1.1){$\hat{a}^{\prime}$}
  \rput[bl]{0}(0.8,-1.1){$\hat{c}^{\prime}$}
 \rput[bl]{0}(-0.4,-0.1){$f$}
  \rput[bl]{0}(-0.5,-2.9){$f^{\prime}$}
  \rput[bl]{0}(-0.5,2.6){$f^{\prime}$}
  \rput[bl]{0}(-0.15,1.4){$\otimes$}
  \rput[bl]{0}(-0.15,-1.6){$\otimes$}
  \rput[bl]{0}(-2.15,1.4){$\otimes$}
  \rput[bl]{0}(-2.15,-1.6){$\otimes$}
\scriptsize
  \rput[bl]{0}(0.1,0.3){$\hat{\mu}$}
  \rput[bl]{0}(0.1,-0.5){$\hat{\mu}^{\prime}$}
  \rput[bl]{0}(0.1,2.5){$\hat{\nu}$}
  \rput[bl]{0}(0.1,-2.6){$\hat{\nu}^{\prime}$}
\endpspicture
}
&=& \sum_{\hat{e}, \hat{\alpha}, \hat{\beta} }
\left( \frac{d_{\hat{a}} d_{\hat{a}^{\prime}} d_{\hat{c}} d_{\hat{c}^{\prime}} }{d_{f^{\prime}}^{2}} \right)^{\frac{1}{2}}
\left[ \left( F_{\hat{a}^{\prime }\hat{c}^{\prime }}^{\hat{a} \hat{c}}\right) ^{-1}\right]_{\left( f,\hat{\mu} ,\hat{\mu}^{\prime } \right) \left( \hat{e},\hat{\alpha} ,\hat{\beta} \right) }
\notag \\
&& \times
\tilde{p}_{\hat{a} \hat{a}^{\prime} \hat{e},B }^{s_{1}} \ldots \tilde{p}_{\hat{a} \hat{a}^{\prime} \hat{e},B }^{s_{N}}
\left[F_{\hat{a}^{\prime } \hat{c}^{\prime }}^{\hat{a} \hat{c}}\right] _{\left( \hat{e}, \hat{\alpha} , \hat{\beta} \right)
\left( f^{\prime }, \hat{\nu} ,\hat{\nu}^{\prime }\right) }
\pspicture[shift=-1.0](-0.6,-1.2)(0.2,1.2)
  \small
  \psset{linewidth=0.9pt,linecolor=black,arrowscale=1.5,arrowinset=0.15}
  \psline(0.0,-1.0)(0.0,1.0)
  \psline{->}(0.0,-0.5)(0.0,0.2)
 \rput[bl]{0}(-0.4,-0.1){$f^{\prime}$}
\endpspicture
\end{eqnarray}
where the resulting probe factors multiplying the corresponding components for the twisted interferometer are given by
\begin{eqnarray}
\tilde{p}_{\hat{a} \hat{a}^{\prime} \hat{e}, b}^{\shortrightarrow } &=&\left| t_{1}\right| ^{2}\left|
r_{2}\right| ^{2}M_{\hat{e} b}+t_{1}r_{1}^{\ast }r_{2}^{\ast }t_{2}^{\ast
}e^{i\left( \theta _{\text{I}}-\theta _{\text{II}}\right) } M_{\hat{a} b}   \notag \\
&&+t_{1}^{\ast }r_{1}t_{2}r_{2}e^{-i\left( \theta _{\text{I}}-\theta _{\text{II}}\right)} M_{\hat{a}^{\prime} b}^{\ast } +\left| r_{1}\right| ^{2}\left| t_{2}\right| ^{2}
\end{eqnarray}%
\begin{eqnarray}
\tilde{p}_{\hat{a} \hat{a}^{\prime} \hat{e}, b}^{\shortuparrow } &=&\left| t_{1}\right| ^{2}\left|
t_{2}\right| ^{2} M_{\hat{e} b} - t_{1}r_{1}^{\ast }r_{2}^{\ast }t_{2}^{\ast
}e^{i\left( \theta _{\text{I}}-\theta _{\text{II}}\right) } M_{\hat{a} b}  \notag \\
&&-t_{1}^{\ast }r_{1}t_{2}r_{2}e^{-i\left( \theta _{\text{I}}-\theta _{\text{II}}\right)
}M_{\hat{a}^{\prime} b}^{\ast } +\left| r_{1}\right| ^{2}\left| r_{2}\right| ^{2}
,
\end{eqnarray}
and
\begin{equation}
\tilde{p}_{\hat{a} \hat{a}^{\prime} \hat{e}, B}^{s} = \sum_{b} \Pr\nolimits_B(b) \tilde{p}_{\hat{a} \hat{a}^{\prime} \hat{e}, b}^{s}
.
\end{equation}
These are identical to the factors that arose in the untwisted interferometer, with the crucial difference that the charge labels have circumflexes, indicating that they have been modified by absorption of the twisting operators.

For the twisted interferometer that utilizes pure braid operations instead of twist operations (Fig.~\ref{fig:twisted_int2}, rather than Fig.~\ref{fig:twisted_int1}), these probe factor terms are modified to incorporate the individual probe anyons' topological spin factors, as previously explained, giving
\begin{eqnarray}
\tilde{p}_{\hat{a} \hat{a}^{\prime} \hat{e}, b}^{\shortrightarrow } &=&\left| t_{1}\right| ^{2}\left|
r_{2}\right| ^{2}M_{\hat{e} b}+t_{1}r_{1}^{\ast }r_{2}^{\ast }t_{2}^{\ast
}e^{i\left( \theta _{\text{I}}-\theta _{\text{II}}\right) } M_{\hat{a} b} \theta_b^{-m}  \notag \\
&&+t_{1}^{\ast }r_{1}t_{2}r_{2}e^{-i\left( \theta _{\text{I}}-\theta _{\text{II}}\right)} M_{\hat{a}^{\prime} b}^{\ast } \theta_b^{m} +\left| r_{1}\right| ^{2}\left| t_{2}\right| ^{2}
\end{eqnarray}%
\begin{eqnarray}
\tilde{p}_{\hat{a} \hat{a}^{\prime} \hat{e}, b}^{\shortuparrow } &=&\left| t_{1}\right| ^{2}\left|
t_{2}\right| ^{2} M_{\hat{e} b} - t_{1}r_{1}^{\ast }r_{2}^{\ast }t_{2}^{\ast
}e^{i\left( \theta _{\text{I}}-\theta _{\text{II}}\right) } M_{\hat{a} b} \theta_b^{-m}  \notag \\
&&-t_{1}^{\ast }r_{1}t_{2}r_{2}e^{-i\left( \theta _{\text{I}}-\theta _{\text{II}}\right)
}M_{\hat{a}^{\prime} b}^{\ast } \theta_b^{m} +\left| r_{1}\right| ^{2}\left| r_{2}\right| ^{2}
.
\end{eqnarray}
This is the only difference in the analysis and results for the twisted interferometer utilizing pure braid operations, as opposed to twist operations. Primarily, this has the effect of modifying the probabilities with which the different possible outcomes may occur, but otherwise is not significant.

Combining all of these steps and putting in the target density matrix's coefficients and appropriate normalization factors, we find that sending $N$ probe anyons through the twisted interferometer will produce a string of measurement outcomes $(s_1 , \ldots , s_N)$ with probability
\begin{eqnarray}
&& \tilde{\Pr} \left(s_1 , \ldots , s_N \right)  =
\sum\limits_{a,c,f,\mu ,\mu ^{\prime}}
\sum_{x,y}
\sum_{\substack{\hat{a},\hat{a}^{\prime },\hat{c},\hat{c}^{\prime }, \hat{\mu},\hat{\mu}^{\prime } \\ \gamma,\gamma^{\prime },\lambda,\lambda^{\prime } } }
\sum_{\substack{ \hat{e}, \hat{\alpha}, \hat{\beta} \\ f^{\prime},\hat{\nu},\hat{\nu}^{\prime },\nu } }
\rho _{\left( a,c;f,\mu \right) \left( a,c;f,\mu^{\prime }\right) }^{AC}
\left(\frac{d_{f^{\prime}}}{ d_f } \right)^{1/2}
\notag \\
&& \times
\left[ \tau^{m} \right]_x \left[ \tau^{-m} \right]_y
\frac{d_{\hat{c}}}{d_c}\frac{d_{\hat{c}^{\prime}}}{d_{c^{\prime}}}
\left[ F^{a x \hat{c} }_{f} \right]^{\ast}_{(\hat{a},\gamma, \hat{\mu}) (c, \lambda, \mu) }
\left[ F^{a y \hat{c}^{\prime} }_{f} \right]_{(\hat{a}^{\prime},\gamma^{\prime}, \hat{\mu}^{\prime}) (c, \lambda^{\prime}, \mu^{\prime})}
\notag \\
&& \times
\left[ \left( F_{\hat{a}^{\prime }\hat{c}^{\prime }}^{\hat{a} \hat{c}}\right) ^{-1}\right]_{\left( f,\hat{\mu} ,\hat{\mu}^{\prime } \right) \left( \hat{e},\hat{\alpha} ,\hat{\beta} \right) }
\tilde{p}_{\hat{a} \hat{a}^{\prime} \hat{e},B }^{s_{1}} \ldots \tilde{p}_{\hat{a} \hat{a}^{\prime} \hat{e},B }^{s_{N}}
\left[F_{\hat{a}^{\prime } \hat{c}^{\prime }}^{\hat{a} \hat{c}}\right] _{\left( \hat{e}, \hat{\alpha} , \hat{\beta} \right)
\left( f^{\prime }, \hat{\nu} ,\hat{\nu}^{\prime }\right) }
\notag \\
&& \times
\left[ F^{a x \hat{c} }_{f^{\prime}} \right]_{(\hat{a},\gamma, \hat{\nu}) (c, \lambda, \nu)}
\left[ F^{a y \hat{c}^{\prime} }_{f^{\prime}} \right]^{\ast}_{(\hat{a}^{\prime},\gamma^{\prime}, \hat{\nu}^{\prime}) (c, \lambda^{\prime}, \nu)}
\end{eqnarray}
for which the target system's density matrix will become
\begin{eqnarray}
&& \tilde{\rho}^{AC} \left(s_1 , \ldots , s_N \right)  =
\sum\limits_{\substack{ a,a^{\prime },c,c^{\prime } \\ f,\mu ,\mu ^{\prime}}}
\sum_{x,y}
\sum_{\substack{\hat{a},\hat{a}^{\prime },\hat{c},\hat{c}^{\prime }, \hat{\mu},\hat{\mu}^{\prime } \\ \gamma,\gamma^{\prime },\lambda,\lambda^{\prime } } }
\sum_{\substack{ \hat{e}, \hat{\alpha}, \hat{\beta} \\ f^{\prime},\hat{\nu},\hat{\nu}^{\prime },\nu ,\nu ^{\prime} } }
\frac{ \rho _{\left( a,c;f,\mu \right) \left( a^{\prime },c^{\prime };f,\mu^{\prime }\right) }^{AC} }{\left( d_f d_{f^{\prime}} \right)^{1/2} }
\notag \\
&& \times
\left[ \tau^{m} \right]_x \left[ \tau^{-m} \right]_y
\frac{d_{\hat{c}}}{d_c}\frac{d_{\hat{c}^{\prime}}}{d_{c^{\prime}}}
\left[ F^{a x \hat{c} }_{f} \right]^{\ast}_{(\hat{a},\gamma, \hat{\mu}) (c, \lambda, \mu) }
\left[ F^{a^{\prime} y \hat{c}^{\prime} }_{f} \right]_{(\hat{a}^{\prime},\gamma^{\prime}, \hat{\mu}^{\prime}) (c^{\prime}, \lambda^{\prime}, \mu^{\prime})}
\notag \\
&& \times
\left[ \left( F_{\hat{a}^{\prime }\hat{c}^{\prime }}^{\hat{a} \hat{c}}\right) ^{-1}\right]_{\left( f,\hat{\mu} ,\hat{\mu}^{\prime } \right) \left( \hat{e},\hat{\alpha} ,\hat{\beta} \right) }
\frac{ \tilde{p}_{\hat{a} \hat{a}^{\prime} \hat{e},B }^{s_{1}} \ldots \tilde{p}_{\hat{a} \hat{a}^{\prime} \hat{e},B }^{s_{N}}}{\tilde{\Pr} \left(s_1 , \ldots , s_N \right)}
\left[F_{\hat{a}^{\prime } \hat{c}^{\prime }}^{\hat{a} \hat{c}}\right] _{\left( \hat{e}, \hat{\alpha} , \hat{\beta} \right)
\left( f^{\prime }, \hat{\nu} ,\hat{\nu}^{\prime }\right) }
\notag \\
&& \times
\left[ F^{a x \hat{c} }_{f^{\prime}} \right]_{(\hat{a},\gamma, \hat{\nu}) (c, \lambda, \nu)}
\left[ F^{a^{\prime} y \hat{c}^{\prime} }_{f^{\prime}} \right]^{\ast}_{(\hat{a}^{\prime},\gamma^{\prime}, \hat{\nu}^{\prime}) (c^{\prime}, \lambda^{\prime}, \nu^{\prime})}
\left| a,c;f^{\prime},\nu \right\rangle \left\langle a^{\prime},c^{\prime};f^{\prime},\nu^{\prime} \right|
.
\end{eqnarray}
While these expressions are rather complicated, they merely express that the interferometry measurement occurs in a twisted basis (as should be clear from the previous steps).

As in the case of the untwisted interferometer, it is useful to ignore the order of measurement outcomes and focus only on the total number $n$ of $s_j = \shortrightarrow$ measurement outcomes, giving
\begin{eqnarray}
&& \tilde{\Pr}_N \left(n \right)  =
\sum\limits_{a,c,f,\mu ,\mu ^{\prime}}
\sum_{x,y}
\sum_{\substack{\hat{a},\hat{a}^{\prime },\hat{c},\hat{c}^{\prime }, \hat{\mu},\hat{\mu}^{\prime } \\ \gamma,\gamma^{\prime },\lambda,\lambda^{\prime } } }
\sum_{\substack{ \hat{e}, \hat{\alpha}, \hat{\beta} \\ f^{\prime},\hat{\nu},\hat{\nu}^{\prime },\nu } }
\rho _{\left( a,c;f,\mu \right) \left( a,c;f,\mu^{\prime }\right) }^{AC}
\left(\frac{d_{f^{\prime}}}{ d_f } \right)^{1/2}
\notag \\
&& \times
\left[ \tau^{m} \right]_x \left[ \tau^{-m} \right]_y
\frac{d_{\hat{c}}}{d_c}\frac{d_{\hat{c}^{\prime}}}{d_{c^{\prime}}}
\left[ F^{a x \hat{c} }_{f} \right]^{\ast}_{(\hat{a},\gamma, \hat{\mu})(c, \lambda, \mu) }
\left[ F^{a y \hat{c}^{\prime} }_{f} \right]_{(\hat{a}^{\prime},\gamma^{\prime}, \hat{\mu}^{\prime}) (c, \lambda^{\prime}, \mu^{\prime}) }
\notag \\
&& \times
\left[ \left( F_{\hat{a}^{\prime }\hat{c}^{\prime }}^{\hat{a} \hat{c}}\right) ^{-1}\right]_{\left( f,\hat{\mu} ,\hat{\mu}^{\prime } \right) \left( \hat{e},\hat{\alpha} ,\hat{\beta} \right) }
W_N \left( n ; \tilde{p}_{\hat{a} \hat{a}^{\prime} \hat{e},B }^{\shortrightarrow} , \tilde{p}_{\hat{a} \hat{a}^{\prime} \hat{e},B }^{\shortuparrow} \right)
\left[F_{\hat{a}^{\prime } \hat{c}^{\prime }}^{\hat{a} \hat{c}}\right] _{\left( \hat{e}, \hat{\alpha} , \hat{\beta} \right)
\left( f^{\prime }, \hat{\nu} ,\hat{\nu}^{\prime }\right) }
\notag \\
&& \times
\left[ F^{a x \hat{c} }_{f^{\prime}} \right]_{(\hat{a},\gamma, \hat{\nu}) (c, \lambda, \nu)}
\left[ F^{a y \hat{c}^{\prime} }_{f^{\prime}} \right]^{\ast}_{(\hat{a}^{\prime},\gamma^{\prime}, \hat{\nu}^{\prime}) (c, \lambda^{\prime}, \nu)}
\end{eqnarray}
and
\begin{eqnarray}
&& \tilde{\rho}^{AC}_N \left(n \right)  =
\sum\limits_{\substack{ a,a^{\prime },c,c^{\prime } \\ f,\mu ,\mu ^{\prime}}}
\sum_{x,y}
\sum_{\substack{\hat{a},\hat{a}^{\prime },\hat{c},\hat{c}^{\prime }, \hat{\mu},\hat{\mu}^{\prime } \\ \gamma,\gamma^{\prime },\lambda,\lambda^{\prime } } }
\sum_{\substack{ \hat{e}, \hat{\alpha}, \hat{\beta} \\ f^{\prime},\hat{\nu},\hat{\nu}^{\prime },\nu ,\nu ^{\prime} } }
\frac{ \rho _{\left( a,c;f,\mu \right) \left( a^{\prime },c^{\prime };f,\mu^{\prime }\right) }^{AC} }{\left( d_f d_{f^{\prime}} \right)^{1/2} }
\notag \\
&& \times
\left[ \tau^{m} \right]_x \left[ \tau^{-m} \right]_y
\frac{d_{\hat{c}}}{d_c}\frac{d_{\hat{c}^{\prime}}}{d_{c^{\prime}}}
\left[ F^{a x \hat{c} }_{f} \right]^{\ast}_{(\hat{a},\gamma, \hat{\mu}) (c, \lambda, \mu) }
\left[ F^{a^{\prime} y \hat{c}^{\prime} }_{f} \right]_{ (\hat{a}^{\prime},\gamma^{\prime}, \hat{\mu}^{\prime}) (c^{\prime}, \lambda^{\prime}, \mu^{\prime})}
\notag \\
&& \times
\left[ \left( F_{\hat{a}^{\prime }\hat{c}^{\prime }}^{\hat{a} \hat{c}}\right) ^{-1}\right]_{\left( f,\hat{\mu} ,\hat{\mu}^{\prime } \right) \left( \hat{e},\hat{\alpha} ,\hat{\beta} \right) }
\frac{W_N \left( n ; \tilde{p}_{\hat{a} \hat{a}^{\prime} \hat{e},B }^{\shortrightarrow} , \tilde{p}_{\hat{a} \hat{a}^{\prime} \hat{e},B }^{\shortuparrow} \right)}{\tilde{\Pr}_N \left(n \right)}
\left[F_{\hat{a}^{\prime } \hat{c}^{\prime }}^{\hat{a} \hat{c}}\right] _{\left( \hat{e}, \hat{\alpha} , \hat{\beta} \right)
\left( f^{\prime }, \hat{\nu} ,\hat{\nu}^{\prime }\right) }
\notag \\
&& \times
\left[ F^{a x \hat{c} }_{f^{\prime}} \right]_{(\hat{a},\gamma, \hat{\nu}) (c, \lambda, \nu) }
\left[ F^{a^{\prime} y \hat{c}^{\prime} }_{f^{\prime}} \right]^{\ast}_{(\hat{a}^{\prime},\gamma^{\prime}, \hat{\nu}^{\prime}) (c^{\prime}, \lambda^{\prime}, \nu^{\prime})}
\left| a,c;f^{\prime},\nu \right\rangle \left\langle a^{\prime},c^{\prime};f^{\prime},\nu^{\prime} \right|
.
\end{eqnarray}

We can similarly define the maximal disjoint subsets $\tilde{\mathcal{C}}_{{\kappa} }$ of charges in the twisted basis that are indistinguishable by the twisted interferometer, in the sense that $\tilde{p}_{\hat{a}\hat{a}0,B}^{\shortrightarrow }=\tilde{p}_{\kappa }$ for all $\hat{a}\in \tilde{\mathcal{C}}_{\kappa }$,
i.e.
\begin{eqnarray}
\tilde{\mathcal{C}}_{{\kappa} } &\equiv &\left\{ \hat{a}\in \mathcal{C}:\tilde{p}_{\hat{a}\hat{a}0,B}^{\shortrightarrow }=\tilde{p}_{\kappa }\right\} \\
\tilde{\mathcal{C}}_{{\kappa} } \cap \tilde{\mathcal{C}}_{\kappa ^{\prime }} &=&\varnothing
\text{ \ \ \ \ for \ }\kappa \neq \kappa ^{\prime } \nonumber \\
\bigcup\limits_{\kappa }\tilde{\mathcal{C}}_{{\kappa} } &=&\mathcal{C} \nonumber
.
\end{eqnarray}%

We recall that, in the $N \rightarrow \infty$ limit, $W_N \left( n ; \tilde{p}_{\hat{a} \hat{a}^{\prime} \hat{e},B }^{\shortrightarrow} , \tilde{p}_{\hat{a} \hat{a}^{\prime} \hat{e},B }^{\shortuparrow} \right)$ only gives non-vanishing contribution if $\tilde{p}_{\hat{a} \hat{a}^{\prime} \hat{e},B }^{\shortrightarrow} = 1- \tilde{p}_{\hat{a} \hat{a}^{\prime} \hat{e},B }^{\shortuparrow}$, $0 \leq \tilde{p}_{\hat{a} \hat{a}^{\prime} \hat{e},B }^{\shortrightarrow} \leq 1$ is real-valued, and $r=n/N$ approaches $\tilde{p}_{\hat{a} \hat{a}^{\prime} \hat{e},B }^{\shortrightarrow}$. These conditions are (generically) only satisfied if $M_{\hat{e}B}=1$ and hence $M_{\hat{a}b}=M_{\hat{a}^{\prime} b}$ for all $b$ with ${\Pr}_{B}(b) \neq 0$, implying $\tilde{p}_{\hat{a} \hat{a}^{\prime} \hat{e},B }^{\shortrightarrow} = \tilde{p}_{\hat{a}\hat{a}0,B}^{\shortrightarrow }$.

With this, we can write the probability distribution for $r=n/N$ (the fraction of probe anyons measured by the $s=\shortrightarrow$ detector) in the $N \rightarrow \infty$ limit
\begin{eqnarray}
&& \tilde{\Pr} \left(r \right)  = \sum_{\kappa} \tilde{\Pr}_{AC} \left(\kappa \right) \delta \left( r - \tilde{p}_{\kappa} \right) \notag \\
&& \tilde{\Pr}_{AC} \left(\kappa \right) = \sum\limits_{a,c,f,\mu ,\mu ^{\prime}}
\sum_{x,y}
\sum_{\substack{\hat{a},\hat{a}^{\prime },\hat{c},\hat{c}^{\prime }, \hat{\mu},\hat{\mu}^{\prime } \\ \gamma,\gamma^{\prime },\lambda,\lambda^{\prime } } }
\sum_{\substack{ \hat{e}, \hat{\alpha}, \hat{\beta} \\ f^{\prime},\hat{\nu},\hat{\nu}^{\prime },\nu } }
\rho _{\left( a,c;f,\mu \right) \left( a,c;f,\mu^{\prime }\right) }^{AC}
\left(\frac{d_{f^{\prime}}}{ d_f } \right)^{1/2}
\notag \\
&& \times
\left[ \tau^{m} \right]_x \left[ \tau^{-m} \right]_y
\frac{d_{\hat{c}}}{d_c}\frac{d_{\hat{c}^{\prime}}}{d_{c^{\prime}}}
\left[ F^{a x \hat{c} }_{f} \right]^{\ast}_{(\hat{a},\gamma, \hat{\mu}) (c, \lambda, \mu)}
\left[ F^{a y \hat{c}^{\prime} }_{f} \right]_{(\hat{a}^{\prime},\gamma^{\prime}, \hat{\mu}^{\prime}) (c, \lambda^{\prime}, \mu^{\prime})}
\notag \\
&& \times
\left[ \left( F_{\hat{a}^{\prime }\hat{c}^{\prime }}^{\hat{a} \hat{c}}\right) ^{-1}\right]_{\left( f,\hat{\mu} ,\hat{\mu}^{\prime } \right) \left( \hat{e},\hat{\alpha} ,\hat{\beta} \right) }
\tilde{\delta}_{\hat{a} \hat{a}^{\prime} \hat{e},B } \left( \tilde{p}_{\kappa} \right)
\left[F_{\hat{a}^{\prime } \hat{c}^{\prime }}^{\hat{a} \hat{c}}\right] _{\left( \hat{e}, \hat{\alpha} , \hat{\beta} \right)
\left( f^{\prime }, \hat{\nu} ,\hat{\nu}^{\prime }\right) }
\notag \\
&& \times
\left[ F^{a x \hat{c} }_{f^{\prime}} \right]_{(\hat{a},\gamma, \hat{\nu}) (c, \lambda, \nu)}
\left[ F^{a y \hat{c}^{\prime} }_{f^{\prime}} \right]^{\ast}_{(\hat{a}^{\prime},\gamma^{\prime}, \hat{\nu}^{\prime}) (c, \lambda^{\prime}, \nu)}
,
\end{eqnarray}
where we have defined
\begin{equation}
\tilde{\delta}_{\hat{a} \hat{a}^{\prime} \hat{e},B } \left( \tilde{p}_{\kappa} \right) =\left\{
\begin{array}{ccl}
1 &  & \text{\ \ \ if } \tilde{p}_{\hat{a} \hat{a}^{\prime} \hat{e},B }^{\shortrightarrow }=1-\tilde{p}_{\hat{a} \hat{a}^{\prime} \hat{e},B }^{\shortuparrow }=\tilde{p}_{\kappa }\text{ and } \hat{a},\hat{a}^{\prime }\in \tilde{\mathcal{C}}_{\kappa } \\
0 &  & \text{\ \ \ otherwise}%
\end{array}%
\right.
.
\end{equation}%
This is similar to the untwisted case, expect we can no longer compute the probability simply by directly applying a projection operator to the $A$ and $A^{\prime}$ anyon charge lines.

Thus, in the $N\rightarrow \infty$ limit, we find $r\rightarrow \tilde{p}_{\kappa}$ with probability $\tilde{\Pr}_{AC} \left(\kappa \right)$, in which case the target system's density matrix will collapse onto the corresponding fixed state of the twisted interferometer, given by
\begin{eqnarray}
&& \tilde{\rho}^{AC}_\kappa =
\sum\limits_{\substack{ a,a^{\prime },c,c^{\prime } \\ f,\mu ,\mu ^{\prime}}}
\sum_{x,y}
\sum_{\substack{\hat{a},\hat{a}^{\prime },\hat{c},\hat{c}^{\prime }, \hat{\mu},\hat{\mu}^{\prime } \\ \gamma,\gamma^{\prime },\lambda,\lambda^{\prime } } }
\sum_{\substack{ \hat{e}, \hat{\alpha}, \hat{\beta} \\ f^{\prime},\hat{\nu},\hat{\nu}^{\prime },\nu ,\nu ^{\prime} } }
\frac{ \rho _{\left( a,c;f,\mu \right) \left( a^{\prime },c^{\prime };f,\mu^{\prime }\right) }^{AC} }{\left( d_f d_{f^{\prime}} \right)^{1/2} }
\notag \\
&& \times
\left[ \tau^{m} \right]_x \left[ \tau^{-m} \right]_y
\frac{d_{\hat{c}}}{d_c}\frac{d_{\hat{c}^{\prime}}}{d_{c^{\prime}}}
\left[ F^{a x \hat{c} }_{f} \right]^{\ast}_{(\hat{a},\gamma, \hat{\mu}) (c, \lambda, \mu) }
\left[ F^{a^{\prime} y \hat{c}^{\prime} }_{f} \right]_{ (\hat{a}^{\prime},\gamma^{\prime}, \hat{\mu}^{\prime}) (c^{\prime}, \lambda^{\prime}, \mu^{\prime})}
\notag \\
&& \times
\left[ \left( F_{\hat{a}^{\prime }\hat{c}^{\prime }}^{\hat{a} \hat{c}}\right) ^{-1}\right]_{\left( f,\hat{\mu} ,\hat{\mu}^{\prime } \right) \left( \hat{e},\hat{\alpha} ,\hat{\beta} \right) }
\tilde{\Delta}_{\hat{a} \hat{a}^{\prime} \hat{e},B } \left( \tilde{p}_{\kappa} \right)
\left[F_{\hat{a}^{\prime } \hat{c}^{\prime }}^{\hat{a} \hat{c}}\right] _{\left( \hat{e}, \hat{\alpha} , \hat{\beta} \right)
\left( f^{\prime }, \hat{\nu} ,\hat{\nu}^{\prime }\right) }
\notag \\
&& \times
\left[ F^{a x \hat{c} }_{f^{\prime}} \right]_{(\hat{a},\gamma, \hat{\nu}) (c, \lambda, \nu) }
\left[ F^{a^{\prime} y \hat{c}^{\prime} }_{f^{\prime}} \right]^{\ast}_{(\hat{a}^{\prime},\gamma^{\prime}, \hat{\nu}^{\prime}) (c^{\prime}, \lambda^{\prime}, \nu^{\prime})}
\left| a,c;f^{\prime},\nu \right\rangle \left\langle a^{\prime},c^{\prime};f^{\prime},\nu^{\prime} \right|
,
\end{eqnarray}
where
\begin{equation}
\tilde{\Delta}_{\hat{a} \hat{a}^{\prime} \hat{e},B } \left( \tilde{p}_{\kappa} \right) = \frac{ \tilde{\delta}_{\hat{a} \hat{a}^{\prime} \hat{e},B } \left( \tilde{p}_{\kappa} \right) }{\tilde{\Pr}_{AC} \left(\kappa \right)}
\end{equation}
similar to the untwisted case.
We reemphasize that the condition that $\tilde{p}_{\hat{a} \hat{a}^{\prime} \hat{e},B }^{\shortrightarrow }=1-\tilde{p}_{\hat{a} \hat{a}^{\prime} \hat{e},B }^{\shortuparrow }=\tilde{p}_{\kappa }$ and $\hat{a},\hat{a}^{\prime }\in \tilde{\mathcal{C}}_{\kappa }$ is equivalent to $M_{\hat{e} B}=1$ (which implies $M_{\hat{a} b}=M_{\hat{a}^{\prime} b}$ for all $b$ with $\Pr_{B}(b) \neq 0$).

We now return to the general case where the twisted interferometer can have probe anyon twisting $\Theta^{m_l}$ and $\Theta^{m_u}$ (or, alternatively, braiding $P^{m_l}$ and $P^{m_u}$), respectively in both the lower and upper paths, and the target anyons $A$ can share entanglement with anyons $C_1$ and $C_2$ that are both below and above the interferometer, respectively. The analysis is straightforward, but produces expressions which are much more complicated than the already very complicated, simplest case examined above. These details are not particularly enlightening, so we will simply state that one must perform a series of $F$-moves (the sequence of which should be clear from the previous analyses) to remove the twists and probes loops. In the case of probe twisting, the resulting factors multiplying the corresponding twisted basis elements of the density matrix are
\begin{eqnarray}
\tilde{p}_{\hat{h}_1 \hat{h}_2 \hat{e}_1 \hat{e}_2 ,b}^{\shortrightarrow } &=&\left| t_{1}\right| ^{2}\left|
r_{2}\right| ^{2}M_{\hat{e}_1 b}+t_{1}r_{1}^{\ast }r_{2}^{\ast }t_{2}^{\ast
}e^{i\left( \theta _{\text{I}}-\theta _{\text{II}}\right) }M_{\hat{h}_1 b }  \notag \\
&&+t_{1}^{\ast }r_{1}t_{2}r_{2}e^{-i\left( \theta _{\text{I}}-\theta _{\text{II}}\right)
}M_{\hat{h}_2 b}^{\ast } +\left| r_{1}\right| ^{2}\left| t_{2}\right| ^{2} M_{\hat{e}_2 b}
\end{eqnarray}%
\begin{eqnarray}
\tilde{p}_{\hat{h}_1 \hat{h}_2 \hat{e}_1 \hat{e}_2 ,b}^{\shortuparrow } &=&\left| t_{1}\right| ^{2}\left|
t_{2}\right| ^{2}M_{\hat{e}_1 b}-t_{1}r_{1}^{\ast }r_{2}^{\ast }t_{2}^{\ast
}e^{i\left( \theta _{\text{I}}-\theta _{\text{II}}\right) }M_{\hat{h}_1 b}   \notag \\
&&-t_{1}^{\ast }r_{1}t_{2}r_{2}e^{-i\left( \theta _{\text{I}}-\theta _{\text{II}}\right)
}M_{\hat{h}_2 b}^{\ast } +\left| r_{1}\right| ^{2}\left| r_{2}\right| ^{2} M_{\hat{e}_2 b}
,
\end{eqnarray}
and
\begin{equation}
\tilde{p}_{\hat{h}_1 \hat{h}_2 \hat{e}_1 \hat{e}_2 ,B}^{s} = \sum_{b} \Pr\nolimits_B(b) \tilde{p}_{\hat{h}_1 \hat{h}_2 \hat{e}_1 \hat{e}_2 ,b}^{s}
,
\end{equation}
where the anyonic charges $\hat{h}_1$, $\hat{h}_2$, $\hat{e}_1$, and $\hat{e}_2$ are the twisted basis analogues of the labels for the diagrams of Eq.~(\ref{eq:rho3_F_moved}).

For the twisted interferometer utilizing pure braid operations, instead of twist operations, these factors are
\begin{eqnarray}
\tilde{p}_{\hat{h}_1 \hat{h}_2 \hat{e}_1 \hat{e}_2 ,b}^{\shortrightarrow } &=&\left| t_{1}\right| ^{2}\left|
r_{2}\right| ^{2}M_{\hat{e}_1 b}+t_{1}r_{1}^{\ast }r_{2}^{\ast }t_{2}^{\ast
}e^{i\left( \theta _{\text{I}}-\theta _{\text{II}}\right) }M_{\hat{h}_1 b } \theta_{b}^{-m_l + m_u}  \notag \\
&&+t_{1}^{\ast }r_{1}t_{2}r_{2}e^{-i\left( \theta _{\text{I}}-\theta _{\text{II}}\right)
}M_{\hat{h}_2 b}^{\ast } \theta_{b}^{m_l - m_u} +\left| r_{1}\right| ^{2}\left| t_{2}\right| ^{2} M_{\hat{e}_2 b}
\end{eqnarray}%
\begin{eqnarray}
\tilde{p}_{\hat{h}_1 \hat{h}_2 \hat{e}_1 \hat{e}_2 ,b}^{\shortuparrow } &=&\left| t_{1}\right| ^{2}\left|
t_{2}\right| ^{2}M_{\hat{e}_1 b}-t_{1}r_{1}^{\ast }r_{2}^{\ast }t_{2}^{\ast
}e^{i\left( \theta _{\text{I}}-\theta _{\text{II}}\right) }M_{\hat{h}_1 b} \theta_{b}^{-m_l + m_u}  \notag \\
&&-t_{1}^{\ast }r_{1}t_{2}r_{2}e^{-i\left( \theta _{\text{I}}-\theta _{\text{II}}\right)
}M_{\hat{h}_2 b}^{\ast } \theta_{b}^{m_l - m_u}+\left| r_{1}\right| ^{2}\left| r_{2}\right| ^{2} M_{\hat{e}_2 b}
.
\end{eqnarray}

We again define modified anyonic charge sets
\begin{equation}
\tilde{\mathcal{C}}_{\kappa } \equiv \left\{ \hat{a}\in \mathcal{C} : \tilde{p}_{\hat{a} \hat{a} 00,B}^{\shortrightarrow }=p_{\kappa }\right\}
.
\end{equation}
There will be a probability $\tilde{\Pr}_{AC}\left( \kappa \right)$ similarly obtained from the initial density matrix, providing the probability that $r = \tilde{p}_{\kappa}$ in the asymptotic limit $N\rightarrow \infty$, resulting in a fixed state density matrix obtained using
\begin{equation}
\tilde{\Delta}_{\hat{h}_1 \hat{h}_2 \hat{e}_1 \hat{e}_2 ,B}\left( \tilde{p}_{\kappa }\right) =\left\{
\begin{array}{ccl}
\frac{1}{\tilde{\Pr}_{AC}\left( \kappa \right) } && \text{if }\tilde{p}_{\hat{a} \hat{a} 00,B}^{\shortrightarrow }=1-\tilde{p}_{\hat{a} \hat{a} 00,B}^{\shortuparrow }=\tilde{p}_{\kappa }\text{ and
}h_1,h_2 \in \tilde{\mathcal{C}}_{\kappa } \\
0 && \text{otherwise}%
\end{array}%
\right.
,
\end{equation}%
which determines the components of the target anyons' density matrix that survive after the interferometry measurement.

As with the untwisted interferometer, we can use $\omega$-loops to compactly express the effect of running the twisted interferometer. The $\omega$-loops occur in the same positions as in the untwisted case, but now have $\tau$-loops linking them in the positions where the probe twisting/braiding occur. Specifically, we can write
\begin{equation}
\tilde{\rho}_{\kappa}^{AC} = \frac{1}{\tilde{\Pr}_{AC}(\kappa)}
\pspicture[shift=-3.5](-2.8,-3.6)(2.8,3.5)
  \small
  \psline(-2.5,-0.5)(-2.5,0.5)
  \psline(-2.5,-0.5)(2.5,-0.5)
  \psline(-2.5,0.5)(2.5,0.5)
  \psline(2.5,-0.5)(2.5,0.5)
  \psset{linewidth=0.9pt,linecolor=black,arrowscale=1.5,arrowinset=0.15}
  \psline(2.0,0.5)(2.0,3.0)
  \psline(0.0,0.5)(0.0,3.0)
  \psline(2.0,-0.5)(2.0,-3.0)
  \psline(0.0,-0.5)(0.0,-3.0)
  \psline(-2.0,0.5)(-2.0,3.0)
  \psline(-2.0,-0.5)(-2.0,-3.0)
  \psline{->}(2.0,1.5)(2.0,2.75)
  \psline{->}(0.0,1.5)(0.0,2.75)
  \psline{-<}(2.0,-1.5)(2.0,-2.75)
  \psline{-<}(0.0,-1.5)(0.0,-2.75)
  \psline{->}(-2.0,1.5)(-2.0,2.75)
  \psline{-<}(-2.0,-1.5)(-2.0,-2.75)
  \rput[bl](1.9,3.05){$C_1$}
  \rput[bl](-0.1,3.1){$A$}
  \rput[bl](1.9,-3.45){$C_1^{\prime}$}
  \rput[bl](-0.1,-3.35){$A^{\prime}$}
  \rput[bl](-2.3,3.05){$C_2$}
  \rput[bl](-2.3,-3.45){$C_2^{\prime}$}
    \psarc[linewidth=0.9pt,linecolor=black]{<-}(-1.0,1.5){0.6}{80}{450}
    \psarc[linewidth=0.9pt,linecolor=black]{<-}(1.0,1.5){0.6}{80}{450}
    \psarc[linewidth=0.9pt,linecolor=black]{<-}(-1.0,-1.5){0.6}{-100}{270}
    \psarc[linewidth=0.9pt,linecolor=black]{<-}(1.0,-1.5){0.6}{-100}{270}
  \psbezier[linewidth=0.9pt,linecolor=black,border=0.05](-0.5,0.0)(-1.25,1.5)(-1.0,2.0)(-0.25,0.5)
  \psbezier[linewidth=0.9pt,linecolor=black,border=0.05](-0.5,0.0)(0.25,-1.5)(1.25,-2.5)(0.25,-0.5)
   \psline{<-}(-0.5,0.0)(-0.55,0.1)
  \psline[linewidth=0.9pt,linecolor=black,border=0.1](-0.4,0.5)(-0.2,0.5)
  \psline[linewidth=0.9pt,linecolor=black,border=0.1](0.4,-0.5)(0.2,-0.5)
  \psbezier[linewidth=0.9pt,linecolor=black,border=0.05](0.5,0.0)(1.5,2.0)(1.0,2.0)(0.1,0.5)
  \psbezier[linewidth=0.9pt,linecolor=black,border=0.05](0.5,0.0)(-0.5,-2.0)(-1.3,-2.0)(-0.4,-0.5)
  \psline[linewidth=0.9pt,linecolor=black,border=0.1](0.4,0.5)(0.05,0.5)
  \psline[linewidth=0.9pt,linecolor=black,border=0.1](-0.5,-0.5)(-0.35,-0.5)
   \psline{->}(0.5,0.0)(0.55,0.1)
  \psellipse[linewidth=0.9pt,linecolor=black,border=0.05](1.5,0.0)(0.3,1.5)
  \psline{->}(1.215,-0.13)(1.215,-0.1)
  \psframe[linewidth=0.9pt,linecolor=white,border=0.1,fillcolor=white,fillstyle=solid](1.7,-0.5)(1.8,0.5)
  \psline(2.0,0.5)(1.5,0.5)
  \psline(2.0,-0.5)(1.5,-0.5)
  \psellipse[linewidth=0.9pt,linecolor=black,border=0.05](-1.5,0.0)(-0.3,1.5)
  \psline{->}(-1.19,-0.13)(-1.19,-0.1)
  \psframe[linewidth=0.9pt,linecolor=white,border=0.1,fillcolor=white,fillstyle=solid](-1.7,-0.5)(-1.8,0.5)
  \psline(-2.0,0.5)(-1.5,0.5)
  \psline(-2.0,-0.5)(-1.5,-0.5)
    \psarc[linewidth=0.9pt,linecolor=black,border=0.05](-1.0,1.5){0.6}{-55}{45}
    \psarc[linewidth=0.9pt,linecolor=black,border=0.05](-1.0,1.5){0.6}{135}{225}
    \psarc[linewidth=0.9pt,linecolor=black,border=0.05](1.0,1.5){0.6}{-45}{45}
    \psarc[linewidth=0.9pt,linecolor=black,border=0.05](1.0,1.5){0.6}{135}{225}
    \psarc[linewidth=0.9pt,linecolor=black,border=0.05](-1.0,-1.5){0.6}{45}{90}
    \psarc[linewidth=0.9pt,linecolor=black,border=0.05](-1.0,-1.5){0.6}{135}{225}
    \psarc[linewidth=0.9pt,linecolor=black,border=0.05](1.0,-1.5){0.6}{-45}{45}
    \psarc[linewidth=0.9pt,linecolor=black,border=0.05](1.0,-1.5){0.6}{90}{155}
  \rput[bl]{0}(-0.3,-0.15){$\rho^{AC}$}
  \scriptsize
  \rput[bl](1.35,-0.4){$\omega_{\mathcal{B}_0}$}
  \rput[bl](0.6,-0.1){$\omega_{\tilde{\mathcal{C}}_\kappa}$}
  \rput[bl](-1.8,-0.4){$\omega_{\mathcal{B}_0}$}
  \rput[bl](-1.0,-0.1){$\omega_{\tilde{\mathcal{C}}_\kappa}$}
  \rput[bl](0.8,2.25){$\tau^{m_l}$}
  \rput[bl](-1.2,2.25){$\tau^{m_u}$}
  \rput[bl](0.75,-2.4){$\tau^{-m_l}$}
  \rput[bl](-1.25,-2.4){$\tau^{-m_u}$}
 \endpspicture
.
\label{eq:twisted_target_projected_omega}
\end{equation}
We emphasize again that the direction of the arrows on the $\tau$-loops are arbitrary. Written this way, it is clear that twisted interferometry has the effect of imposing projections of the anyonic charge along loops (i.e. the $\omega$-loops) that are twisted with each other (by the $\tau$-loops).

\section{Topological Descriptions}
\label{sec:topological_descriptions}

\begin{figure}[t!]
  \labellist
  \pinlabel $H_2$ at 350 650
  \pinlabel $\omega_a$ at 50 500
  \pinlabel $\omega_a$ at -50 450
  \pinlabel $\omega_0$ at 720 320
  \pinlabel $\gamma$ at 680 430
  \pinlabel $\bar{\gamma}$ at 620 210
  \pinlabel $\text{(a)}$ at 350 -110
  \pinlabel $H_3$ at 1000 700
  \pinlabel $\omega_a$ at 1400 700
  \pinlabel $\omega_0$ at 900 330
  \pinlabel $\omega_0$ at 1830 330
  \pinlabel $\omega_a$ at 1370 -20
  \pinlabel $\beta$ at 930 480
  \pinlabel $\bar{\beta}$ at 930 180
  \pinlabel $\gamma$ at 1780 480
  \pinlabel $\bar{\gamma}$ at 1780 180
  \pinlabel $\text{(b)}$ at 1370 -110
  \endlabellist
  \centering
  \includegraphics[width=0.8\textwidth]{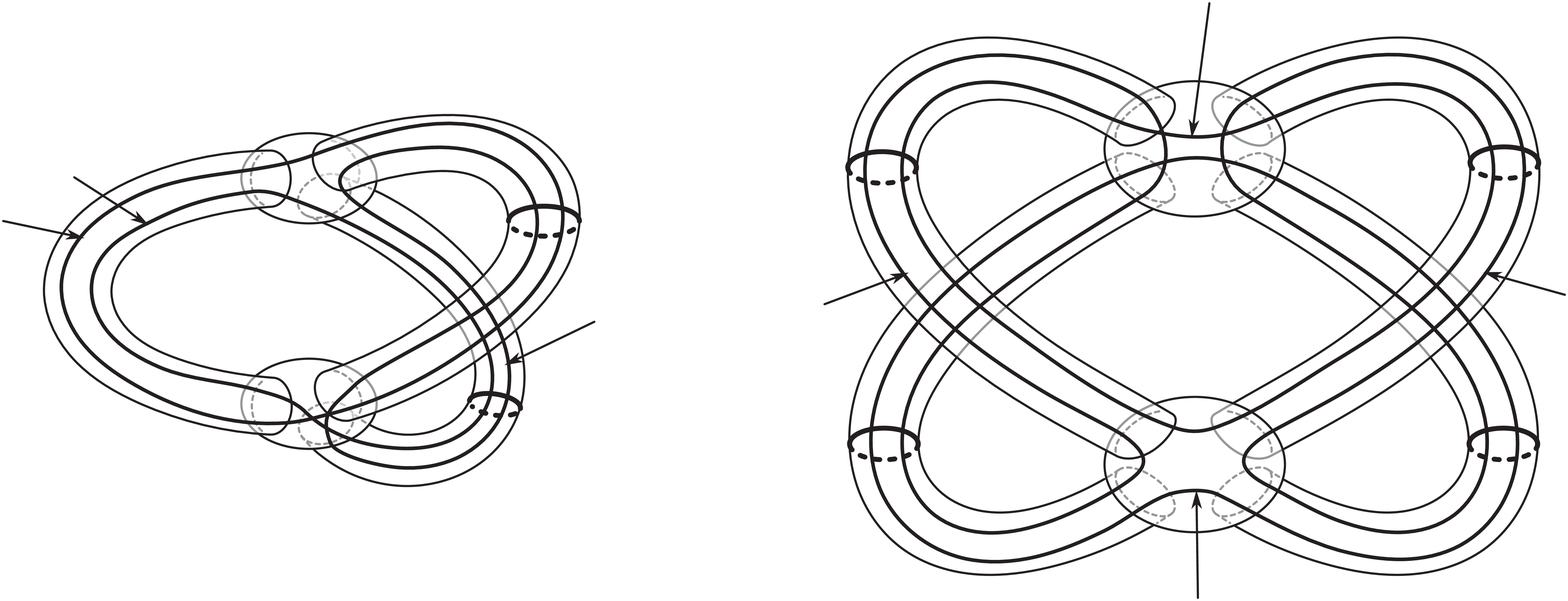}\vspace{0.75cm}
  \caption{(a) The genus $2$ handle-body $H_2$ used for an anyonic interferometer with $C_2 = 0$, $m_u=0$, and probe anyons that can distinguish all anyonic charge types. $m_l$ Dehn twists are applied along the cycle $\gamma$ and $-m_l$ along $\bar{\gamma}$. (b) The genus $3$ handle-body $H_3$ for a general anyonic interferometer with probe anyons that can distinguish all anyonic charge types. $m_l$, $-m_l$, $m_u$, and $-m_u$ Dehn twists are applied along the cycle $\gamma$, $\bar{\gamma}$, $\beta$, and $\bar{\beta}$, respectively.}
  \label{fig:handle_bodies}
\end{figure}

\begin{figure}[t!]
  \labellist
  \pinlabel $\omega_a$ at 50 500
  \pinlabel $\omega_a$ at -50 450
  \pinlabel $\omega_0$ at 720 320
  \pinlabel $\text{(a)}$ at 350 -110
  \pinlabel $\omega_a$ at 1400 700
  \pinlabel $\omega_0$ at 900 330
  \pinlabel $\omega_0$ at 1830 330
  \pinlabel $\omega_a$ at 1370 -20
  \pinlabel $\text{(b)}$ at 1370 -110
  \endlabellist
  \centering
  \includegraphics[width=0.8\textwidth]{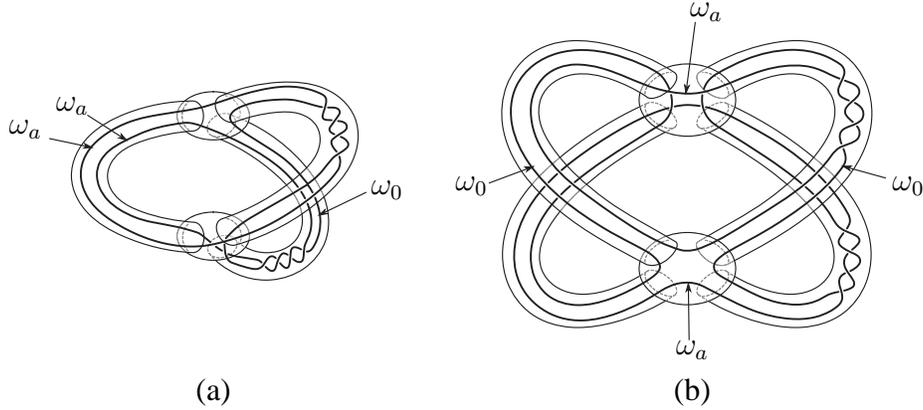}\vspace{0.75cm}
  \caption{The handle-bodies $H_2$ and $H_3$ from Fig.~\ref{fig:handle_bodies} after $m_l = 2$ Dehn twists have been applied along $\gamma$ and $-m_l=-2$ along $\bar{\gamma}$.}
  \label{fig:handle_bodies_twisted}
\end{figure}

The operations described in this paper can also be approached from a more topological formulation involving surgeries of $3$-manifolds. In particular, one can treat the diagrammatic anyonic charge lines as Wilson lines of a TQFT embedded in a $3$-manifold. Having expressed operations in terms of $\omega$-loops and $\tau$-loops allows a straightforward translation to the surgery formulation. In particular, an anyonic charge loop may be introduced into a $3$-manifold by cutting a solid torus out of the manifold and then gluing back in a solid torus that contains the appropriate Wilson loop. The twist operations and $\tau$-loops described in this paper are directly related to the Dehn twist surgery. In particular, cutting a handle-body out of the manifold and then gluing it back in with a $2\pi$ twist along a cycle of one of the handles is equivalent to inserting a $\tau$-loops along that cycle. Thus, if we start from a manifold that is occupied by the Wilson line representation of the initial target system's density matrix, we can obtain the effect of interferometry through surgery. The effect of running an untwisted interferometer may be obtained by surgeries that glue in Wilson loops representing the probe anyons or, for the asymptotic limit $N \rightarrow \infty$, that glue in $\omega$-loops. The effect of running a twisted interferometer can then be obtained in the same way, but by also performing Dehn twist surgeries for the twists. This can be represented by a single surgery, cutting a genus $2$ or $3$ handle-body out of the manifold and gluing back in the handle-body containing Wilson loops or $\omega$-loops, possibly with Dehn twists performed on cycles of the handles. Figs.~\ref{fig:handle_bodies} and \ref{fig:handle_bodies_twisted} displays these handle-bodies enclosing the probe anyons' Wilson loops and an example of Dehn surgery's effect on them.

We discuss the $3$-manifold surgery perspective for twisted and untwisted interferometry in greater detail in a companion paper~\cite{Bonderson13c}.

\section{Twisted Interferometry with Ising Anyons}
\label{sec:Ising}

Ising-type anyons are perhaps the most promising type of non-Abelian anyons in terms of physical realization. In quantum Hall systems, the Moore-Read (MR) Pfaffian~\cite{Moore91} and anti-Pfaffian~\cite{Lee07,Levin07} states are the strongest candidates for the $\nu=5/2$ quantum Hall state~\cite{Willett87,Pan99,Eisenstein02} and the Bonderson-Slingerland (BS) hierarchy states~\cite{Bonderson07d,Bonderson09a} over these provides a strong candidate for the $\nu=12/5$ state~\cite{Xia04,Kumar10}. Recent experiments provide evidence best supporting the anti-Pfaffian state as the description of the $\nu=5/2$ state~\cite{Radu08,Willett09a,Willett09b,Bid10a}. Additionally, the MR state may arise in rotating Bose condensates~\cite{Cooper01a,Cooper04a}. These candidate quantum Hall states all possess quasiparticle excitations that are Ising-type anyons~\cite{Bonderson11}. Ising-type anyons also arise in the form of Majorana zero modes occurring in 2D topological ($p_x +i p_y$) superfluids and superconductors~\cite{Read00,Volovik99}. Such topological superconductors are believed to be realized in strontium ruthenate (Sr$_2$RuO$_4$), and there have recently been several promising proposals to synthesize topological superconductors in heterostructures of more mundane materials~\cite{Fu08,Sau10a,Alicea10a,Qi10a}. Another possible (though less practical to implement) realization of Ising anyons is in Kitaev's honeycomb model~\cite{Kitaev06a}.

The braiding operations of Ising anyons are known to generate a subset of the Clifford gates. As such, they are not computationally universal. Supplementing the braiding operations of Ising anyons with the ability to perform (untwisted) interferometry measurements of anyonic charge expands the set of topologically protected computational gates to the entire Clifford gate set, which is generated by the gates
\begin{equation}
H=\frac{1}{\sqrt{2}} \left[
\begin{array}{rr}
1  &  1 \\
1  & -1
\end{array}
\right]
,
\quad
P=R_{\frac{\pi}{2}} =\left[
\begin{array}{rr}
1  &  0 \\
0  &  i
\end{array}
\right]
,
\quad \text{and} \quad
\text{CNOT} = \left[
\begin{array}{rrrr}
1  &  0  &  0  &  0 \\
0  &  1  &  0  &  0 \\
0  &  0  &  0  &  1 \\
0  &  0  &  1  &  0
\end{array}
\right]
.
\end{equation}
(We note that the Clifford gates include the Pauli gates $\sigma^z = P^2$, $\sigma^x = H \sigma^z H$, and $\sigma^y = i \sigma^x \sigma^z $.)
These are still not computationally universal, but can be made universal if supplemented by the $\pi/8$-phase gate
\begin{equation}
T = R_{\frac{\pi}{4}} =\left[
\begin{array}{rr}
1  &  0 \\
0  &  e^{i \pi /4}
\end{array}
\right]
\end{equation}
(or any $\theta/2$-phase gate $R_{\theta} = \text{diag}[1,e^{i \theta}]$ with $\theta \neq n \pi/2$) or, equivalently, the ability to produce ``magic states,'' such as
\begin{equation}
\left| \mathcal{B}_{-\frac{\pi}{4} } \right\rangle =  H R_{\frac{\pi}{4}} H \left| 0 \right\rangle= \cos(\pi/8) \left| 0 \right\rangle -i \sin(\pi/8) \left| 1 \right\rangle
,
\end{equation}
or any state obtained from this one by application of single-qubit Clifford gates.

Here, we demonstrate explicitly that twisted interferometry (together with braiding and/or topological charge measurement of pairs of anyons) is capable of generating such magic states. This should not be surprising now that we have seen how twisted interferometry is related to Dehn twists, and recalling that Ref.~\cite{Bravyi00-unpublished} also demonstrated a method of generating topologically-protected $\pi/8$-phase gates for the Ising TQFT using Dehn surgery (though the specific surgery utilized there is distinct from the one to which twisted interferometry is related).

For convenience, we recall the fusion and braiding properties of the Ising MTC
\begin{equation*}
\begin{tabular}{|l|l|}
\hline
\multicolumn{2}{|l|}{$\mathcal{C}=\left\{I,\sigma,\psi \right\}, \quad I\times a=a,\quad \sigma \times \sigma =I+\psi,\quad \sigma \times \psi=\sigma,\quad \psi \times \psi=I$}
\\ \hline
\multicolumn{2}{|l|}{$\left[ F_{\sigma}^{\sigma \sigma \sigma}\right] _{ef}=
\left[ F_{\sigma \sigma}^{\sigma \sigma}\right] _{ef}=
\left[
\begin{array}{rr}
\frac{1}{\sqrt{2}} & \frac{1}{\sqrt{2}} \\
\frac{1}{\sqrt{2}} & \frac{-1}{\sqrt{2}}%
\end{array}\right] _{ef}^{\phantom{T}}$} \\
\multicolumn{2}{|l|}{$\left[ F_{\psi}^{\sigma \psi \sigma}\right] _{\sigma \sigma}=%
\left[ F_{\sigma}^{\psi \sigma \psi}\right] _{\sigma \sigma_{\phantom{j}}}\!\!=
\left[ F_{\psi \sigma}^{\sigma \psi}\right] _{\sigma \sigma}=
\left[ F_{\sigma \psi}^{\psi \sigma}\right] _{\sigma \sigma}=-1 $} \\ \hline
\multicolumn{2}{|l|}{
$R_{I}^{\sigma \sigma}=e^{-i\frac{\pi }{8}},\quad R_{\psi}^{\sigma \sigma}=e^{i\frac{3\pi }{8}},
\quad R_{\sigma}^{\sigma \psi}=R_{\sigma}^{\psi \sigma}=e^{-i\frac{\pi }{2}},\quad R_{I}^{\psi \psi}=-1$} \\ \hline
$S=\frac{1}{2}\left[
\begin{array}{rrr}
1 & \sqrt{2} & 1 \\
\sqrt{2} & 0 & -\sqrt{2} \\
1 & -\sqrt{2} & 1%
\end{array}%
\right]^{\phantom{T}}_{\phantom{j}} $ & $M=\left[
\begin{array}{rrr}
1 & 1 & 1 \\
1 & 0 & -1 \\
1 & -1 & 1%
\end{array}%
\right] $ \\ \hline
$d_{I}=d_{\psi}=1,\quad d_{\sigma_{\phantom{j}}}\!\!=\sqrt{2}, \quad \mathcal{D}=2$ & $\theta _{I}=1,\quad \theta
_{\sigma}=e^{i\frac{\pi }{8}},\quad \theta _{\psi}=-1$ \\ \hline
\end{tabular}%
\end{equation*}%
where $I$ is the vacuum charge (previously denoted by $0$), $e,f\in \left\{ I,\psi\right\} $, and the Greek symbols labeling fusion vertices are omitted because there are trivially determined as there are no fusion multiplicities ($N_{ab}^{c}=0$ or $1$). The $F$-symbols and $R$-symbols not listed here are trivial, meaning they are equal to $1$ if allowed by the fusion rules.

We note that this gives
\begin{equation}
S T^{m} S^{\dagger} =
\left[
\begin{array}{ccc}
\frac{\theta_{\sigma}^{m}}{2} & \frac{\sqrt{2}}{2} & -\frac{\theta_{\sigma}^{m}}{2} \\
\frac{\sqrt{2}}{2} & 0 & \frac{\sqrt{2}}{2} \\
-\frac{\theta_{\sigma}^{m}}{2} & \frac{\sqrt{2}}{2} & \frac{\theta_{\sigma}^{m}}{2}
\end{array}
\right]
\, \,  \text{ for $m$ odd,}
\quad
\left[
\begin{array}{ccc}
\frac{1+\theta_{\sigma}^{m}}{2} & 0 & \frac{1-\theta_{\sigma}^{m}}{2} \\
0 & 1 & 0 \\
\frac{1-\theta_{\sigma}^{m}}{2} & 0 & \frac{1+\theta_{\sigma}^{m}}{2}
\end{array}
\right]
\,\, \text{ for $m$ even.}
\end{equation}
and that $\frac{1+\theta_{\sigma}^{m}}{2} = e^{i m \pi / 16} \cos(m\pi/16)$ and $\frac{1-\theta_{\sigma}^{m}}{2} = -i e^{i m \pi / 16} \sin(m\pi/16)$.

We consider a twisted interferometer with $m_u = 0$ and $m_l = m$, using probe anyons with definite charge $b=\sigma$, so that they are capable of distinguishing all charge types and $\tilde{\mathcal{C}}_{a} = \left\{ a \right\}$ each contains only a single charge type. In this case, there will be at most three possible values that the fraction $r=n/N$ of probe anyons measured at the $s=\shortrightarrow$ detector can approach in the asymptotic limit $N\rightarrow \infty$. For twisted interferometers that utilizes twist operator, these are given by
\begin{eqnarray}
\tilde{p}_{I} &=& \tilde{p}_{III, \sigma}^{\shortrightarrow } =\left| t_{1}\right| ^{2}\left|
r_{2}\right| ^{2}+\left| r_{1}\right| ^{2}\left| t_{2}\right| ^{2}+ 2 \text{Re} \left\{ t_{1}r_{1}^{\ast }r_{2}^{\ast }t_{2}^{\ast
}e^{i\left( \theta _{\text{I}}-\theta _{\text{II}}\right) }  \right\} \\
\tilde{p}_{\sigma} &=& \tilde{p}_{\sigma \sigma I, \sigma}^{\shortrightarrow } =\left| t_{1}\right| ^{2}\left|
r_{2}\right| ^{2}+\left| r_{1}\right| ^{2}\left| t_{2}\right| ^{2} \\
\tilde{p}_{\psi} &=& \tilde{p}_{\psi \psi I, \sigma}^{\shortrightarrow } =\left| t_{1}\right| ^{2}\left|
r_{2}\right| ^{2}+\left| r_{1}\right| ^{2}\left| t_{2}\right| ^{2} - 2 \text{Re} \left\{ t_{1}r_{1}^{\ast }r_{2}^{\ast }t_{2}^{\ast
}e^{i\left( \theta _{\text{I}}-\theta _{\text{II}}\right) }  \right\}
,
\end{eqnarray}
while for twisted interferometers that utilize pure-braid operators, they are
\begin{eqnarray}
\tilde{p}_{I} &=& \tilde{p}_{III, \sigma}^{\shortrightarrow } =\left| t_{1}\right| ^{2}\left|
r_{2}\right| ^{2}+\left| r_{1}\right| ^{2}\left| t_{2}\right| ^{2}+ 2 \text{Re} \left\{ t_{1}r_{1}^{\ast }r_{2}^{\ast }t_{2}^{\ast
}e^{i\left( \theta _{\text{I}}-\theta _{\text{II}}\right) } \theta_\sigma^{-m} \right\} \\
\tilde{p}_{\sigma} &=& \tilde{p}_{\sigma \sigma I, \sigma}^{\shortrightarrow } =\left| t_{1}\right| ^{2}\left|
r_{2}\right| ^{2}+\left| r_{1}\right| ^{2}\left| t_{2}\right| ^{2} \\
\tilde{p}_{\psi} &=& \tilde{p}_{\psi \psi I, \sigma}^{\shortrightarrow } =\left| t_{1}\right| ^{2}\left|
r_{2}\right| ^{2}+\left| r_{1}\right| ^{2}\left| t_{2}\right| ^{2} - 2 \text{Re} \left\{ t_{1}r_{1}^{\ast }r_{2}^{\ast }t_{2}^{\ast
}e^{i\left( \theta _{\text{I}}-\theta _{\text{II}}\right) } \theta_\sigma^{-m} \right\}
.
\end{eqnarray}

We are most interested in applying twisted interferometry to topological qubits. In the ``standard'' encoding, a qubit is encoded in four $\sigma$ anyons which have collective charge $I$ such that the basis states $\left|0 \right\rangle$ and $\left|1 \right\rangle$ of the qubit are given by the states in which the $\sigma$ anyons 1 and 2 fuse to $I$ and $\psi$, respectively. (Since the collective charge of the four anyons is $I$, the fusion channel of anyons 3 and 4 is equal to that of 1 and 2.) We place anyons 1 and 2 inside an interferometer (making them the $A$ anyons with allowed collective charge values $a=I$ and $\psi$) and anyons 3 and 4 in the region below the interferometer (making them the $C_1$ anyons and there be no $C_2$ anyons). This is described by the corresponding initial target system density matrix
\begin{equation}
\rho^{AC} = \sum\limits_{a,a^{\prime } = I,\psi } \rho _{\left( a,a;I \right) \left( a^{\prime },a^{\prime };I\right)}^{AC}
 \left| a,a;I \right\rangle \left\langle a^{\prime },a^{\prime };I \right| = \left[
\begin{array}{rr}
\rho _{00}  &  \rho _{01} \\
\rho _{10}  &  \rho _{11 }
\end{array}
\right]
=\rho
,
\label{eq:Ising_initial_state}
\end{equation}
which has $f=I$, $c=a$, and $c^{\prime}=a^{\prime}$. The expression in terms of the qubit density matrix $\rho$ uses the translation $\left| 0 \right\rangle = \left| I,I;I \right\rangle$ and $\left| 1 \right\rangle = \left| \psi,\psi;I \right\rangle$.

For $m$ odd, performing the twisted interferometry measurement on the initial density matrix of Eq.~(\ref{eq:Ising_initial_state}) may result in three possible outcomes, with $r$ approaching one of $\tilde{p}_{I}$, $\tilde{p}_{\sigma}$, or $\tilde{p}_{\psi}$ with corresponding probabilities
\begin{equation}
\tilde{\Pr}_{AC} \left( I \right) = 1/4, \qquad
\tilde{\Pr}_{AC} \left( \sigma \right) = 1/2, \qquad
\tilde{\Pr}_{AC} \left( \psi \right) = 1/4
\end{equation}
and resulting (fixed state) density matrices
\begin{equation}
\tilde{\rho}_{I} =
\left[
\begin{array}{rr}
\rho _{00}  &  -\rho _{01} \\
-\rho _{10}  &  \rho _{11 }
\end{array}
\right]
, \quad
\tilde{\rho}_{\sigma} =
\left[
\begin{array}{rr}
\rho _{00}  &  \rho _{01} \\
\rho _{10}  &  \rho _{11 }
\end{array}
\right]
, \quad
\tilde{\rho}_{\psi} =
\left[
\begin{array}{rr}
\rho _{00}  &  -\rho _{01} \\
-\rho _{10}  &  \rho _{11 }
\end{array}
\right]
.
\end{equation}
We notice that the $I$ and $\psi$ outcomes have the effect of applying a $\sigma^{x}$ gate to the qubit, while the $\sigma$ outcome leaves the target system unchanged (reflecting the fact that the $\sigma$ charge in the twisted basis does not distinguish between $I$ and $\psi$ charges in the untwisted basis).

For $m$ even, performing the twisted interferometry measurement on the initial density matrix of Eq.~(\ref{eq:Ising_initial_state}) can result in two possible outcomes, with $r$ approaching one of $\tilde{p}_{I}$ or $\tilde{p}_{\psi}$ with corresponding probabilities
\begin{eqnarray}
\tilde{\Pr}_{AC} \left( I \right) &=& \cos^2 \left( m\pi/16 \right) \rho_{00} +  \sin^2 \left( m\pi/16 \right) \rho_{11},
\label{eq:prob_I} \\
\tilde{\Pr}_{AC} \left( \psi \right) &=& \sin^2 \left( m\pi/16 \right) \rho_{00} +  \cos^2 \left( m\pi/16 \right) \rho_{11}
\label{eq:prob_psi}
\end{eqnarray}
and resulting (fixed state) density matrices
\begin{equation}
\tilde{\rho}_{I} = \frac{1}{\tilde{\Pr}_{AC} \left( I \right)}
\left[
\begin{array}{cc}
\cos^2 \left( \frac{m \pi}{16} \right) \rho _{00}  &  i \cos \left(  \frac{m \pi}{16} \right) \sin \left(  \frac{m \pi}{16} \right) \rho _{01} \\
-i \cos \left(  \frac{m \pi}{16} \right) \sin \left(  \frac{m \pi}{16} \right) \rho _{10}  & \sin^2 \left(  \frac{m \pi}{16} \right)   \rho _{11 }
\end{array}
\right]
,
\label{eq:rho_I}
\end{equation}
\begin{equation}
\tilde{\rho}_{\psi} = \frac{1}{\tilde{\Pr}_{AC} \left( \psi \right)}
\left[
\begin{array}{cc}
\sin^2 \left(  \frac{m \pi}{16} \right) \rho _{00}  &  -i \cos \left(  \frac{m \pi}{16} \right) \sin \left(  \frac{m \pi}{16} \right) \rho _{01} \\
i \cos \left(  \frac{m \pi}{16} \right) \sin \left(  \frac{m \pi}{16} \right) \rho _{10}  & \cos^2 \left(  \frac{m \pi}{16} \right)   \rho _{11 }
\end{array}
\right]
.
\label{eq:rho_psi}
\end{equation}

We can now easily see that if we start from the (qubit) state $\left| \Psi_{H} \right\rangle = \frac{1}{\sqrt{2}} \left( \left| 0 \right\rangle + \left| 1 \right\rangle \right) = H \left| 0 \right\rangle$ (which can be generated by braiding) and perform a twisted interferometry measurement with $m=2$, we will obtain a magic state. In particular, for the measurement outcome $\kappa_I$ (which will occur with probability $1/2$), we obtain the post-measurement state $\left| \Psi_{\mathcal{M}} \right\rangle =  \cos(\pi/8) \left| 0 \right\rangle -i \sin(\pi/8) \left| 1 \right\rangle$. For the measurement outcome $\kappa_\psi$ (which will occur with probability $1/2$), we obtain the state $\left| \Psi_{\mathcal{M}^{\prime}} \right\rangle =  -i \sin(\pi/8) \left| 0 \right\rangle + \cos(\pi/8) \left| 1 \right\rangle = \sigma^{x} \left| \Psi_{\mathcal{M}} \right\rangle$.

One might also be interested to know the effect of twisted interferometry on the target state described by
\begin{equation}
\rho^{AC} = \left| \sigma,\sigma;I \right\rangle \left\langle \sigma,\sigma;I \right|
.
\end{equation}
For $m$ even, applying twisted interferometry to this state has only one possible measurement outcome with $r$ approaching $\tilde{p}_{\sigma}$ and a post-measurement state of
\begin{equation}
\tilde{\rho}^{AC}_{\sigma} = \frac{1}{2} \left[ \left| \sigma,\sigma;I \right\rangle \left\langle \sigma,\sigma;I \right| + \left| \sigma,\sigma;\psi \right\rangle \left\langle \sigma,\sigma;\psi \right| \right]
.
\end{equation}
For $m$ odd, there are three possible measurement outcomes, with $r$ approaching one of $\tilde{p}_{I}$, $\tilde{p}_{\sigma}$, or $\tilde{p}_{\psi}$ with corresponding probabilities
\begin{equation}
\tilde{\Pr}_{AC} \left( I \right) = 1/4, \qquad
\tilde{\Pr}_{AC} \left( \sigma \right) = 1/2, \qquad
\tilde{\Pr}_{AC} \left( \psi \right) = 1/4
\end{equation}
and resulting (fixed state) density matrices
\begin{eqnarray}
\tilde{\rho}^{AC}_{I} &=& \tilde{\rho}^{AC}_{\psi} =  \left| \sigma,\sigma;I \right\rangle \left\langle \sigma,\sigma;I \right|
\\
\tilde{\rho}^{AC}_{\sigma} &=& \left| \sigma,\sigma;\psi \right\rangle \left\langle \sigma,\sigma;\psi \right|
.
\end{eqnarray}
We note that the $m$ even result is the same as in the untwisted case.

\subsection{Ising vs. Ising-type}

At this point, we must emphasize a subtle, but important distinction between Ising-type theories. In particular, the $\nu=1/2$ MR quantum Hall state is not described by a purely Ising TQFT, nor even the direct product of an Ising TQFT with a U$(1)_2$ charge sector. Rather, it is the restricted product of an Ising and a U$(1)_2$ theory, where the restriction is such that the $I$ and $\psi$ Ising charges are paired with integer valued U$(1)_2$ vortices, while the $\sigma$ Ising charge is paired with half-integer valued U$(1)_2$ vortices. Hence, the MR state has $S$-matrix and $T$-matrix that are distinct from those of an Ising theory or those of a direct product of an Ising theory with another TQFT. In fact, the $\nu=1/2$ MR state, or any fermionic quantum Hall state for that matter, is not a modular theory, as it has a degenerate $S$-matrix. This means the previous analysis cannot be applied, because Eq.~(\ref{eq:omega_a_projection}) does not hold.

However, the situation can be partially salvaged because these fermionic quantum Hall states can be treated as a $\mathbb{Z}_2$-graded TQFT or a spin field theory~\cite{Dijkgraaf90}, where each anyonic charge type forms a $\mathbb{Z}_2$-doublet with the charge obtained from it by fusion with an electron. In other words, we want to treat the electron as a trivial object, (in some sense) equivalent to the ``vacuum'' of the theory, in order to remove the degeneracy of the $S$-matrix. Of course, we cannot simply identify them, because the trivial/vacuum topological charge has bosonic exchange statistics while the electron has fermionic exchange statistics, so we must introduce a $\mathbb{Z}_2$-grading. By grouping charges into $\mathbb{Z}_2$ doublets this way, the resulting $S$-matrix becomes unitary, but it leads to another issue. The two topological charges within a $\mathbb{Z}_2$-doublet necessarily have topological spins that are negative that of each other, e.g. $\theta_{I}=1$ and $\theta_{e^{-}}=-1$. Thus, there is no well-defined $T$-matrix for the $\mathbb{Z}_2$-graded theory. There is, however, a well-defined $T^2$-matrix and the $\mathbb{Z}_2$-graded TQFT is well-defined if one restricts to the subset of modular transformations which also preserve spin structures of the manifold (hence the name ``spin field theory''). In this case, Eq.~(\ref{eq:omega_a_projection}) can be used with the understanding that it projects onto $\mathbb{Z}_2$-doublets. This causes no problems for applications in planar systems, since one cannot have superpositions between topological charges corresponding to different electrical charge values anyway. For use in twisted interferometry, this also causes no problems as long as one restricts to operations involving even numbers of twists.

Hence, one can salvage the previous analysis (with slight modifications) for fermionic quantum Hall states, as long as we use the allowed $\mathbb{Z}_2$-graded operations, meaning one must use only even values of twists $m_{l}$ and $m_{u}$. Doing this for the $\nu=1/2$ MR state results in the $\mathbb{Z}_2$-graded matrices
\begin{equation}
S=\frac{1}{\sqrt{8}}\left[
\begin{array}{cccccc}
1 & 1 & \sqrt{2} & 1 & 1 & \sqrt{2} \\
1 & 1 & -\sqrt{2} & 1 & 1 & -\sqrt{2} \\
\sqrt{2} & -\sqrt{2} & 0 & i\sqrt{2} & -i\sqrt{2} & 0 \\
1 & 1 & i\sqrt{2} & -1 & -1 & -i\sqrt{2} \\
1 & 1 & -i\sqrt{2} & -1 & -1 & i\sqrt{2} \\
\sqrt{2} & -\sqrt{2} & 0 & -i\sqrt{2} & i\sqrt{2} & 0 \\
\end{array}%
\right]
\end{equation}
and
\begin{equation}
T^2=\text{diag} \left[1,1,i,-1,-1,i \right]
,
\end{equation}
where the column/row order of charge type doublets is $(I_0, \psi_2)$, $(\psi_0, I_2)$, $(\sigma_{1/2}, \sigma_{5/2})$, $(I_{1}, \psi_{3})$, $(\psi_{1}, I_{3})$, and $(\sigma_{3/2}, \sigma_{7/2})$, and the subscript on the topological charges indicate the U$(1)_2$ flux values. For these, we find that twisted interferometry and/or Dehn surgery cannot be used to generate magic states for topological qubits in the $\nu=1/2$ MR state\footnote{
To see why the Dehn surgery protocol of Ref.~\cite{Bravyi00-unpublished} fails to produce magic states for the $\nu=1/2$ MR state, one also needs the ``spectacles diagram''
$S^{(\psi_0)} = \frac{1}{\sqrt{2}} \left[
\begin{matrix}
1 & i \\
i & 1
\end{matrix}
\right]$, where the columns/rows of this matrix take the charge type doublet values $(\sigma_{1/2}, \sigma_{5/2})$ and $(\sigma_{3/2}, \sigma_{7/2})$ only (i.e. the matrix elements are zero for other values).}.
The same inability to produce magic states is true for the anti-Pfaffian state and any other $\nu = p/q$ Ising-type quantum Hall state with even denominator $q$.

For $\nu = p/q$ Ising-type quantum Hall states with odd denominator $q$, it is possible to generate magic states using twisted interferometry (or Dehn surgery), though it may require the use of $m = 2q$ twists.
This includes the bosonic $\nu=1$ MR state, which may arise in rotating bose condensates and whose TQFT is SU$(2)_2$. It also includes the BS hierarchy states built over the MR or anti-Pfaffian states, that have filling fractions with even numerator $p$. Most notably, these include the $\nu=2/5$ and $2/3$ BS states, which are candidates for the experimentally observed $\nu=12/5$ and $8/3$ quantum Hall plateaus. In fact, these BS states can be written as a $\mathbb{Z}_2$-graded TQFT which is the direct product of a $\mathbb{Z}_q$ TQFT with one of the eight Galois conjugates of the Ising TQFT, which describes the neutral sector of one of these even numerator BS states. The Galois conjugates of the Ising TQFT have the same fusion rules as Ising, but may have different $F$-matrices and $R$-matrices. They all have the same $S$-matrix as the Ising TQFT, as well as the same topological spins $\theta_I=1$ and $\theta_{\psi}=-1$, but different values of topological spin $\theta_{\sigma}=e^{i \frac{2n+1}{8} \pi}$ for integers $n$ (mod 8), which uniquely identifies the particular Galois conjugate TQFT. In particular, the Ising TQFT has $\theta_{\sigma}=e^{i \frac{\pi}{8}}$, the SU$(2)_2$ TQFT has $\theta_{\sigma}=e^{i \frac{3\pi}{8}}$, and the neutral sector of the $\nu=2/5$ and $2/3$ BS states have $\theta_{\sigma}=e^{-i \frac{3\pi}{8}}$ and $\theta_{\sigma}=e^{i \frac{5\pi}{8}}$, respectively. In terms of which computational gates may be obtained via braiding and/or modular transformations, the Galois conjugates of the Ising TQFT are all equivalent to Ising (certain topological operations may produce different gates in the different TQFTs, but they will still generate the same gate set).

Similarly, the time-reversal invariant systems built from superconductors and topological insulators~\cite{Fu08} are Ising-type theories that are not able to generate magic states or $\pi/8$-phase gates using twisted interferometry and/or modular transformations. Chiral $p_x + i p_y$ superconductor systems~\cite{Read00,Volovik99,Sau10a,Alicea10a,Qi10a} are Ising theories, but these suffer from a different problem that makes them unsuitable for twisted interferometry and modular operations, which is that they (and the superconductor heterostructure systems) are actually quasi-topological phases~\cite{Bonderson12d}, as we will discuss in Section~\ref{sec:engineer_twist}.

\section{Partial Interferometry and Faking the Twist for Ising-type Anyons}
\label{sec:fake_twist}

We now return to the ordinary untwisted anyonic interferometer and explore some potentially useful, interesting properties of Ising-type anyons.

We first recall from Refs.~\cite{Bonderson07b,Bonderson07c} that if the beam splitters are such that $|t_1|^2=|t_2|^2=|r_1|^2 =|r_2|^2=\frac{1}{2}$, then we tune the interferometer so that it will perfectly distinguish between the charges $a=I$ and $\psi$ using only a \emph{single} $\sigma$ probe anyon. In particular, if we tune the interferometer's parameters so that $ t_{1}r_{1}^{\ast }r_{2}^{\ast }t_{2}^{\ast}e^{i\left( \theta _{\text{I}}-\theta _{\text{II}}\right) } = \frac{1}{4}$, then we find that
\begin{eqnarray}
{p}_{I} &=& {p}_{III, \sigma}^{\shortrightarrow } = 1 \\
{p}_{\psi} &=& {p}_{\psi \psi I, \sigma}^{\shortrightarrow } = 0
,
\end{eqnarray}
so if the single probe anyon is measured at the $s=\shortrightarrow$ detector, then the state is completely projected onto charge $a=I$, and if it is measured at the $\shortuparrow$ detector, then the state is completely projected onto charge $a=\psi$ [assuming the state of the target anyons was in a superposition of only these two charge values, as it is for the topological qubit describe in Eq.~(\ref{eq:Ising_initial_state})].

We now notice that if we use the same configuration of the (untwisted) interferometer, but instead tune the parameters so that $ t_{1}r_{1}^{\ast }r_{2}^{\ast }t_{2}^{\ast}e^{i\left( \theta _{\text{I}}-\theta _{\text{II}}\right) } = \frac{\theta_{\sigma}^{m} }{4} = \frac{e^{i m \pi /8}}{4} $, we find that
\begin{eqnarray}
p_{III,\sigma}^{\shortrightarrow } &=& \frac{1}{2} \left[ 1 + \cos\left( m \pi / 8 \right) \right] = \cos^2 \left( m \pi / 16 \right) \\
p_{I \psi \psi,\sigma}^{\shortrightarrow } &=& \frac{i}{2} \sin \left( m \pi / 8 \right) = i \cos \left( m \pi / 16 \right) \sin \left( m \pi / 16 \right)  \\
p_{ \psi I \psi,\sigma}^{\shortrightarrow } &=& -\frac{i}{2} \sin \left( m \pi / 8 \right) = -i \cos \left( m \pi / 16 \right) \sin \left( m \pi / 16 \right) \\
p_{\psi \psi I,\sigma}^{\shortrightarrow } &=& \frac{1}{2} \left[ 1 - \cos\left( m \pi / 8 \right) \right] = \sin^2 \left( m \pi / 16 \right)
.
\end{eqnarray}
If we consider the target system to be a topological qubit, as in Eq.~(\ref{eq:Ising_initial_state}), and send a single $\sigma$ probe anyon through the interferometer, we find that the probe anyon will be measured at the $s=\shortrightarrow$ and $\shortuparrow$ detectors with respective probabilities
\begin{eqnarray}
{\Pr} \left( \shortrightarrow \right) &=& \cos^2 \left( m\pi/16 \right) \rho_{00} +  \sin^2 \left( m\pi/16 \right) \rho_{11},
\label{eq:prob_right} \\
{\Pr} \left( \shortuparrow \right) &=& \sin^2 \left( m\pi/16 \right) \rho_{00} +  \cos^2 \left( m\pi/16 \right) \rho_{11}
\label{eq:prob_up}
\end{eqnarray}
and the corresponding post-measurement target system density matrices will be
\begin{equation}
{\rho} \left( \shortrightarrow \right) = \frac{1}{{\Pr}\left( \shortrightarrow \right)}
\left[
\begin{array}{cc}
\cos^2 \left( \frac{m \pi}{16} \right) \rho _{00}  &  i \cos \left(  \frac{m \pi}{16} \right) \sin \left(  \frac{m \pi}{16} \right) \rho _{01} \\
-i \cos \left(  \frac{m \pi}{16} \right) \sin \left(  \frac{m \pi}{16} \right) \rho _{10}  & \sin^2 \left(  \frac{m \pi}{16} \right)   \rho _{11 }
\end{array}
\right]
,
\label{eq:rho_right}
\end{equation}
\begin{equation}
{\rho}\left( \shortuparrow \right) = \frac{1}{{\Pr} \left( \shortuparrow \right)}
\left[
\begin{array}{cc}
\sin^2 \left(  \frac{m \pi}{16} \right) \rho _{00}  &  -i \cos \left(  \frac{m \pi}{16} \right) \sin \left(  \frac{m \pi}{16} \right) \rho _{01} \\
i \cos \left(  \frac{m \pi}{16} \right) \sin \left(  \frac{m \pi}{16} \right) \rho _{10}  & \cos^2 \left(  \frac{m \pi}{16} \right)   \rho _{11 }
\end{array}
\right]
.
\label{eq:rho_up}
\end{equation}
Comparing to Eqs.~(\ref{eq:prob_I},\ref{eq:prob_psi},\ref{eq:rho_I},\ref{eq:rho_psi}), we see that this single probe untwisted interferometry measurement with tuned parameters exactly replicates the effect of the ($m$ even) twisted interferometer on the topological qubit and can similarly be used to generate magic states! Of course, this single probe operation is not topologically protected as it requires fine-tuning of the interferometer's parameters. Moreover, it requires the ability to send precisely one probe anyon through the interferometer, which may be difficult depending on the system. Thus, this method, which bears a number of similarities to that of Refs.~\cite{Bonderson10c,Clarke10a}, can only be viewed as a topologically unprotected method of producing $\pi/8$-phase gates, which will require error-correction. Fortunately, it was shown that if one has access to topologically protected Clifford gates (as one does for Ising anyons), then magic states can be error-corrected using the ``magic state distillation'' protocol of Ref.~\cite{Bravyi05}, which has a remarkably high error threshold of approximately $0.14$. It is worth emphasizing that this method works for any Ising-type system, not just for pure Ising TQFTs.

It should be clear that there was nothing special about the phase $\theta_{\sigma}^{m}$ that was required for this to work. In particular, we could have set $ t_{1}r_{1}^{\ast }r_{2}^{\ast }t_{2}^{\ast}e^{i\left( \theta _{\text{I}}-\theta _{\text{II}}\right) } = \frac{ e^{i \phi} }{4}$ and the result of the single probe measurement would have probabilities
\begin{eqnarray}
{\Pr} \left( \shortrightarrow \right) &=& \cos^2 \left( \phi / 2 \right) \rho_{00} +  \sin^2 \left( \phi / 2 \right) \rho_{11},
\label{eq:prob_right_phi} \\
{\Pr} \left( \shortuparrow \right) &=& \sin^2 \left( \phi / 2 \right) \rho_{00} +  \cos^2 \left( \phi / 2 \right) \rho_{11}
\label{eq:prob_up_phi}
\end{eqnarray}
and corresponding post-measurement target system density matrices
\begin{equation}
{\rho} \left( \shortrightarrow \right) = \frac{1}{{\Pr}\left( \shortrightarrow \right)}
\left[
\begin{array}{cc}
\cos^2 \left( \frac{\phi}{2} \right) \rho _{00}  &  i \cos \left(  \frac{\phi}{2} \right) \sin \left(  \frac{\phi}{2} \right) \rho _{01} \\
-i \cos \left(  \frac{\phi}{2} \right) \sin \left(  \frac{\phi}{2} \right) \rho _{10}  & \sin^2 \left(  \frac{\phi}{2} \right)   \rho _{11 }
\end{array}
\right]
,
\label{eq:rho_right_phi}
\end{equation}
\begin{equation}
{\rho}\left( \shortuparrow \right) = \frac{1}{{\Pr} \left( \shortuparrow \right)}
\left[
\begin{array}{cc}
\sin^2 \left(  \frac{\phi}{2} \right) \rho _{00}  &  -i \cos \left(  \frac{\phi}{2} \right) \sin \left(  \frac{\phi}{2} \right) \rho _{01} \\
i \cos \left( \frac{\phi}{2} \right) \sin \left(  \frac{\phi}{2} \right) \rho _{10}  & \cos^2 \left( \frac{\phi}{2} \right)   \rho _{11 }
\end{array}
\right]
.
\label{eq:rho_up_phi}
\end{equation}
As such, this is a topologically unprotected method of generating arbitrary phase gates for Ising-type anyons. In particular, one must fine-tune the interferometer's parameters to produce a gate with a specific value of $\phi$, and the corrections due to imprecise tuning are not exponentially suppressed.

It is worth considering how the single probe method worked and how generally it can be applied. A key property of this method was that the single probe measurement took a initial pure state to a final pure state. For this to be possible, we had to be able to write the factors $p_{aa^{\prime }e,b}^{s}$ of the pertinent charge values as product. This is actually the case for Ising-type anyons when $a,a^{\prime} = I$ and $\psi$, regardless of the values of the interferometer's parameters. In particular, for arbitrary parameter values, we can write
\begin{equation}
\label{eq:p_product}
p_{aa^{\prime }e,b}^{s}= \mathcal{A}^{s}_{a} \mathcal{A}^{s \ast}_{a^{\prime}}
,
\end{equation}
(for $a,a^{\prime} = I$ and $\psi$ and $b=\sigma$) where
\begin{eqnarray}
\mathcal{A}^{\shortrightarrow}_{I} &=& t_{1}r_{2}^{\ast } e^{i \theta _{\text{I}} }+ r_{1} t_{2} e^{i\theta _{\text{II}} } \\
\mathcal{A}^{\shortrightarrow}_{\psi} &=& - t_{1}r_{2}^{\ast } e^{i \theta _{\text{I}}} + r_{1} t_{2} e^{i\theta _{\text{II}} } \\
\mathcal{A}^{\shortuparrow}_{I} &=& -t_{1} t_{2}^{\ast } e^{i \theta _{\text{I}}} + r_{1} r_{2} e^{i\theta _{\text{II}} } \\
\mathcal{A}^{\shortuparrow}_{\psi} &=&  t_{1}t_{2}^{\ast } e^{i \theta _{\text{I}}} + r_{1} r_{2} e^{i\theta _{\text{II}} }
.
\end{eqnarray}
Thus, applying a single $\sigma$ probe measurement with outcome $s$ to a topological qubit in an arbitrary initial pure state $\left| \Psi \right\rangle =  \Psi_0 \left| 0 \right\rangle + \Psi_1 \left| 1 \right\rangle$ results in the post-measurement state
\begin{equation}
\left| \Psi \right\rangle \mapsto \left| \Psi(s) \right\rangle =  \frac{ \mathcal{A}^{s}_{I} \Psi_0 \left| 0 \right\rangle + \mathcal{A}^{s}_{\psi} \Psi_1 \left| 1 \right\rangle}{\left[ \left| \mathcal{A}^{s}_{I} \Psi_0  \right|^2 + \left| \mathcal{A}^{s}_{\psi} \Psi_1  \right|^2 \right]^{1/2} }
.
\end{equation}

Similarly, if we send (a finite number) $N$ probes through the interferometer and $n$ of them are measured with outcome $s = \shortrightarrow$, the post-measurement state will be
\begin{equation}
\left| \Psi \right\rangle \mapsto \left| \Psi_{N}(n) \right\rangle =  \frac{ \left( \mathcal{A}^{\shortrightarrow}_{I} \right)^{n} \left( \mathcal{A}^{\shortuparrow}_{I} \right)^{N-n} \Psi_0 \left| 0 \right\rangle + \left( \mathcal{A}^{\shortrightarrow}_{\psi} \right)^{n} \left( \mathcal{A}^{\shortuparrow}_{\psi} \right)^{N-n} \Psi_1 \left| 1 \right\rangle}{\left[ \left| \left( \mathcal{A}^{\shortrightarrow}_{I} \right)^{n} \left( \mathcal{A}^{\shortuparrow}_{I} \right)^{N-n} \Psi_0  \right|^2 + \left| \left( \mathcal{A}^{\shortrightarrow}_{\psi} \right)^{n} \left( \mathcal{A}^{\shortuparrow}_{\psi} \right)^{N-n} \Psi_1  \right|^2 \right]^{1/2} }
.
\end{equation}
We call this ``partial interferometry,'' since the post-measurement target system's state is not necessarily in a fixed state of definite charge $a = I$ or $\psi$ (i.e. having the qubit projected onto either $\left| 0 \right\rangle$ or $\left| 1 \right\rangle$), as would be the case resulting in the $N\rightarrow \infty$.

Similar to the single probe measurement, one could use partial interferometry for Ising anyons to generate qubit states that can not be obtained using only Clifford gate operations. This is, however, not a deterministic process, as each probe sent through will have some probability of being found measured at either detector [with the probabilities of the outcomes given in Eq.~(\ref{eq:prob_n_N})], so it will require adaptive post-measurement processing. Again, this is a topologically unprotected method of generating computational gates for Ising-type anyons, because it requires fine-tuning of the interferometer's parameters.

Finally, we consider how generally this method can be applied for other types of anyons. It is easy to see that for $p_{aa^{\prime }e,b}^{s}$ to take the product form in Eq.~(\ref{eq:p_product}), one must have $\left| M_{ab} \right| = \left| M_{a^{\prime}b} \right| = 1$, $a = a^{\prime} \times e$, and $M_{eb}= \pm 1$. These conditions are generically not satisfied by a general anyon model. In particular, they are not satisfied by the Fibonacci anyons or SU$(2)_k$ for $k>2$. Ising-type anyons are among this special class of anyons for which partial interferometry can take pure states to pure states.

\section{Engineering the Twist}
\label{sec:engineer_twist}

Having determined the functional effect of twisted interferometry and established that it could prove useful, particularly in the context of Ising anyons, we now consider the possible implementations of twisted interferometers in physical systems. We focus on quantum Hall systems and 2D topological ($p_x \pm i p_y$) superconductors, including their synthetic realizations, since these offer the most likely physical realizations. In these systems, the target anyons are bulk non-Abelian quasiparticles and the probe anyons are excitations of the gapless edge modes that carry the appropriate non-Abelian anyonic charge. We propose two different methods of implementing probe twisting or braiding operations in their paths through the interferometer.

\begin{figure}[t!]
\begin{center}
  \includegraphics[scale=0.6]{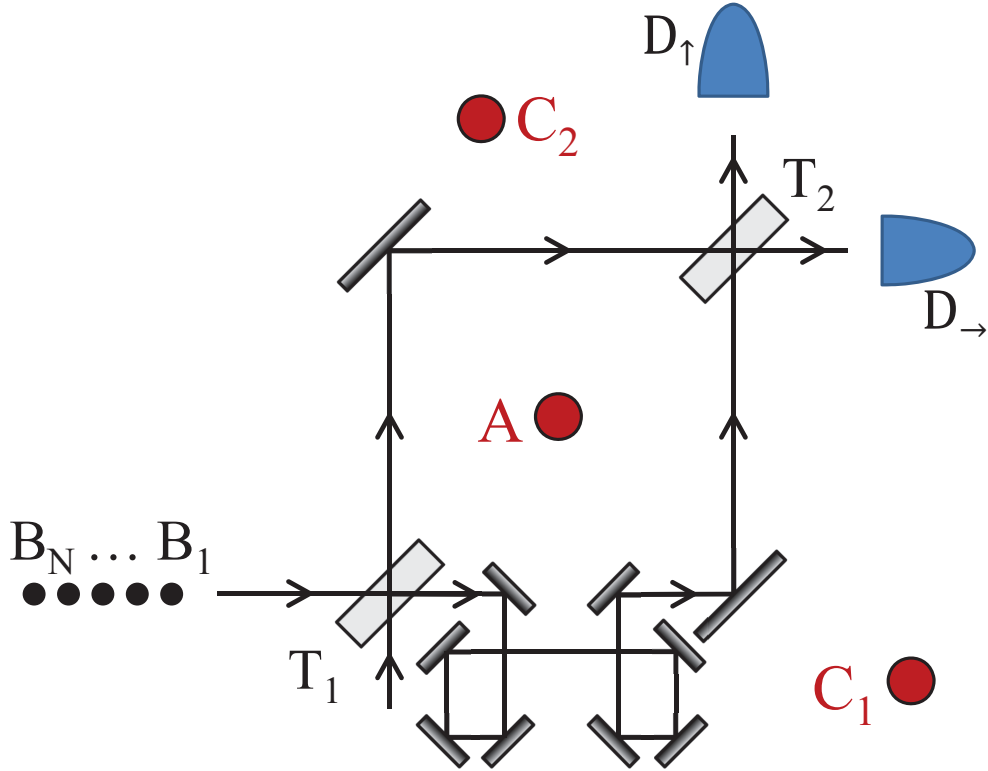}
  \caption{An idealized Mach-Zehnder anyonic interferometer with a doubly twisted path in its lower leg.}
  \label{fig:Double_Twisted_int}
\end{center}
\end{figure}

\subsection{Twisted Track}
\label{sec:twisted_track}

The first proposed method of implementation is to very literally build the interferometer with twists in one or both legs of the interferometer, which we call a ``twisted track.'' We showed how to construct such twists for an idealized anyonic system in Fig.~\ref{fig:Double_Twist}, and illustrate the addition of $m=2$ twists in the lower leg of an ideal Mach-Zehnder anyonic interferometer in Fig.~\ref{fig:Double_Twisted_int}.

\begin{figure}[t!]
\begin{center}
  \includegraphics[scale=0.5]{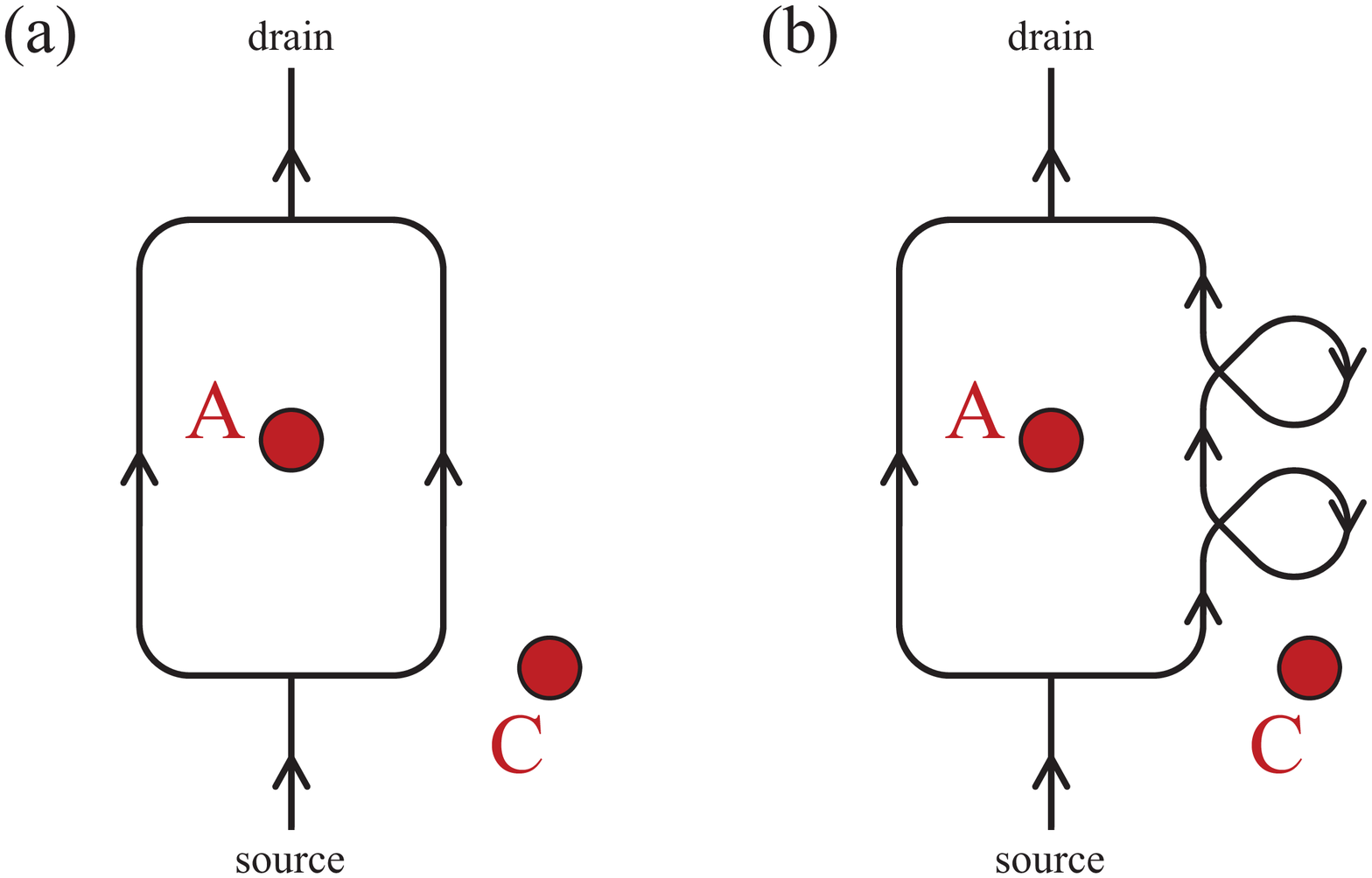}
  \caption{(a) A transmission interferometer for 2D $p_x \pm i p_y$ superconductors and Majorana heterostructures. Channels (indicated by black lines) for Josephson vortices ($\sigma$ anyons) are created between superconducting regions using insulators. The vortices are driven through the channels by applying a transverse supercurrent, which induces a Magnus force on the vortices. (b) A transmission interferometer with a doubly twisted track built in one path around the interference region.}
  \label{fig:transmission_int}
\end{center}
\end{figure}

\begin{figure}[t!]
\begin{center}
\includegraphics[scale=0.8]{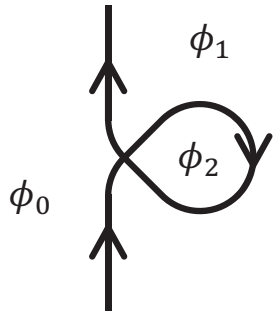}
\caption{Relative superconducting phases $0 < \phi_0 < \phi_1 < \phi_2 \ll \pi$ used to properly orient the self-crossing junction in a track for Josephson vortices.}
\label{fig_phi_states}
\end{center}
\end{figure}

This idea of constructing twisted tracks can be implemented in some physically realistic anyonic interferometers, such as the transmission interferometers proposed in Ref.~\cite{Grosfeld11a} for topological superconductors and Majorana heterostructures, as illustrated in Fig.~\ref{fig:transmission_int}. These transmission interferometers involve a branching ``track'' of insulator a few nanometers wide in a 2D topological superconductor or Majorana heterostructure. The track is essentially a long Josephson junction. The probe quasiparticles in this device are Josephson vortices or ``fluxons,'' which essentially behave as Ising $\sigma$ non-Abelian anyons. By controlling the relative superconducting phase across the track, a transverse supercurrent can be established. This will propel Josephson vortices along the track via the Magnus force. (For many more details on the physics of such systems, we refer the reader to~\cite{Wallraff-PHD2000,Wallraff03} and references therein.) At the branching of the track, the vortices will have amplitudes $t_L$ and $t_R$ for traversing the left or right path, respectively, through the interferometer. The twisting is added to the interferometer by looping the track around and allowing it to cross itself, as shown in Fig.~\ref{fig:transmission_int}(b).
The track may cross itself, provided that superconducting phase differences are arranged so that supercurrent is always tunneling from left to right as seen from the frame of the propagating Josephson vortex. The desired phase differences may be achieved by producing phases $0 < \phi_0 < \phi_1 < \phi_2 \ll \pi$ in the three complementary regions, as indicated in Fig.~\ref{fig_phi_states}, effectively orienting the track so the Josephson vortices travel with the arrow of orientation. It is important that the vortices pass straight through the self-crossing junction and not make angled turns that bypass or make multiple passes through the twisting loop.

\subsection{Track Switching}
\label{sec:track_switching}

\begin{figure}[t!]
\begin{center}
  \includegraphics[scale=0.4]{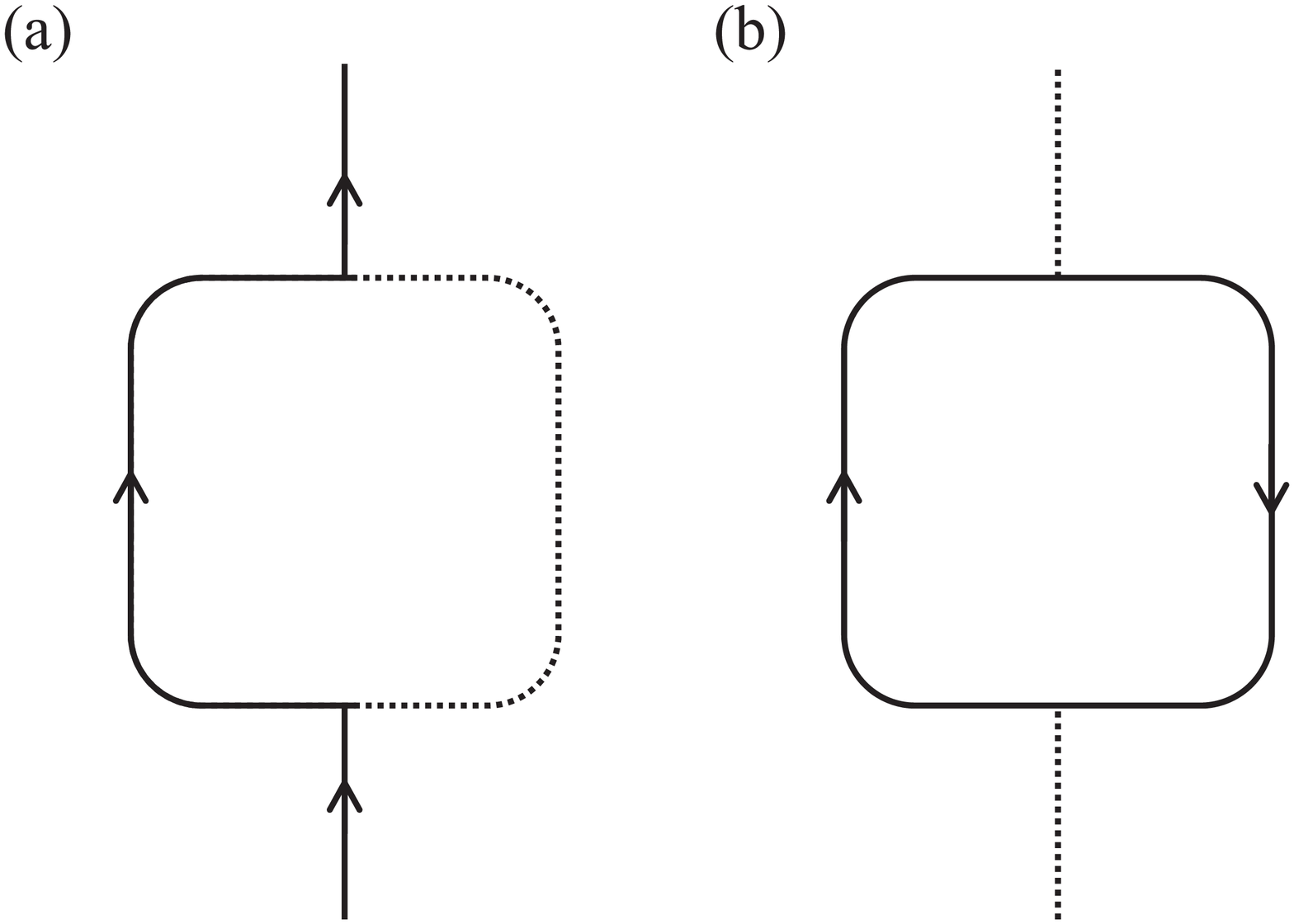}
  \caption{Twisting or braiding of anyons traveling along a track can be introduced by dynamically switching the track configuration (e.g. using gates with appropriate timing) from (a) a straight-through path to (b) a closed-loop path, and then back to the straight-through path. Solid lines indicate segments of the track permits anyons to flow through, while dotted lines indicate segments where the track is effectively blocked.}
  \label{fig:track_switching}
\end{center}
\end{figure}

The second proposed method involves ``track switching'' to divert the flow of probe anyons onto a looped track, where they can twist/braid around each other before redirecting them to continue their path through the interferometer. This is shown schematically in Fig.~\ref{fig:track_switching}. This method requires one to dynamically alter the track in a manner that is appropriately timed to allow all the probe anyons passing through a given path to be diverted onto a closed looped track, as in Fig.~\ref{fig:track_switching}(b), until they have executed the desired number of twists/pure-braids, and finally altered to allow all of them to resume their course through the interferometer.

It is clear how to include such a track switching configuration into one (or both) of the legs of the transmission interferometers shown in Fig.~\ref{fig:transmission_int}(a). The dynamical switching of the track can be controlled by using gates, currents, and tuning the phases of the superconducting order parameters of different regions.

\begin{figure}[t!]
\begin{center}
  \includegraphics[scale=0.26]{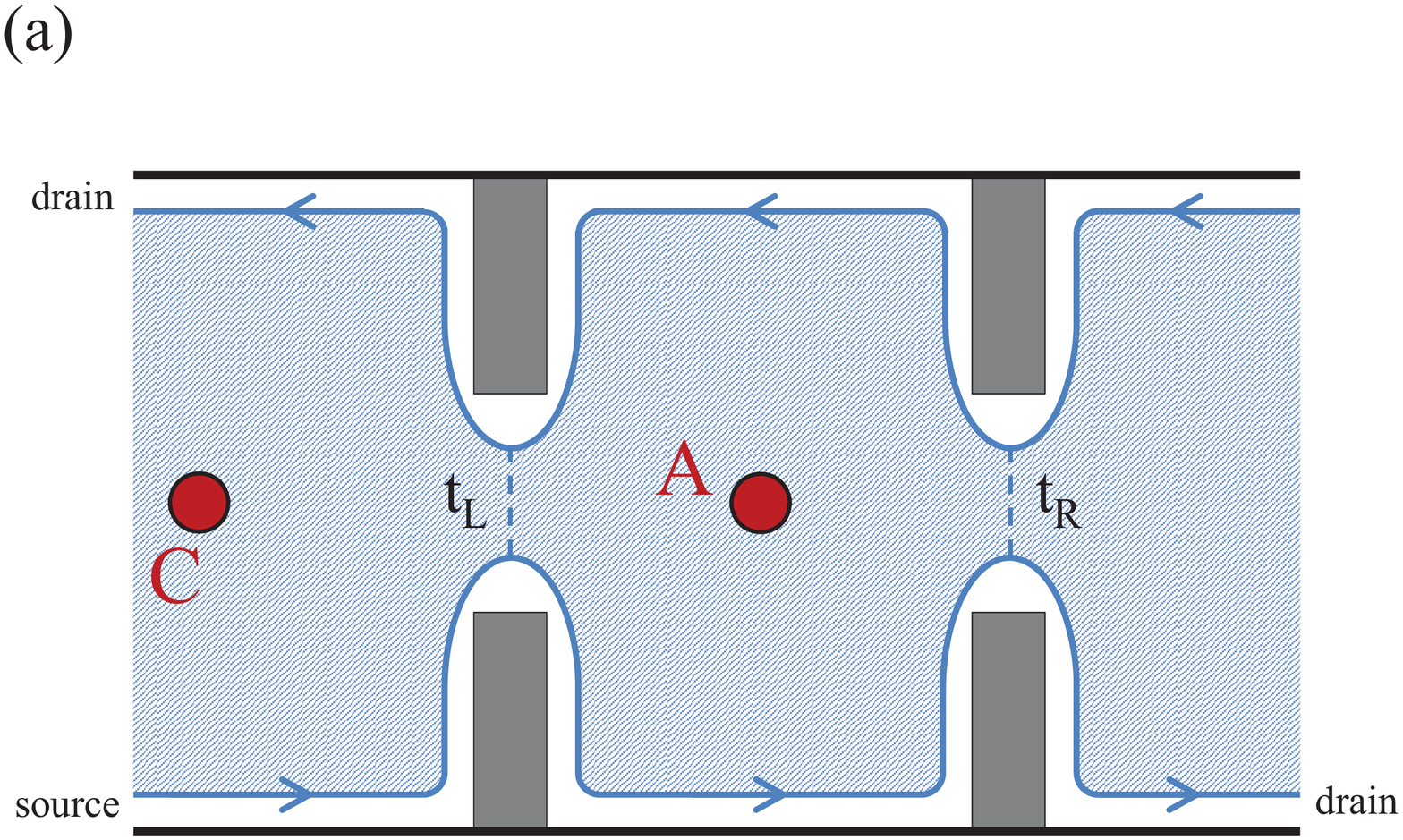}
  \includegraphics[scale=0.26]{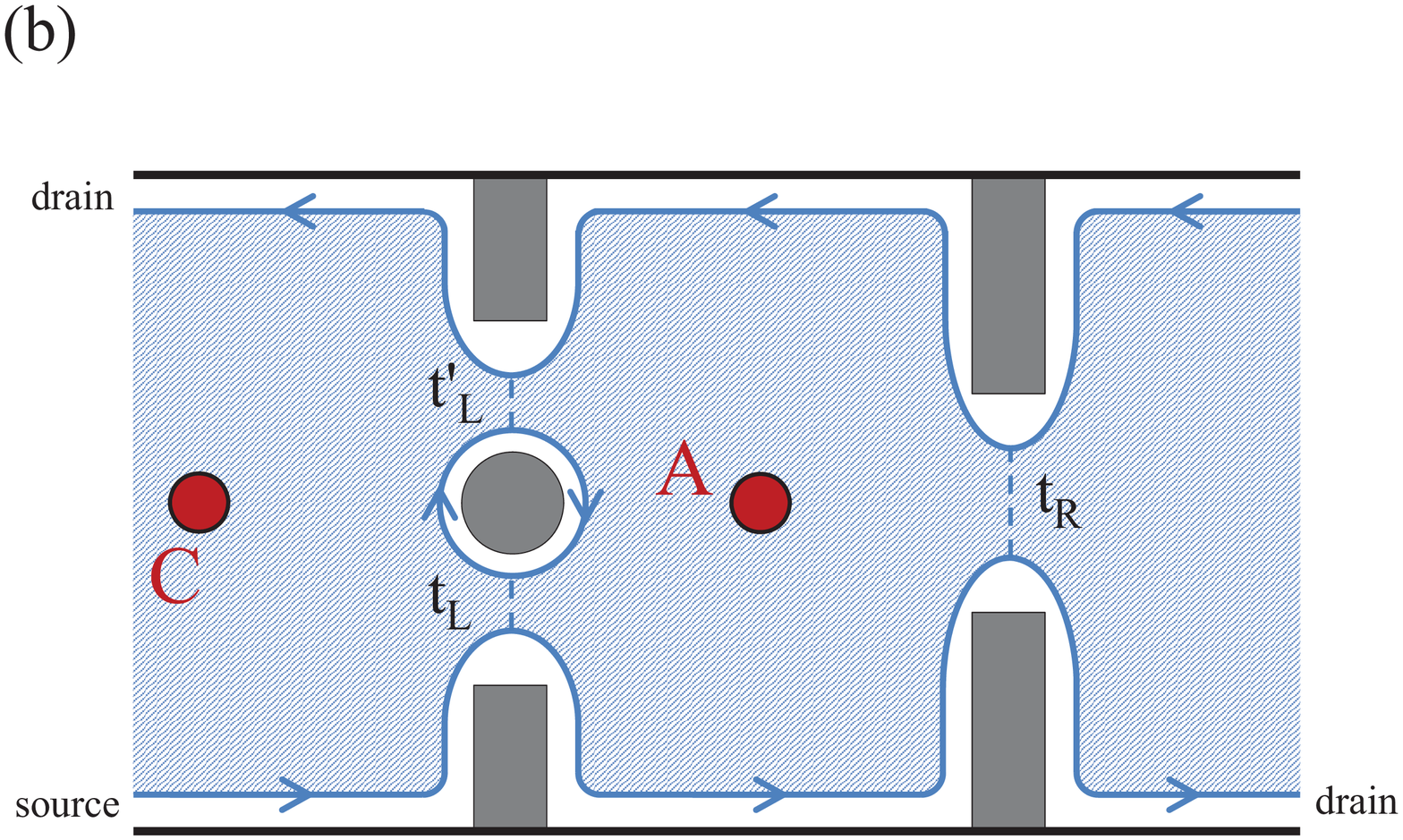}
  \caption{(a) A Fabry-P\'{e}rot type double point contact interferometer for quantum Hall states or 2D $p_x \pm i p_y$ superconductors and Majorana heterostructures. The bulk topological phase is indicated by the shaded blue region. Edge modes, which provide the ``beam of probe anyons,'' are indicated by solid blue lines with arrows indicating their chirality. Two junctions (point contacts) between edges are created with tunneling amplitudes $t_L$ and $t_R$, respectively, using electrostatic gates, indicated by the solid grey bars. (b) The double point contact interferometer is modified to include twisting in one of the paths by introducing an intermediate island in one of the point contacts using an electrostatic gate. By dynamically controlling the tunneling amplitudes $t_L$ onto the island from the lower edge and $t_{L}^{\prime}$ off of the island onto the upper edge, one can allow a specified number of probe quasiparticles to tunneling onto the island and then circulate around the island (and hence each other) a desired number of times equal to the twisting number for the path before tunneling off the island to the other edge.}
  \label{fig:DPC_int}
\end{center}
\end{figure}

The track switching method of generating probe twisting/braiding can also be introduced in other physically realistic interferometers, such as Fabry-P\'{e}rot type double point contact interferometers, which have been proposed for use in quantum Hall systems, topological superconductors, Majorana heterostructures, and any 2D topological phase with gapless edge modes~\cite{Chamon97,Fradkin98,DasSarma05,Stern06a,Bonderson06a,Bonderson06b,Bonderson07c,Akhmerov09a}, and have been experimentally realized in quantum Hall systems~\cite{Willett09a,Willett09b}. We illustrate such an interferometer in Fig.~\ref{fig:DPC_int}(a). The ``beam'' of probe anyons is provided by the gapless edge modes of the topological system, with edge excitations carrying non-Abelian topological charge being the probe anyons. These probe quasiparticles are injected by an edge current source, which could be a source for electric current, or possibly another form of current, such as heat current, that is transported by the edge. ``Beam-splitters'' are provided by point contacts, which are formed by pinching in opposite edges of the system to induce quasiparticle tunneling between them (with amplitudes $t_L$ and $t_R$, respectively). Probe outcome measurements are conducted by measuring current at the designated drains. The functioning of these Fabry-P\'{e}rot type interferometers is essentially the same as the idealized Mach-Zehnder anyonic interferometer in the limit where multiple passes of a given probe anyon through the Fabry-P\'{e}rot interferometer are rare, i.e. when the tunneling amplitudes $t_L$ and $t_R$ are small. (For topological superconductors and Majorana heterostructures, this may involve tunneling of Abrikosov vortices. So far, quantum behavior, e.g. tunneling, has not been experimentally observed for Abrikosov vortices and they are likely too massive, due to mini-gap states, for practical application in this regard.)

\begin{figure}[t!]
\begin{center}
  \includegraphics[scale=0.26]{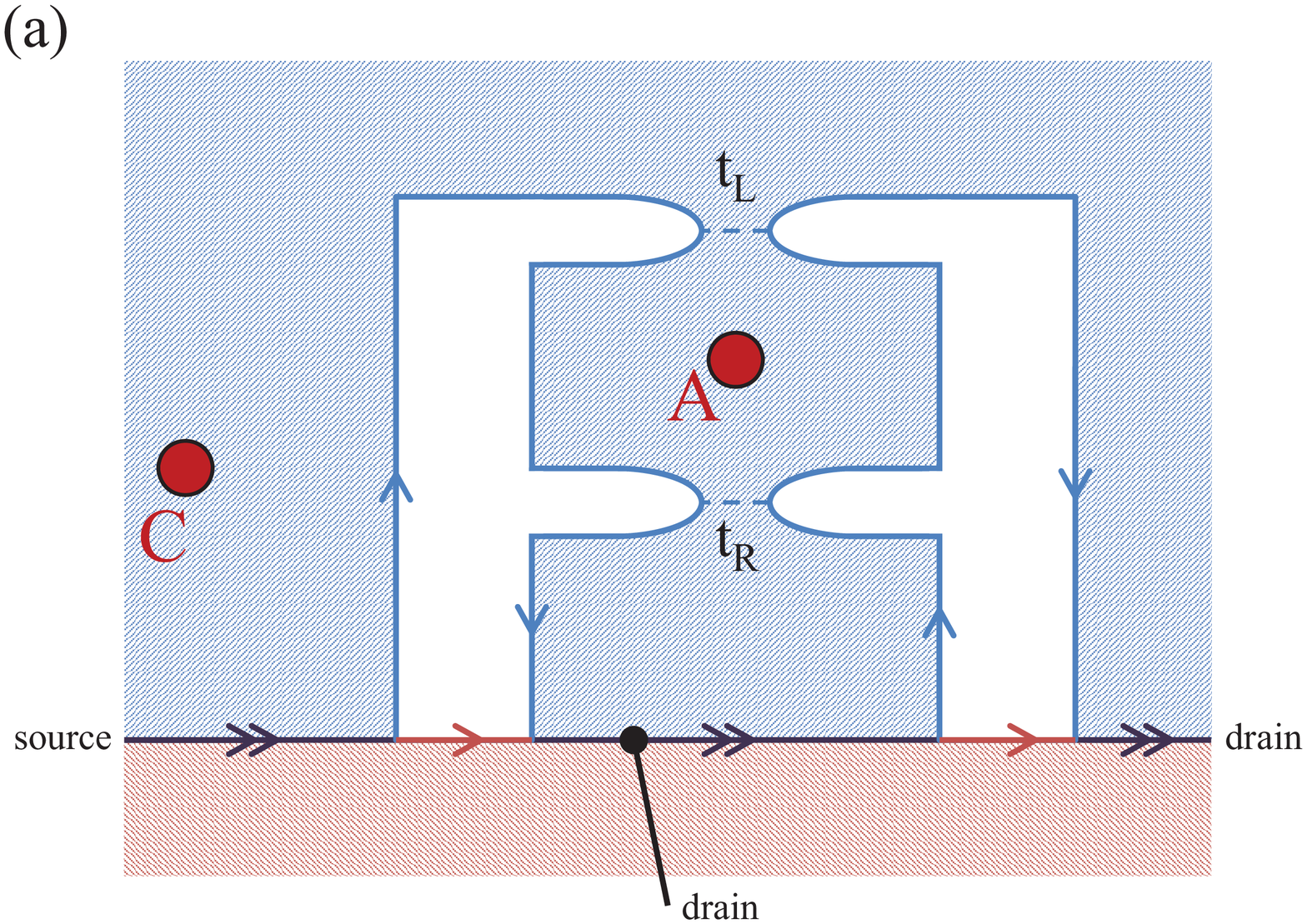}
  \includegraphics[scale=0.26]{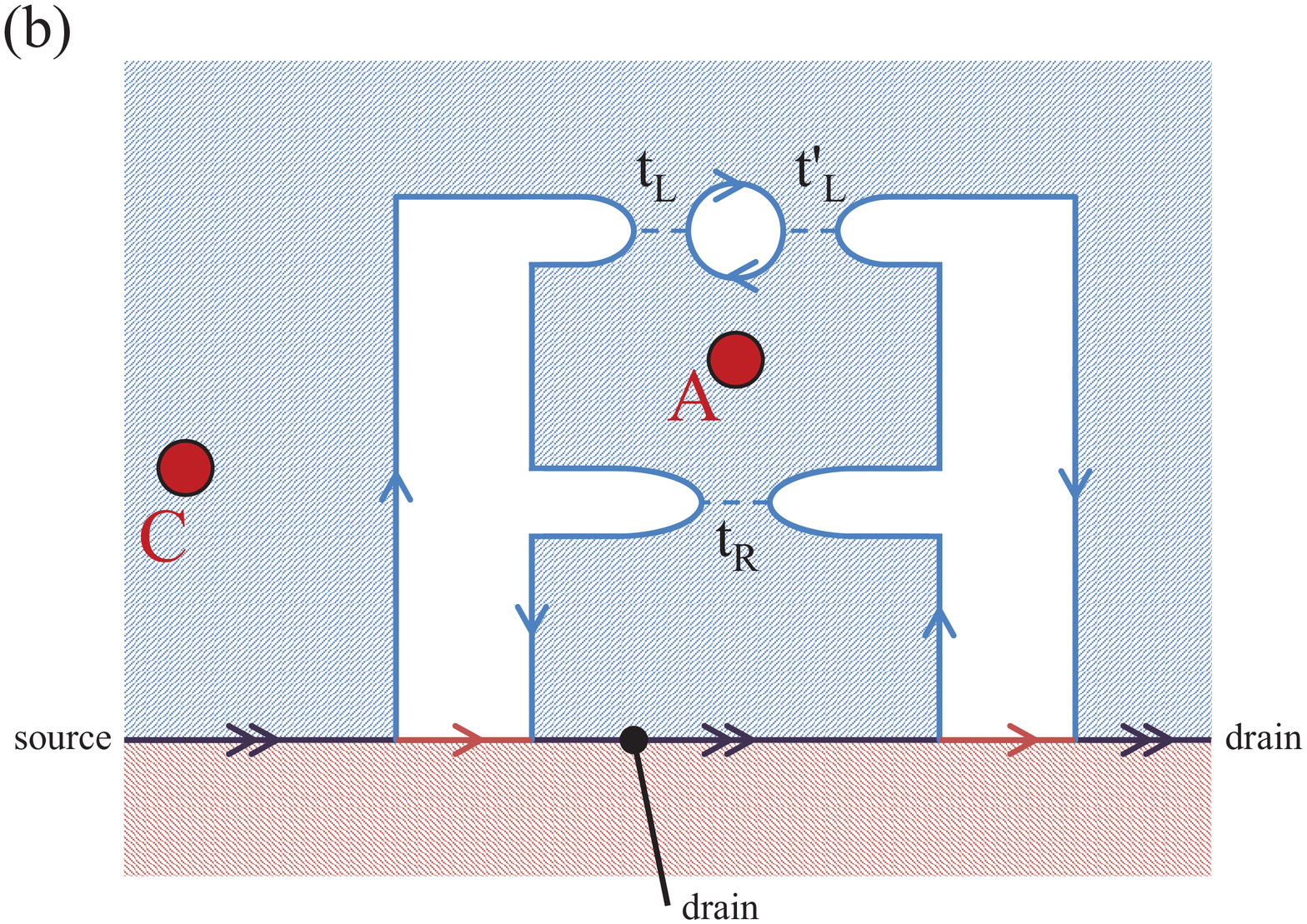}
  \caption{(a) A Fabry-P\'{e}rot type double point contact interferometer for 2D $p_x + i p_y$ superconductors and Majorana heterostructures that utilizes a Dirac edge mode. The bulk $p_x +ip_y$ topological phase is indicated by the shaded blue region. A $p_x - i p_y$ topological phase, indicated by the shaded red region, is created adjacent to portions of the $p_x +ip_y$ phase in order to give rise to Dirac edge modes. The white regions are ``vacuum'' (topologically trivial phases). Majorana edge modes of the $p_x +ip_y$ and $p_x -ip_y$ phases are indicated by solid blue and red lines, respectively, with arrows indicating their respective chiralities. Solid purple lines with double arrows indicate Dirac edge modes formed from the combination of the two Majorana edge modes. Two junctions (point contacts) between edges are created with tunneling amplitudes $t_L$ and $t_R$, respectively, using electrostatic gates. (b) The double point contact interferometer is modified to include twisting in one of the paths by introducing an intermediate island in one of the point contacts using an electrostatic gate. By dynamically controlling the tunneling amplitudes $t_L$ onto the island and $t_{L}^{\prime}$ off of the island, one can allow a specified number of probe quasiparticles to tunneling onto the island and then circulate around the island (and hence each other) a desired number of times equal to the twisting number for the path before tunneling off the island to the other edge.}
  \label{fig:DPC_int_pwave}
\end{center}
\end{figure}

The track switching method of generating twisting/braiding can be incorporated in a Fabry-P\'{e}rot type double point contact interferometers using the track switching method by introducing a twisting island into one of the point contacts, as illustrated in Fig.~\ref{fig:DPC_int_pwave}(b). For this configuration with a twisting island in the left point contact, one must dynamically control the tunneling amplitudes $t_L$ between the lower edge of the system and the island and $t_L^{\prime}$ between the island and the upper edge in the following manner. First, one should start with $t_L \neq 0$ and $t_L^{\prime}=0$ (up to exponentially suppressed corrections) to allow the desired number $N$ of quasiparticles to tunnel from the lower edge onto the twisting island. Next, one should turn off the tunneling ($t_L=0$ and $t_L^{\prime}=0$) for an amount of time that allows the $N$ quasiparticles to circulate the twisting island $m$ times, generating the $m$ twists/pure braids of these quasiparticles with each other. Finally, one should turn on $t_L^{\prime} \neq 0$ (while leaving $t_L=0$), so that the $N$ quasiparticles on the twisting island can tunnel off to the upper edge of the system. The tunneling amplitudes may be controlled, for example, by using gates with appropriate timing.

For $p_x + i p_y$ superconductors and Majorana heterostructures, the Fabry-P\'{e}rot type double point contact interferometer of Fig.~\ref{fig:DPC_int} may be difficult to implement, because the edge modes are electrically neutral and would require more difficult heat current measurements. To circumvent this difficulty, one can juxtapose an opposite chirality $p_x - i p_y$ version of the system with the $p_x + i p_y$ system in order to create a Dirac edge mode from the two co-propagating Majorana edge modes. These Dirac edge modes can then be coupled to electrical current and utilized to measure the edge currents for a similarly designed interferometer in the $p_x + i p_y$ system, as shown in Fig.~\ref{fig:DPC_int_pwave}. Track switching and twisting/braiding of probe quasiparticles can be incorporated in precisely the same manner described in the previous paragraph.

\subsection{Implementation Issues and Obstacles}
\label{sec:implmentation}

There are a number of issues and obstacles that must be addressed in any physical implementation of a twisted interferometer that result both from the nature of the twisted interferometry operation itself and the details of the physical system in which it is implemented. Perhaps the most imposing requirement of the twisted interferometry operation is that all the probe anyons passing through a twisted path must do so concurrently, in order for them to all twist or braid with each other, as previously described. This leads to many of the implementation issues and obstacles that we discuss here.

We first point out that the twisted interferometry operation is not topologically protected in the same sense as standard untwisted anyon interferometry. In particular, unlike an untwisted interferometer, one must fix the number of probe anyons $N$ that are sent through the twisted interferometer at the start of the operation and cannot simply run the interferometer multiple times to improve the convergence of the measurement. Running a twisted interferometer multiple times does not generate the same operation as running it once with a greater number of probes, since the probe quasiparticles of different runs cannot twist/braid with each other in the same way as those in a single run. Prior to running a twisted interferometry operation, one must determine a sufficient value for $N$, based on the interferometer's parameters (in particular, the probabilities $\tilde{p}_\kappa$) and the desired level of confidence $1-\alpha$ of the operation. If $N$ is not chosen to be sufficiently large for a given application of twisted interferometry, the result cannot be salvaged. Fortunately, the convergence of the interferometry process is exponentially fast in number of probes sent through an interferometer~\cite{Bonderson07b,Bonderson07c}, so twisted interferometry is still topologically protected in the sense that the error of the operation will be exponentially suppressed as $N$ is increased. Under good circumstances, $N$ will not need to be excessively large, but there will be a lower bound $N \gtrsim \left( \frac{z^{\ast}_{\alpha/2}}{ \Delta \tilde{p}} \right)^2$, where $z^{\ast}_{\alpha/2} = \sqrt{2} \text{erf}^{-1}(1-\alpha)$ and $\Delta \tilde{p} = \left| \tilde{p}_\kappa - \tilde{p}_{\kappa^{\prime}} \right|$  is the difference between the values of probabilities for $\kappa$ and $\kappa^{\prime}$ that must be distinguished~\cite{Bonderson07b,Bonderson07c}. The realistic functioning of an interferometer will involve many sources of dephasing and loss of coherence (some of which are discussed below), which reduce the visibility of quantum interference. This effectively multiplies the interference terms and $\Delta \tilde{p}$ by a suppressive factor $Q < 1$, which impairs the distinguishability of measurement outcomes and increases the number of probe anyons needed.

It is also important to avoid operation errors that may arise from probe-probe interactions. This means the interferometers cannot operate in the plane wave limit, but, rather, will require well-localized probe quasiparticles that are sufficiently separated, so that the effect of probe-probe interactions do not cause significant errors in the functioning of the interferometer. Probe-probe interactions can cause problematic effects in a number of ways, such as topological interactions, which generically split the degeneracy of the topological state space non-Abelian anyons~\cite{Bonderson09}, or attractions/repulsions between probes quasiparticles that effectively alter the transmission/reflection amplitudes in a way that can differ from one probe quasiparticle to another. The separation $\ell$ between probe quasiparticles is bounded by the correlation length $\xi$ for quasiparticles, above which the effect of interactions are negligible, i.e. $\ell \gtrsim \xi$.

As previously described, all probe quasiparticles must enter the twisting loop before any of them exit the twisting loop. For a twisting loop of length $L_{\text{t}}$, this means the separation between probe quasiparticles on the loop will be roughly $\ell \lesssim L_{\text{t}}/N$. Thus, there will also be an upper bound $N \lesssim L_{\text{t}}/\xi$ on the number of probe anyons. This, together with the velocity at which probe quasiparticles travel through the interferometer, will also determine the size of the time interval for which the probes should be sent into the interferometer. One can mitigate this upper bound, in part, by increasing the length of the twisting loop. However, this length is also bounded. Obviously, the length of the loop will also be limited by the system size, together with the constraint that the path not approach within a correlation length $\xi$ of itself (to avoid interactions). Likely more limiting is the constraint that the length of the loop should be shorter than the coherence length $L_{\phi}$ of interference for the probe quasiparticle, which is dictated, for example, by the temperature and properties of the edge modes that provide the beam of probe anyons~\cite{Bishara08a}. Thus, the number of probe anyons will have an upper bound that depends on the details of the system and interferometer geometry, in addition to the lower bound dictated by the desired operation precision.

Since the probe quasiparticles must be well-localized and sent into the interferometer concurrently, the device should be designed so that the lengths of the two paths through the interferometer are approximately equal, with reasonably high precision (assuming equivalent average probe velocities for the two paths). This is required to ensure that the predominant contribution is due to processes in which each probe anyon is interfering with itself traveling through the two paths, rather than other probe anyons. Otherwise, the contribution from topologically distinct processes would result in errors in the twisted interferometry operation.

For twisted track implementations of probe anyon twisting/braiding, errors will be introduced if probe anyons go the wrong way at the self crossings of the track. These self crossing junctions must be engineered so that the amplitude of transmission (going straight through the junction) is close to $1$ and that of making turns are highly suppressed. For track switching implementations of probe anyon twisting/braiding, precision is required for switching gates' timing and one must be careful to ensure that all probes anyons exit the twisting loop, as leaving any behind will introduce errors. Additionally, it is important that the switching gates be adequately isolated from from relevant computational and probe quasiparticles, so that the sharply pulsed application of gates does not generate errors. It is also important that the tracks, in particular the twisting loop or island, not detect the presence or absence of probe quasiparticles, through energetic or other means, otherwise it will suppress quantum interference.

\begin{figure}[t!]
\begin{center}
\labellist \normalsize\hair 2pt
  \pinlabel $\text{storage ring}$ at 85 115
  \pinlabel $\phi_0$ at 90 80
  \pinlabel $\phi_1$ at -40 80
  \pinlabel $\text{top gate}$ at 115 190
  \pinlabel $\text{$\longleftarrow$ top gate}$ at 238 77
  \pinlabel $\text{$\longleftarrow$ track to}$ at 225 165
  \pinlabel $\text{interferometer}$ at 260 150
  \pinlabel $\downarrow$ at 112 180
\endlabellist
\includegraphics[scale=0.8]{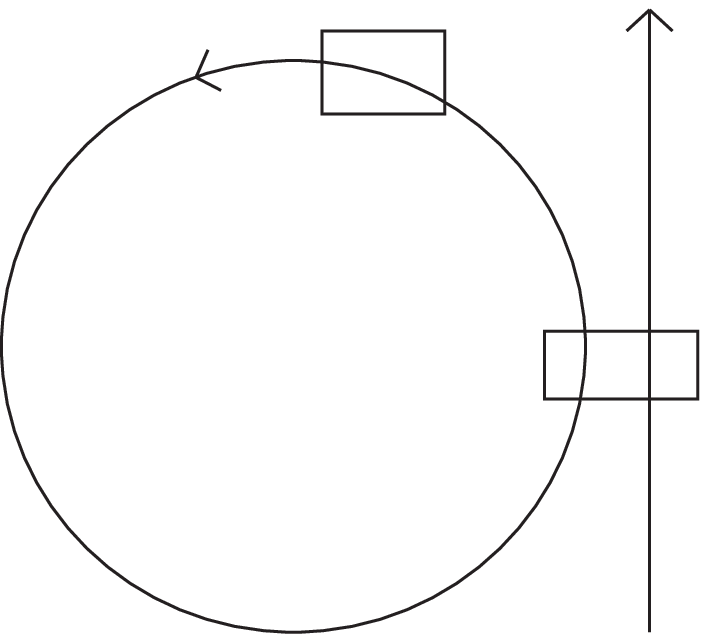}
  \caption{A Josephson vortex storage ring for 2D $p_x + i p_y$ superconductors and Majorana heterostructures. Josephson vortices can be generated in the ring and stored circulating around the ring until ready for use. Application of track switching gates sends the Josephson vortices to an interferometer for use as probe quasiparticles.}
\label{fig_storage_ring}
\end{center}
\end{figure}

The creation and control of concentrated bursts of well-localized and well-separated probe anyons will also pose a substantial technological challenge. This will certainly involve fairly precise pulsing of gates and/or current sources and likely require further methods to ensure the desired non-Abelian quasiparticles are being produced. We speculate that it may be possible to do this in some systems, such as quantum Hall states, by adapting ideas similar to quantum-dot turnstiles~\cite{Kouwenhoven91} to (fractional) anyonic quasiparticles.
For 2D $p_x + i p_y$ topological superconductors and Majorana heterostructure, quantum tunneling of Abrikosov vortices has not been experimentally observed, so they are likely impractical for use as probe quasiparticles in an interferometer. On the other hand, quantum tunneling has been observed for Josephson vortices~\cite{Wallraff-PHD2000,Wallraff03}, so these should be an acceptable choice for probe quasiparticles. Josephson vortices can be generated in the Josephson junction tracks by applying pulsed currents, as experimentally demonstrated in~\cite{Sakai84,Sakai87}.
Given the ability to generate these vortices with sufficient control, it may be useful to also utilize a sort of Josephson vortex storage ring, as shown schematically in Fig.~\ref{fig_storage_ring}. In this setup, a internal/external phase differential $\phi_1>\phi_0$ will keep Josephson vortices in circulation around the circular track. Then, when one is ready to use the Josephson vortices, one can apply track switching, using a pair of precisely timed electrostatic top gates to break the ring and lower the tunneling barrier to a second track leading to the interferometer.

As previously mentioned, interactions between probe quasiparticles are an important source of errors that must be avoided. This is a particularly difficult problem for physically realistic topological systems, such as $p_x +ip_y$ topological superconductors, Majorana heterostructures, and even (electronic) quantum Hall states, since they are actually quasi-topological phases of matter~\cite{Bonderson12d}. True topological phases of matter are gapped and the interactions between quasiparticles decay exponentially as their separation $\ell$ increases, giving $O(e^{-\ell/\xi})$ errors. Quasi-topological phases, however, include gapless degrees of freedom, which, depending on how they couple to the topological degrees of freedom, may preserve or destroy their topological properties and protection. Consequently, quasiparticles of quasi-topological phases will generally experience interactions with each other that falls off as a power law $O(\ell^{-z})$, rather than exponentially. In the more fortunate cases (which are what people usually refer to as ``topological phases''), the gapless excitations couple to the topological degrees of freedom in a seemingly benign way that does not destroy the topological properties of bulk quasiparticles, i.e. the topological state space and braiding statistics of non-Abelian quasiparticles are preserved. However, even in these cases, there can be simple attractive/repulsive interactions, which, even though they do not differentiate between different topological states, will still causes problems for implementing twisted interferometry (as previously discussed). For example, Josephson vortices in topological superconductors and Majorana heterostructures couple to electromagnetism, which results in $1/\ell$ interactions between them in a planar geometry. Electrically charged quasiparticles in quantum Hall states will experience a similar electromagnetic interaction. Some quantum Hall states are electrically neutral or have an electrically neutral subsector that could potentially be employed, but figuring out how to use this (and not the electrically charged quasiparticles) also poses a challenge. In practice, it may be possible to make power law interactions sufficiently small, e.g. if probe quasiparticle separations can be made large enough, if the coefficient of interaction is or can be made small, and/or if the power of interaction can somehow be increased to make the falloff faster. However, this depends on specific details of the device and microscopic properties of the particular system being used, and thus involves significant non-topological physics.

In summary, there are a number of crucial engineering issues and obstacles for implementing twisted interferometry that must be addressed more adequately than we have done here. We also leave detailed analysis of the functioning of twisted interferometers under non-ideal conditions as a topic beyond the scope of this paper.

\section*{Acknowledgements}

We thank R. Lutchyn, C. Nayak, K. Shtengel, and J. Slingerland for illuminating discussions. P.~B. and M.~F. thank the Aspen Center for Physics for hospitality and support under the NSF Grant No. 1066293.


\end{document}